\documentclass[12pt]{article}
\pdfoutput=1
\usepackage[margin=1in]{geometry} 
\usepackage{natbib}

\usepackage{appendix,soul}
	\renewcommand{\appendixpagename}{Appendices -- for online publication}
\usepackage{graphicx}
\usepackage{tabularx,booktabs,array,dcolumn}
    \newcolumntype{L}[1]{>{\raggedright\let\newline\\\arraybackslash\hspace{0pt}}m{#1}}
    \newcolumntype{C}[1]{>{\centering\let\newline\\\arraybackslash\hspace{0pt}}m{#1}}
    \newcolumntype{R}[1]{>{\raggedleft\let\newline\\\arraybackslash\hspace{0pt}}m{#1}}
    \newcolumntype{d}[1]{D{.}{.}{0}}
\usepackage{caption,subcaption}
\usepackage{chngpage}
\usepackage{float,rotating,placeins}
\usepackage{color,xcolor}
\usepackage{amsmath,amsthm,amssymb,bbold,amsfonts,mathtools}
    \allowdisplaybreaks
\usepackage{setspace}
\usepackage[bottom]{footmisc}

\usepackage[pdftex,
			pdfauthor={Anisha Ghosh, Alexandros Theloudis},
            pdftitle={Consumption Partial Insurance in the Presence of Tail Income Risk},
            pdfsubject={Consumption Partial Insurance in the Presence of Tail Income Risk},
            pdfkeywords={Income risk, skewness, kurtosis, consumption, partial insurance, PSID},
			hyperfootnotes=false,bookmarksopen=true]{hyperref}
\hypersetup{colorlinks=true,allcolors=blue}

\DeclareMathOperator*{\E}{\mathbb{E}}
\DeclareMathOperator*{\Var}{\mathrm{Var}}
\DeclareMathOperator*{\Cov}{\mathrm{Cov}}

\DeclareMathOperator*{\Skew}{\mathrm{Skew}}
\DeclareMathOperator*{\Kurt}{\mathrm{Kurt}}

\newcommand{\nsz}{\normalsize}
\newcommand{\fsz}{\footnotesize}
\newcommand{\ssz}{\scriptsize}

\newcommand{\mcl}{\multicolumn}

\title{\textbf{Consumption Partial Insurance\\in the Presence of Tail Income Risk}\thanks{We are grateful to the editor, Greg Kaplan, and several referees for detailed comments. We also thank Manuel Arellano, Richard Blundell, Jeff Campbell, George Constantinides, Mariacristina De Nardi, Dmytro Hryshko, Christian Julliard, Luigi Pistaferri, Chris Telmer, and audiences in Antwerp, Cambridge, Freiburg, ULB, Zaragoza, the Structural Econometrics Group at Tilburg, the Barcelona Summer Forum (Macroeconomics and Social Insurance), the Econometric Society winter meeting in Manchester, the UCL Alumni conference, the New Advances in Family Economics workshop in Paris, and the Midwest Macro meeting in Richmond for many helpful comments. We finally thank Ties Schalij for excellent research assistance.}}

\author{
Anisha Ghosh\thanks{Desautels Faculty of Management, McGill University; email: \href{mailto:anisha.ghosh@mcgill.ca}{anisha.ghosh@mcgill.ca}.} 
\and 
Alexandros Theloudis\thanks{Department of Econometrics \& Operations Research, Tilburg University; email: \href{mailto:a.theloudis@gmail.com}{a.theloudis@gmail.com}.}
}
\date{November 11, 2025}

\begin{document}
\maketitle

\begin{abstract}
We propose a simple framework to measure consumption insurance against income shocks that accounts for higher-order moments of the income distribution. We derive a nonlinear consumption function in which the extent of insurance varies with both the sign and magnitude of shocks. Using recent PSID data, we estimate an asymmetric pass-through of bad versus good permanent shocks -- 17\% of a $3\sigma$ negative shock transmits to consumption versus 9\% of an equal-sized positive shock -- with greater pass-through as shocks worsen. These consumption transmission rates further vary by age, wealth, and along the income distribution. Our results align with survey responses to hypothetical events and suggest that tail risk matters substantially for consumption. \\[5pt]

\end{abstract}


\pagebreak
\section{Introduction}

How does consumption respond to income shocks? This question is central to understanding the welfare effects of shifts in the income distribution, i.e., the link between income and consumption inequality, how families cope after adverse events, government insurance design, and the dynamics of business cycles. A consistent empirical finding is that consumption is partially insured against income shocks, so income fluctuations translate into consumption fluctuations less than one-to-one \citep[e.g.,][]{Blundell_Pistaferri_Preston2008,Heathcote_Storesletten_Violante2014}.\footnote{The study of consumption insurance dates back to tests of the permanent income \citep[][]{Deaton_Paxson1994} and complete markets hypotheses \citep[][]{Cochrane1991_TestConsumptionInsurance,AttanasioDavis1996_RelativeWageMovements,Hayashi_Altonji_Kotlikoff1996,BravConstantinidesGeczy_2002_AssetPricing}. \citet{JappelliPistaferri2010Review} and \citet{CrawleyTheloudis2024_IncomeShocks} review this literature.} With a few exceptions discussed subsequently, most measures of consumption insurance do not distinguish between good and bad shocks, or between small and large ones. However, recent evidence shows that income risk has a \emph{long and fat left} tail; that is, there is an asymmetrically large probability that households face some really bad income shocks. This paper proposes a simple framework to measure the consumption response to income shocks, precisely when the distribution of said shocks may exhibit long and fat tails.

In a seminal paper, \cite{Blundell_Pistaferri_Preston2008} -- henceforth BPP -- introduce an empirical methodology to measure the pass-through of income shocks of varying persistence into consumption. Using Panel Study of Income Dynamics (PSID) and Consumer Expenditure Survey (CEX) data over 1980--1992, BPP find evidence of partial insurance to permanent shocks and full insurance to transitory ones. BPP rely on a workhorse covariance between income and consumption growth, namely a \emph{second} moment of their joint distribution. The idea is that the extent to which income growth (driven by income shocks) varies with consumption growth reflects the strength of the transmission of income shocks into consumption. 

In a separate strand of literature, \citet{Guvenen_Karahan_Ozkan_Song2021} use data from the U.S. Social Security Administration and establish important new facts about the distribution of unexplained income growth, namely the distribution of income shocks. They show that income growth is very negatively skewed, i.e., it has a long left tail, so far more people experience large negative than large positive shocks. Moreover, income growth exhibits excess kurtosis, i.e., fat tails, so far more people face either small or large shocks, than moderate ones, relative to a Gaussian density. These are now benchmark facts about income dynamics, observed in many countries \citep[e.g.,][compare the U.S. and the Netherlands]{DeNardi_EtAl2021_Netherlands_US}. 

If most income shocks away from the mean are large and bad shocks, their effect on consumption might, intuitively, differ from that of small/moderate shocks. In other words, left skewness and excess kurtosis, which are statements about the third and fourth moments of the income distribution respectively, may matter for the pass-through of shocks into consumption. This is precisely what we investigate in this paper. Specifically, we measure the degree of consumption insurance to income shocks of varying persistence, accounting not only for second- but also higher-order moments of their distribution (i.e., tail shocks). 

We offer three main contributions: 

First, we re-estimate BPP using recent PSID data over 1999--2019, targeting the original second moments of income and consumption, and compare the results with those based on imputed consumption from the contemporaneous CEX. The PSID previously lacked detailed consumption data, so BPP relied on an imputation from the CEX. However, such data are now internally available in the PSID. We subsequently use this data to study tail income risk, so it is natural to start with a `replication' of BPP using the updated PSID. This exercise is insightful because of the different time period and the comparison with the CEX. 

Second, we introduce into BPP the third and fourth moments of the joint distribution of income and consumption, in addition to the second moments above. This is our first attempt to go beyond the workhorse income-consumption covariance and measure partial insurance targeting additional information in the tails of the distribution.

Third, we generalize the method of measuring consumption insurance to explicitly permit a role for tail income risk. BPP log-linearize the policy rule of a lifecycle consumption/savings problem to obtain a consumption function that depends \emph{linearly} on income shocks. Our first two contributions above revolve firmly around this linear framework. However, linearity may be a poor approximation to behavior when income is subject to tail shocks. We thus derive a higher-order (quadratic) consumption function that allows for \emph{nonlinear} transmission of shocks. We characterize and empirically measure the consumption pass-through of income shocks, which now depends explicitly on the shocks' \emph{magnitude} and \emph{sign}. While a nonlinear consumption function has been estimated before -- most notably in \citet{Arellano_Blundell_Bonhomme2017}, henceforth ABB -- our simple framework offers key advantages: it separates the nature of income risk from the nature of the consumption response as distinct drivers of consumption growth along the income, age, and wealth distributions; it is far less data demanding (second moments of consumption and up to fourth moments of income suffice, with no need to observe the full consumption-income joint distribution or wealth); and it links identification and estimation transparently, aiding interpretation. We elaborate on these points below.

Our core empirical exercises use income and consumption panel data from the PSID from 1999 to 2019. We focus on a representative sample of stable married couples that is similar to the sample in BPP. We show that unexplained income growth in this sample features large negative skewness and excess kurtosis, qualitatively and numerically close to the skewness and kurtosis in the administrative data of \citet{Guvenen_Karahan_Ozkan_Song2021}. For the re-estimation of BPP in the recent years, we also use a comparable sample from the CEX.

We obtain four main findings:

First, permanent and transitory shocks to disposable household income are negatively skewed, with skewness coefficients at $-0.83$ and $-1.43$, respectively, and highly leptokurtic, with kurtosis coefficients at $44.39$ and $49.23$, respectively. Left skewness implies that a negative shock is typically more unsettling than a positive shock, because it is farther away from the zero mean. Excess kurtosis implies that more shocks are concentrated either in the middle of the distribution or far out in the tails, relative to the shoulders. Taken together, large left skewness and kurtosis suggest that for most households the realizations of income shocks tend to be either small (i.e., close to the zero mean) or far out in the left tail. However, these features vary by age and by the household's position in the income distribution.

Second, in estimating BPP using the updated PSID in recent years, we find full insurance to transitory shocks and a small pass-through of permanent shocks at 0.15 versus 0.64 in BPP. A 10\% permanent income cut thus reduces consumption by only 1.5\%, vis-\`a-vis 6.4\% in BPP. In other words, we find much larger partial insurance to permanent shocks in recent years than originally found in 1980--1992.\footnote{Using recent PSID data but a different method, ABB and \citet{Arellano_Blundell_Bonhomme_Light2021} find a similarly low pass-through of persistent shocks. We discuss this further when we present our empirical results.} We attribute about a third of the gap from BPP to the consumption imputation from the CEX that BPP use. The imputation brings in large amounts of measurement error, inflating the covariance between income and consumption growth. Another third of the gap arises because the modern PSID provides data biennially, versus annually previously, which tends to mute the transmission of higher frequency shocks. The last third reflects the different calendar times.

Third, introducing the third and fourth moments of income and consumption growth into BPP's linear consumption setting does not change our estimates of partial insurance. This is counterintuitive in light of the literature that highlights, in various settings, the large welfare costs of tail risk \citep[e.g.,][]{Barro2009,Busch_Ludwig2024}. One would thus expect that targeting higher-order moments reduces the degree of partial insurance, which is not what we find. This result is suggestive of a potential misspecification in the linear consumption function, which motivates our generalization of the methodology through the introduction of a higher-order consumption specification subsequently.

Fourth, in estimating our quadratic consumption function, we find an \emph{asymmetry} in the pass-through of large negative shocks versus positive or smaller ones. In this environment, the pass-through depends on the magnitude and sign of shocks; and targeting higher-order income moments enables the identification of the transmission parameters. We draw three main conclusions: \emph{1.)} Bad \emph{permanent} shocks are more unsettling than equal-sized good ones because they have larger transmission rates into consumption -- a three standard deviations negative shock (a 50\% permanent income cut) has a pass-through of $0.17$ (17\% of the shock transmits into consumption), while a positive shock of similar magnitude has a pass-through of $0.09$. \emph{2.)} The pass-through of bad shocks increases with the severity of the shock. Bad permanent shocks are thus double hurtful: not only are they more extreme than good ones (they lie farther out in the left tail), but they also transmit more strongly as their magnitude worsens. \emph{3.)} \emph{Transitory} shocks are fully insured in the full sample, at least from a statistical point of view, although the point estimates of their transmission parameters suggest that large transitory windfalls shift consumption considerably, consistent with quasi-experimental evidence of large responses to transitory gains \citep[e.g.,][]{FagerengEtAl_2021LotteryWins}. These and previous findings are robust to allowing for shocks that have linearly persistent, though not necessarily permanent, effects on income. Overall, the results are consistent with an environment in which households save a good income shock but dissave in the presence of a bad one. There is clearly a limit to how much one can dissave or borrow, so a large negative shock ends up affecting consumption by relatively more.

The asymmetric transmission of bad versus good permanent shocks varies with age and the household's position in the income distribution. While older households insure \emph{average} permanent shocks better than younger ones, they respond more strongly to \emph{larger} negative shocks. Moreover, bad permanent shocks in the lower tail of income risk cause more harm to the rich (households in the top of the income distribution) than to the poor (households in the bottom); the converse is true for good shocks. 

How a shock from a given position in the distribution affects consumption depends on both the sign/size of the shock \emph{and} the extent of its transmission. A key advantage of our moments-based approach is that it separates income risk and consumption transmission, allowing us to analyze each independently. Among the rich, the asymmetry is driven by an increasing severity of downside risk along the income distribution, rather than an increasing consumption transmission. Among the old, however, the asymmetry is because both the rate of transmission \emph{and} the severity of downwards income risk increase with age.  

The asymmetric transmission of bad versus good permanent shocks also varies with wealth and education. Transmission of a given average or tail shock decreases as we move from the least to the most wealthy/educated, so wealthier and highly educated households are better insured to all types of permanent shocks. However, the nature of income risk differs vastly across groups: the wealthy/educated face more severe downwards income risk, ultimately leading to larger consumption declines when bad shocks hit.

With left skewness and excess kurtosis, large negative shocks are more likely and more severe than positive ones, so households would rather remove such tail risk from their portfolio. A back-of-the-envelop calculation suggests that households facing large downwards risk would give up 4\%-8\% of lifetime consumption when income risk transmits linearly and rises to 6\%-12\%, almost up to 1/8 of lifetime consumption, in the presence of nonlinear transmission. There are several estimates of the welfare costs of income risk in the literature. Two settings in which the baseline environment concerns idiosyncratic tail income risk of similar nature and severity to ours are \citet{Busch_Ludwig2024} and \citet{GuvenenOzkanMadera2024NonGaussianRisk}. Agents in the former paper sacrifice between 0.5\% and 14\% of lifetime consumption to live in a world with Gaussian risk (a similar counterfactual to ours), while households in the latter sacrifice almost 22\% of consumption to live in a world \emph{without} risk.\footnote{These papers calculate welfare costs using full structural models in which the consumption responses depend on endogenously accumulated wealth. By contrast, we rely on approximate consumption rules, which take wealth as given. We show in appendix \ref{Appendix::WelfareCosts} that this is unlikely to be quantitatively important.}\textsuperscript{,}\footnote{Welfare cost estimates of \emph{aggregate} risk include \citet{Barro2009}, who find that agents can give up about 20\% of GDP each year to eliminate rare disasters, and \citet{Georgarakos2025HowCostlyBusinessCycle}, who find that households can sacrifice 5\%-6\% of lifetime consumption to eliminate business cycle fluctuations.}

There is a large literature in macro and labor economics that studies the transmission of income shocks into consumption, i.e., the link between income and consumption inequality. There are two distinct strands in this literature: a quasi-reduced form branch that \emph{measures} the degree of partial insurance in the data without fully specifying the underlying insurance mechanisms, and a structural branch that fully \emph{models} specific insurance channels. The present paper falls in the first branch.

The seminal paper in the first branch, namely BPP, sparked a large volume of research.\footnote{\cite{Hall_Mishkin1982}, \citet{Deaton_Paxson1994} and \citet{Blundell_Preston1998} characterize the empirical joint distribution of consumption and income, so they are early precursors in this literature.} \citet{Blundell_Pistaferri_Saporta-Eksten2016} measure partial insurance against wage risk, \citet{Theloudis2021} allows for preference heterogeneity in partial insurance, and \cite{Commault2022} allows consumption to respond to past transitory shocks and finds a strong response similar to what experimental studies find. \cite{Hryshko_Manovskii2022} identify sets of households in the PSID with different degrees of insurance based on certain features of their incomes, and \cite{Crawley_Kuchler2023} address neglected time aggregation in the PSID data. These papers restrict income shocks to transmit linearly into consumption. ABB and \citet{Arellano_Blundell_Bonhomme_Light2021} generalize the income and consumption processes by letting income feature nonlinear persistence, which varies with the level of income and its shock, and nonlinear transmission that is more flexible than ours. Using recent PSID data and quantile methods, they measure the marginal propensity to consume from income along the income distribution. To quantify the pass-through of income \emph{shocks}, they simulate the model and find asymmetries in the transmission of bad and good shocks, or between rich and poor households, similar to ours. 

Our third contribution shares ABB's goal of measuring nonlinear transmission, but differs in three key dimensions. First, ABB compare consumption growth across quantiles of income and income shocks, without distinguishing between the nature of income risk and that of the consumption response. For example, the rich appear to update consumption more strongly than the poor after a lower-tail shock, but this can reflect more severe downside risk or/and a stronger transmission independent of the shock. Our moments-based approach separates these channels, allowing us to analyze each independently. In this example, we find richer households respond less to a given shock (\emph{more} insurance), but face worse downside risk, leading to larger overall consumption changes. Second, the quantile methods in ABB require observing the full consumption-income joint distribution, and wealth, while our approach requires only second moments of consumption and up to fourth moments of income. This makes our framework less data demanding, appealing both on its own and because, unlike income, we cannot benchmark higher-order consumption moments against population data. The price of our econometrically more accessible approach is that, unlike ABB, we maintain a linear income process (but subject to tail shocks). Yet, we find that nonlinear persistence is not crucial for the main consumption results, i.e., the asymmetry across different shocks and income levels. Third, our consumption insurance parameters are directly estimable in the data without simulations, and tightly linked to their identification. They are comparable to the existing literature on partial insurance and `portable' as they describe transparently how shocks of different signs/magnitudes shift consumption.

The second branch of the literature studies specific insurance mechanisms using calibrated consumption lifecycle models.\footnote{\cite{Gourinchas_Parker2002} and \citet{Krueger_Perri2006} are early precursors in this strand.} Early works focus on self-insurance: \citet{Kaplan_Violante2010} find less insurance than BPP, while \cite{Guvenen_Smith2014}, who jointly estimate income risk and self-insurance, find more. Recent studies examine the insurance implications of labor supply \citep{Heathcote_Storesletten_Violante2014}, durables \citep{Madera2019_ConsumptionResponseTailShocks}, health \citep{Blundell_Borella_Commault_DeNardi2020}, and added workers \citep{WuKrueger2021_ConsumptionInsurance}. Except for \citet{Madera2019_ConsumptionResponseTailShocks}, these papers abstract from higher income moments due to computational reasons. \citet{DeNardi_Fella_PazPardo2019} embed a non-Gaussian income process into a consumption/savings model and show that tail risk strengthens precautionary motives and increases partial insurance. In a similar model, \citet{Busch_Ludwig2024} target income skewness and kurtosis, distinguishing good from bad shocks and showing the latter are worse insured. \citet{GuvenenOzkanMadera2024NonGaussianRisk} find that tail shocks are worse insured than average ones but BPP's benchmark method overstates insurance against them. Unlike this quantitative literature, we take no stance on specific insurance mechanisms; we measure aggregate insurance across all mechanisms while accounting for higher-order moments and keeping our model parametrically unspecified.

The asymmetric response to bad and good shocks is recently found in surveys that elicit responses to hypothetical events. \cite{Bunn_LeRoux_Reinold_Surico2018} find that the marginal propensity to consume from bad shocks is larger than from good shocks; \citet{ChristelisEtAl_2019_AsymmetricConsumptionEffects} and \cite{Fuster_Kaplan_Zafar2021} find that responses to losses are larger than to gains, and the response increases with the severity of the loss. These papers consider observed hypothetical transitory shocks, while we document similar patterns in response to unobserved \emph{permanent} shocks.\footnote{Given our finding of heterogeneous responses to shocks, our paper is also broadly related to the literature that studies MPC heterogeneity \citep[e.g.,][]{Lewis_Melcangi_Pilossoph2022,Crawley_Kuchler2023}.}

The paper is finally related to the newer income dynamics literature that focuses on non-Gaussian aspects. \cite{Geweke_Keane2000} and \cite{Bonhomme_Robin2010} use the PSID to document, respectively, left skewness and excess kurtosis in earnings growth, while \cite{Guvenen_Karahan_Ozkan_Song2021} provide richer evidence from administrative data. \cite{Guvenen_Ozkan_Song2014} and \cite{Busch_Domeij_Guvenen_Madera2018} study the cyclical nature of skewness in several countries. While these papers are mostly descriptive of the higher-order features of income, we further document the implications for consumption. We thus contribute to a growing literature on the implications of tail risk, e.g., for growth \citep{Barro2009}, asset prices \citep{Backus_Chernov_Martin2011,Julliard_Ghosh2012,Constantinides_Ghosh2017}, business cycles \citep{McKay2017}, and labor market dynamics \citep{Ai_Bhandari2021}.

In the rest of the paper, section \ref{Sec::Model} introduces the income and consumption processes and section \ref{Sec::Identification} discusses their identification. Section \ref{Sec::Empirical_Implementation} presents the empirical implementation, section \ref{Sec::Results} discusses the results, section \ref{Sec::Income_Persistence} extends the main analysis to a more general income process, and section \ref{Sec::Conclusion} concludes. The appendix has additional derivations and results.

\section{An organizing framework: a lifecycle model with tail income risk}\label{Sec::Model}

The framework through which we organize the discussion of partial insurance to income shocks is a consumption lifecycle model with idiosyncratic income risk.

Household $i$ chooses a sequence of consumption $\{C_{it}\}_{t=0}^{T}$ and savings $\{A_{it+1}\}_{t=0}^{T}$ to maximize its expected discounted utility over its lifecycle. The problem is formulated as
\begin{equation}\label{Eq::Household_Problem}
    \max_{\{C_{it},A_{it+1}\}_{t=0}^{T}} {\E}_{0} \sum_{t=0}^{T} \beta^{t} U(C_{it};\mathbf{Z}_{it}),
\end{equation}
subject to the lifetime budget constraint
\begin{equation}\label{Eq::Budget_Constraint}
    A_{i0} + {\E}_{0} \sum_{t=0}^{T} \frac{Y_{it}}{(1+r)^t} = {\E}_{0} \sum_{t=0}^{T} \frac{C_{it}}{(1+r)^t},
\end{equation}
where $U(\cdot;\mathbf{Z}_{it})$ is utility over consumption, $\mathbf{Z}_{it}$ is a vector of observed taste shifters (e.g., age, education), $\beta$ is the discount factor, $r$ is the deterministic interest rate, $A_{i0}$ is initial financial wealth, ${\E}_{0}$ denotes expectations over uncertain future income $Y_{it}$, and $T$ is the end of the horizon. We assume that utility satisfies standard regularity conditions, in particular it is continuously twice differentiable, but we do not otherwise parameterize it.

\subsection{Income process}\label{SubSec::Model_Income_Process}

Income is the only source of (idiosyncratic) uncertainty that the household faces. Its logarithm is given by
\vspace{-1ex}
\begin{equation*}
    \ln Y_{it} = \mathbf{X}_{it}^\prime \boldsymbol{\delta} + P_{it} + v_{it},
\end{equation*}
where $\mathbf{X}_{it}$ is a vector of characteristics in the $t=0$ information set of the household that drive the deterministic profile of income over the lifecycle (e.g., age, education, race) and $\boldsymbol{\delta}$ is its loading factor. The remaining terms make up the stochastic component of income, consisting of persistent income $P_{it}$ and transitory shock $v_{it}$. 

We assume that $P_{it}$ has a unit root, namely $P_{it} = P_{it-1} + \zeta_{it}$, so in fact $P_{it}$ is permanent income and $\zeta_{it}$ is the permanent income shock. This modeling choice gives rise to the familiar \emph{permanent}-transitory formulation
\vspace{-1ex}
\begin{equation}\label{Eq::Income_Process}
	\Delta y_{it} =  \zeta_{it} + \Delta v_{it},
\vspace{-1ex}
\end{equation}
where $\Delta y_{it} = \Delta \ln Y_{it} - \Delta \mathbf{X}_{it}^\prime \boldsymbol{\delta}$ is income growth net of the deterministic profile, i.e., unexplained income growth, and $\Delta$ is the first difference operator.\footnote{See \citet{MeghirPistaferri2011} for a comprehensive review of modern income processes.} Section \ref{Sec::Income_Persistence} verifies our main results using a more general income process that relaxes the unit root assumption.

We let the permanent and transitory income shocks, $\zeta_{it}$ and $v_{it}$, be non-Gaussian, with first four central moments given by
\vspace{-1ex}
\begin{equation*}
    {\E}(\zeta_{it}^{m}) =
        \begin{cases}
            0  					    & \text{for } m=1 \\[-2pt]
            \sigma_{\zeta_t}^2      & \text{for } m=2 \\[-2pt]
            \gamma_{\zeta_t}        & \text{for } m=3 \\[-2pt]
            \kappa_{\zeta_t}        & \text{for } m=4 \\[-2pt]
        \end{cases} \qquad \text{and} \qquad
    {\E}(v_{it}^{m}) =
        \begin{cases}
            0  					    & \text{for } m=1 \\[-2pt]
            \sigma_{v_t}^2  	    & \text{for } m=2 \\[-2pt]
            \gamma_{v_t}  	        & \text{for } m=3 \\[-2pt]
            \kappa_{v_t}  	        & \text{for } m=4.\\[-2pt]
        \end{cases}
\vspace{-1ex}
\end{equation*}
$\sigma^2$ describes the dispersion of the distribution about its mean; the mean is zero by construction. $\gamma$ denotes the skewness/asymmetry of the distribution, i.e., the relative length of the upper and lower tails. Finally, $\kappa$ denotes the kurtosis, i.e., the tendency of the distribution to amass away from the middle $[-\sigma,\sigma]$, and thus characterizes the thickness of the tails.

We allow the central moments of shocks to depend on $t$ to reflect that different stages of the lifecycle may differ in the amount and nature of income risk; \citet{Guvenen_Karahan_Ozkan_Song2021} provide some evidence for that. Otherwise, shocks across households are identically distributed given $t$. Permanent and transitory shocks are mutually independent and independent over time.

As discussed in the introduction, several recent studies document that the cross-sectional distribution of income growth exhibits left skewness and excess kurtosis, i.e., a long and fat left tail. Our specification of the moments of income shocks enables our income process to be consistent with these facts, allowing us to subsequently decompose the various income moments into the relative contributions of permanent and transitory shocks.

\subsection{Linear consumption function}\label{SubSec::Model_Linear_C_function}

In line with BPP, we first consider a linear specification of the household consumption function, given by
\begin{equation}\label{Eq::Consumption_Function_Linear}
	\Delta c_{it} = \xi_{it} + \phi_{it}^{(1)}\zeta_{it} + \psi_{it}^{(1)}v_{it},
\end{equation}
where $\Delta c_{it}$ is the residual from a regression of real consumption growth, $\Delta \ln C_{it}$, on the taste observables $\mathbf{Z}_{it}$ (i.e., consumption growth net of its predictable profile). The linear function \eqref{Eq::Consumption_Function_Linear} is consistent with a log-linearization (i.e., a first-order Taylor series expansion) of the problem's first-order conditions and the household budget constraint, as we show in appendix \ref{Appendix::Derivation_Consumption_Function}. These approximations are independent of the distributions of income shocks and mimic similar derivations in BPP (who abstract from higher-order moments).

Permanent and transitory income shocks drive unexplained consumption growth in \eqref{Eq::Consumption_Function_Linear}. The transmission parameters $\phi_{it}^{(1)}=\partial \Delta c_{it}/\partial \zeta_{it}$ and $\psi_{it}^{(1)}=\partial \Delta c_{it}/\partial v_{it}$ reflect the pass-through of permanent and transitory shocks, respectively, into consumption. Adopting BPP's terminology, $1-\phi_{it}^{(1)}$ is then the extent of insurance to permanent shocks (i.e., the fraction of a permanent shock that does \emph{not} translate into a consumption fluctuation), while $1-\psi_{it}^{(1)}$ is the extent of insurance to transitory shocks. Therefore, $\phi_{it}^{(1)}=\psi_{it}^{(1)}=0$ implies that consumption is \emph{fully} insured to income shocks, while $\phi_{it}^{(1)}=\psi_{it}^{(1)}=1$ implies \emph{no insurance}. By contrast, values $\phi_{it}^{(1)},\psi_{it}^{(1)}\in(0,1)$ reflect \emph{partial} insurance: a given income shock translates into a consumption fluctuation, albeit less than one-to-one. 

The transmission parameters vary, in principle, with $i$ (household) and $t$ (age), reflecting heterogeneity in financial wealth, among other things.\footnote{Households in \citet{Blundell_Pistaferri_Saporta-Eksten2016} differ in their access to external insurance and insurance through family labor supply, while households in \citet{Theloudis2021} differ in their preferences.} The idea is that a household can use wealth to smooth consumption when income shocks hit; and the more wealth one holds, the better they can self-insure. We establish this link from wealth to $\phi_{it}^{(1)}$, $\psi_{it}^{(1)}$ in appendix \ref{Appendix::Derivation_Consumption_Function}. So while the pass-through of shocks depends on wealth, it does not depend on the magnitude or sign of the shock itself. Function \eqref{Eq::Consumption_Function_Linear} thus implies that a household can insure big or small, good or bad shocks similarly, a point to which we return subsequently.

In addition to income shocks, unexplained consumption growth is also driven by $\xi_{it}$. This reflects unobserved (but non-stochastic) consumption taste heterogeneity across households, independent of the income shocks. We let the distribution of $\xi_{it}$ across households have zero mean and variance $\sigma_{\xi_t}^2$, as in BPP. We also let its third and fourth central moments be $\gamma_{\xi_t}$ and $\kappa_{\xi_t}$ respectively; these will appear subsequently in expressions for the skewness and kurtosis of consumption growth.

\subsection{Quadratic consumption function}\label{SubSec::Model_Quadratic_C_function}

The linear function offers an approximation to the true consumption rule when income shocks do not move the household too far away from its deterministic consumption profile. Function \eqref{Eq::Consumption_Function_Linear} is consistent with a linearization of marginal utility around that deterministic profile; but such a linearization will be inaccurate in the presence of tail income risk, e.g., when large shocks may induce large shifts in consumption as a consequence of concave utility.\footnote{This echoes \citet{Carroll2001_Death}'s critique of approximate Euler equations. He argues that the -then- popular log-linear Euler equation, and to a lesser extent its quadratic counterpart, imply consumption behavior that deviates from that in the true underlying model. However, \citet{Kaplan_Violante2010} and \citet{BlundellPrestonLow2013decomposing} assess the linear approximations in BPP across alternative scenarios, though not in one with respect to the severity of income shocks, and they find that they perform reasonably well. The second-order approximations herein aim at improving the inference drawn from these previously established approximations.}

To address this issue, we generalize the consumption function to a nonlinear (quadratic) specification, given by
\begin{equation}\label{Eq::Consumption_Function_Quadratic}
	\Delta c_{it} = \xi_{it} + \phi_{it}^{(1)}\zeta_{it} + \psi_{it}^{(1)}v_{it} + \phi_{it}^{(2)}\zeta_{it}^2 + \psi_{it}^{(2)}v_{it}^2 + \omega_{it}^{(22)}\zeta_{it}v_{it},
\end{equation}
where $\Delta c_{it}$ and $\xi_{it}$ are defined as previously. The quadratic function \eqref{Eq::Consumption_Function_Quadratic} is consistent with a second-order Taylor approximation to the problem's first-order conditions and the household budget constraint, as we show in appendix \ref{Appendix::Derivation_Consumption_Function}. These operations and resulting expressions generalize the original approximations in BPP and in several papers thereafter.\footnote{We do not go beyond a quadratic approximation for reasons of tractability and because higher-order terms in a regression of $\Delta c_{it}$ on a polynomial in $\Delta y_{it}$ are imprecisely estimated.} 

The quadratic function is characterized by several parameters. As in the linear case, $\phi_{it}^{(1)}$ and $\psi_{it}^{(1)}$ are the transmission parameters of the linear part of a polynomial in income shocks.\footnote{$\phi_{it}^{(1)}$ is analytically the same between linear and quadratic approximations. The same is true for $\psi_{it}^{(1)}$. Appendix \ref{Appendix::Derivation_Consumption_Function} provides clarity on this point.} By extension, $\phi_{it}^{(2)}$, $\psi_{it}^{(2)}$, and $\omega_{it}^{(22)}$ are the transmission parameters of the quadratic part of the same polynomial. In the context of long and fat tails in income risk, $\phi_{it}^{(1)}$ and $\psi_{it}^{(1)}$ can be thought of as the transmission parameters of the \emph{average} shock (a shock close to the mean of the distribution), while $\phi_{it}^{(2)}$, $\psi_{it}^{(2)}$, $\omega_{it}^{(22)}$ can be seen as the transmission parameters of a shock \emph{away from the middle} of the distribution. The parameters may again vary with $i$ (household) and $t$ (age), reflecting heterogeneity in financial wealth among other things.

In this generalized framework, $\phi_{it}^{(1)}$ no longer represents the pass-through of permanent shocks to consumption. Instead, the pass-through is given by
\begin{equation}\label{Eq::Quadratic_C_function_Phi}
	\partial \Delta c_{it}/\partial \zeta_{it} = \phi_{it}^{(1)} + 2\phi_{it}^{(2)}\zeta_{it} + \omega_{it}^{(22)}v_{it},
\end{equation}
which depends on multiple parameters \emph{and} on the magnitude and sign of the income shocks. As such, there is an asymmetric pass-through of good versus bad shocks into consumption. Suppose, for example, that $\phi^{(1)}>0$ and $\phi^{(2)}<0$, as our results subsequently suggest. At $v_{it}=0$ (the average transitory shock), a \emph{negative} permanent shock has a larger pass-through to consumption relative to a \emph{positive} shock of the same magnitude. The pass-through of negative shocks would thus be underestimated (i.e., excessive insurance concluded) if the econometrician assessed the extent of insurance on the basis of $\phi^{(1)}$ alone, ignoring the non-linearity in the consumption response. The size of this bias increases with the magnitude of $\phi^{(2)}$ and the size of the negative permanent shock (e.g., the extent of left skewness).

A similar asymmetry exists in the transmission of good versus bad transitory shocks. In this case, the pass-through into consumption is given by
\begin{equation}\label{Eq::Quadratic_C_function_Psi}
	\partial \Delta c_{it}/\partial v_{it} = \psi_{it}^{(1)} + 2\psi_{it}^{(2)}v_{it} + \omega_{it}^{(22)}\zeta_{it},
\end{equation}
which again depends on multiple parameters \emph{and} on the magnitude and sign of the income shocks. Importantly, this generalized pass-through does not rule out the possibility of full insurance to transitory shocks, a recurrent empirical finding in the literature. Suppose, for example, that $\psi^{(1)}\approx\psi^{(2)}\approx0$; then $\partial \Delta c_{it}/\partial v_{it}\approx0$ at $\zeta_{it}=0$ (the average permanent shock), indicating full insurance to transitory shocks. 

A final remark is needed here. The specific lifecycle model \eqref{Eq::Household_Problem}-\eqref{Eq::Budget_Constraint} is not the only structure that gives rise to the quadratic specification \eqref{Eq::Consumption_Function_Quadratic} or the subsequent econometric framework. Several models, e.g., models with liquidity constraints \citep{CampbellHercowitz2019LiquidityConstraints} or models for the wealthy hand-to-mouth \citep{KaplanViolante2014ConsumptionResponseFiscalStimulus}, to offer just two examples, imply consumption policy rules of the form $C_{it} = g_t (Y_{it}, \dots)$, motivating our empirical specification through a quadratic approximation to $g_t$. This light dependence on specific theory is why we used the term `quasi-reduced form' to name our framework.

\section{Identification}\label{Sec::Identification}

Our specification of the income and consumption processes imposes restrictions on the second and higher-order moments of income and consumption that can be used to identify the various parameters. These include the variance ($\sigma_{\zeta_t}^2$, $\sigma_{v_t}^2$), skewness ($\gamma_{\zeta_t}$, $\gamma_{v_t}$), and kurtosis ($\kappa_{\zeta_t}$, $\kappa_{v_t}$) of the distributions of income shocks,\footnote{With a slight abuse of terminology, we use the terms skewness and kurtosis to refer to the third and fourth central moments. We will be clear whenever those terms refer to \emph{standardized} moments.} the partial insurance parameters in the linear ($\phi^{(1)}_t$, $\psi^{(1)}_t$) and quadratic consumption functions ($\phi^{(1)}_t$, $\psi^{(1)}_t$, $\phi^{(2)}_t$, $\psi^{(2)}_t$, $\omega^{(22)}_t$), and the moments of unobserved taste heterogeneity ($\sigma_{\xi_t}^2$, $\gamma_{\xi_t}$, $\kappa_{\xi_t}$). We first restrict the partial insurance parameters to not vary across households but we subsequently explore their variation across certain demographic groups and by the household's position in the income distribution.

For identification, we assume existence of up to fourth moments of the joint distribution of income and consumption growth. In the case of the quadratic consumption function, we also assume that income shocks are \emph{not} Bernoulli. For consistent estimation of such moments, particularly over demographic groups subsequently, we further assume non-vanishing shares in each group asymptotically. For inference, we assume existence of up to eighth moments.

Two important aspects of the data must be addressed before proceeding to identification: the biennial nature of the modern PSID and measurement error.

As we describe in section \ref{Sec::Empirical_Implementation}, the PSID provides data only every two years within our time period. This necessitates to recast the income and consumption growth processes of section \ref{Sec::Model} to this lower frequency, a straightforward operation that we carry out in appendices \ref{Appendix::Derivation_Consumption_Function} and \ref{Appendix::Identification}. Nevertheless, for consistency with the earlier literature, we provide the identifying statements subsequently \emph{as if} the data come at a yearly frequency. The complete identifying statements at the lower frequency of the modern PSID appear in appendix \ref{Appendix::Identification}. 

Measurement error in income and consumption is more challenging. In the permanent-transitory specification of income, the moments of classical income error are not separately identifiable from the moments of the transitory shock. This necessitates that we restrict the income error to be Gaussian and retrieve its variance from validation studies of the PSID, as in \citet{MeghirPistaferri2004IncomeVariance}. We return to this in section \ref{Sec::Empirical_Implementation}. By contrast, the moments of classical consumption error are identifiable; this introduces three additional parameters for the variance, skewness, and kurtosis of consumption error ($\sigma_{u^c_t}^2$, $\gamma_{u^c_t}$, $\kappa_{u^c_t}$). For simplicity, however, we subsequently show identification \emph{as if} income and consumption errors are not present; we provide our full identifying statements with measurement error in appendix \ref{Appendix::Identification}.

\subsection{Income process parameters}\label{SubSec::Identification_Income}

Given the income process in \eqref{Eq::Income_Process} and the properties of shocks, the second and higher-order moments of the cross-sectional distribution of income growth are given by
\begin{align*}
    {\E} ((\Delta y_{it})^{2}) &= \sigma_{\zeta_t}^2 + \sigma_{v_t}^2 + \sigma_{v_{t-1}}^2, \\
    {\E} ((\Delta y_{it})^{3}) &= \gamma_{\zeta_t} + \gamma_{v_t} - \gamma_{v_{t-1}}, \\
    {\E} ((\Delta y_{it})^{4}) &= \kappa_{\zeta_t} + \kappa_{v_t} + \kappa_{v_{t-1}} + 6\sigma_{\zeta_t}^{2} (\sigma_{v_t}^{2}+\sigma_{v_{t-1}}^{2}) + 6\sigma_{v_{t}}^{2}\sigma_{v_{t-1}}^{2}.
\end{align*}
The second moment depends on the variance of permanent shocks at $t$ and the sum of variances of transitory shocks at $t$ and $t-1$; it thus reflects the dispersion of one permanent and two transitory shocks. The fourth moment reflects, similarly, the dispersion and kurtosis of both types of shocks. By contrast, the third moment is driven by skewness in the permanent shock. Barring strong non stationarities in the distribution of transitory shocks, the difference between $\gamma_{v_t}$ and $\gamma_{v_{t-1}}$ will be small (zero if the distribution is stationary), leaving $\gamma_{\zeta_t}$, the skewness in permanent shocks, to drive the third moment of income.

The above expressions convey information on the parameters of the income process. Formally, the variance of shocks is identified as
\begin{align*}
    \sigma_{\zeta_t}^{2}  &=  {\E}(\Delta y_{it} \times (\Delta y_{it-1}+\Delta y_{it}+\Delta y_{it+1})), \\
    \sigma_{v_t}^2        &= -{\E}(\Delta y_{it} \times \Delta y_{it+1}).
\end{align*}
The long sum in the first line strips income growth at $t$ of its contemporary transitory shock; its covariance with $\Delta y_{it}$ thus picks up the variance of the permanent shock. The covariance of two consecutive income growths in the second line identifies the variance of the transitory shocks due to mean reversion in the shock. In a similar way, skewness is identified from 
\begin{align*}
    \gamma_{\zeta_t}    &=  {\E}((\Delta y_{it})^{2} \times (\Delta y_{it-1}+\Delta y_{it}+\Delta y_{it+1})), \\
    \gamma_{v_t}        &= -{\E}((\Delta y_{it})^{2} \times \Delta y_{it+1}),
\end{align*}
while, conditional on identification of the second moments, kurtosis is jointly identified from
\begin{align*}
	\kappa_{\zeta_t}    &= {\E}((\Delta y_{it})^4) - \kappa_{v_t} - \kappa_{v_{t-1}} - 6\sigma_{\zeta_t}^{2} (\sigma_{v_t}^{2}+\sigma_{v_{t-1}}^{2}) - 6\sigma_{v_{t}}^{2}\sigma_{v_{t-1}}^{2}, \\
	\kappa_{v_t}        &= {\E}((\Delta y_{it})^2 \times (\Delta y_{it+1})^2) - \sigma_{\zeta_t}^2 \sum_{\tau=0}^{1}\sigma_{v_{t+\tau}}^2 - \sigma_{\zeta_{t+1}}^2 \sum_{\tau=-1}^{0}\sigma_{v_{t+\tau}}^2 - \sigma_{v_{t-1}}^2 \sum_{\tau=0}^{1}\sigma_{v_{t+\tau}}^2 
                        - \sigma_{\zeta_t}^2\sigma_{\zeta_{t+1}}^2 - \sigma_{v_{t}}^2\sigma_{v_{t+1}}^2.
\end{align*}
There are several over-identifying restrictions for both skewness and kurtosis.

\subsection{Linear consumption function parameters}\label{SubSec::Identification_Linear_Consumption_Function}

Given the linear consumption function in \eqref{Eq::Consumption_Function_Linear} and the properties of shocks, the second and higher-order moments of the distribution of consumption growth are given by
\begin{align*}
    {\E} ((\Delta c_{it})^{2}) &= (\phi_{t}^{(1)})^{2} \sigma_{\zeta_t}^2 + (\psi_{t}^{(1)})^{2} \sigma_{v_t}^2 + \sigma_{\xi_t}^2, \\
    {\E} ((\Delta c_{it})^{3}) &= (\phi_{t}^{(1)})^{3} \gamma_{\zeta_t}   + (\psi_{t}^{(1)})^{3} \gamma_{v_t}   + \gamma_{\xi_t}, \\
    {\E} ((\Delta c_{it})^{4}) &= (\phi_{t}^{(1)})^{4} \kappa_{\zeta_t}   + (\psi_{t}^{(1)})^{4} \kappa_{v_t}   + \kappa_{\xi_t} + 6(\phi_{t}^{(1)})^{2}(\psi_{t}^{(1)})^{2}\sigma_{\zeta_t}^2 \sigma_{v_t}^2 \\
                               &+ 6(\phi_{t}^{(1)})^{2} \sigma_{\zeta_t}^2 \sigma_{\xi_t}^2 + 6(\psi_{t}^{(1)})^{2} \sigma_{v_t}^2 \sigma_{\xi_t}^2.
\end{align*}
The moments of the distribution of consumption growth reflect the underlying distributions of income shocks, as well as the distribution of unobserved taste heterogeneity. In general, larger dispersion, skewness, or kurtosis in the distributions of income shocks imply larger dispersion, skewness, or kurtosis in the distribution of consumption growth. The link between them depends on the degree of insurance with respect to these shocks, so these moments provide restrictions that help identify the parameters of partial insurance.

BPP discuss identification in the linear function at length; we provide a summary here while we relegate the details to appendix \ref{SubAppendix::Identification_LinearConsumptionF}. Formally, the transmission parameter of permanent shocks is identified as
\begin{equation*}
    \phi^{(1)}_t = {\E}(\Delta c_{it} \times (\Delta y_{it-1} + \Delta y_{it} + \Delta y_{it+1}))/\sigma_{\zeta_t}^2, 
\end{equation*}
while the transmission parameter of transitory shocks is identified as
\begin{equation*}
    \psi^{(1)}_t = - {\E}(\Delta c_{it} \times \Delta y_{it+1})/\sigma_{v_t}^2;
\end{equation*}
in both cases, the variance of shocks is given by the earlier expressions.\footnote{The variance of consumption taste heterogeneity is further given by $\sigma_{\xi_t}^2 = {\E}((\Delta c_{it})^{2}) - (\phi_t^{(1)})^{2} \sigma_{\zeta_t}^2 - (\psi_t^{(1)})^{2} \sigma_{v_t}^2$. Its skewness and kurtosis are similarly identified from higher moments of consumption growth.} These statements reflect the intuition behind identification in BPP: the \emph{covariance} between consumption and (appropriately defined) income growth captures the strength of the relation between the two series and, consequently, the pass-through of income shocks into consumption.\footnote{An equivalent view is that the partial insurance parameters are identified from a regression of consumption growth on income growth, using permanent or future income as instruments for income growth.} 

This workhorse covariance, however, is a \emph{second} moment of the joint distribution of consumption and income and, as such, neglects information in the tails. If most income shocks are either zero or rather bad, identifying partial insurance based on second moments alone may be misleading. However, the partial insurance parameters are heavily over-identified by higher-order moments of the marginal and joint distributions of consumption and income; inspect, for example, the expressions for ${\E} ((\Delta c_{it})^{3})$ or ${\E} ((\Delta c_{it})^{4})$. We will thus subsequently exploit these higher-order moments in the context of the linear consumption function in our first attempt to understand the implications of non-Gaussianity for partial insurance.

\subsection{Quadratic consumption function parameters}\label{SubSec::Identification_Quadratic_Consumption_Function}

As per the quadratic specification in \eqref{Eq::Consumption_Function_Quadratic} and the properties of shocks, the second moment of consumption growth is given by
\begin{align*}
	{\E}((\Delta c_{it})^{2}) 
	&= (\phi_{t}^{(1)})^{2} \sigma_{\zeta_t}^2 + (\psi_{t}^{(1)})^{2} \sigma_{v_t}^2 
	 + 2\phi_{t}^{(1)}\phi_{t}^{(2)}\gamma_{\zeta_t} + 2\psi_{t}^{(1)}\psi_{t}^{(2)}\gamma_{v_t} 
	 + (\phi_{t}^{(2)})^{2}\kappa_{\zeta_t} + (\psi_{t}^{(2)})^{2}\kappa_{v_t} \\
	&+ (\omega_{t}^{(22)})^{2} \sigma_{\zeta_t}^2\sigma_{v_t}^2 + 2\phi_{t}^{(2)}\psi_{t}^{(2)}\sigma_{\zeta_t}^2\sigma_{v_t}^2 + \sigma_{\xi_t}^2.
\end{align*}
This depends not only on the variance of permanent and transitory income shocks, as in the linear case, but also on their skewness and kurtosis. Contrary to the linear case, however, the higher-order moments ${\E}((\Delta c_{it})^{3})$ and ${\E}((\Delta c_{it})^{4})$ depend on higher than fourth-order moments of the shocks, which we do not model. 

Identification of the partial insurance parameters comes from the joint moments of consumption and income, conceptually similar to identification in the linear case. Multiple moments contribute to the identification of any single parameter. Formally, the parameters are given by 
\begin{equation}\label{Eq::Quadratic_C_function_identification}
	\left(
		\begin{array}{l}
			\phi^{(1)}_t\\
			\psi^{(1)}_t\\
			\phi^{(2)}_t\\
			\psi^{(2)}_t\\
			\omega^{(22)}_t
		\end{array}\right) = \mathbf{A}^{-1}
	\left(
		\begin{array}{l}
			{\E}\left(\Delta c_{it}  \times \Delta y_{it}\right) \\
			{\E}\left(\Delta c_{it}  \times (\Delta y_{it})^2\right) \\
			{\E}\left(\Delta c_{it} \times \Delta y_{it+1}\right) \\
			{\E}\left(\Delta c_{it} \times (\Delta y_{it+1})^2\right) \\
			{\E}\left(\Delta c_{it}  \times \Delta y_{it}  \times \Delta y_{it+1}\right) \\
		\end{array}\right),
\end{equation}
where square matrix $\mathbf{A}$ depends exclusively on the second, third, and fourth moments of income shocks.\footnote{The matrix of coefficients is given by
\begin{equation*}
\mathbf{A} = 
\left(
\begin{array}{lllll}
	\sigma_{\zeta_t}^2 	& \sigma_{v_t}^2 	& \gamma_{\zeta_t} 																& \gamma_{v_t} 							& 0 \\
	\gamma_{\zeta_t} 	& \gamma_{v_t} 		& \kappa_{\zeta_t} + \sigma_{\zeta_t}^2 (\sigma_{v_{t}}^2+\sigma_{v_{t-1}}^2)	& \kappa_{v_t} + \sigma_{v_t}^2(\sigma_{\zeta_{t}}^2+\sigma_{v_{t-1}}^2)	   & 2\sigma_{\zeta_t}^2 \sigma_{v_t}^2 \\
	0 					& -\sigma_{v_t}^2 	& 0 																			& -\gamma_{v_t} 						& 0 \\ 
	0 					& \gamma_{v_t} 		& \sigma_{\zeta_t}^2 (\sigma_{\zeta_{t+1}}^2+\sigma_{v_{t}}^2+\sigma_{v_{t+1}}^2) 	& \kappa_{v_t} + \sigma_{v_t}^2(\sigma_{\zeta_{t+1}}^2+\sigma_{v_{t+1}}^2)		& 0 \\ 	
	0 					& -\gamma_{v_t} 	& -\sigma_{\zeta_t}^2 \sigma_{v_t}^2											& -\kappa_{v_t} 	 					& -\sigma_{\zeta_t}^2 \sigma_{v_t}^2 \\
\end{array}\right).
\end{equation*}}\textsuperscript{,}\footnote{The variance of consumption taste heterogeneity is identified from ${\E}((\Delta c_{it})^2)$. Its skewness and kurtosis enter the third and fourth moments of consumption growth. Those depend on higher than fourth-order moments of income, which we do not model, so they are generally not identified in the current setting.} We establish in appendix \ref{SubAppendix::Identification_QuadraticConsumptionF} that matrix $\mathbf{A}$ is full rank --and hence invertible-- for \emph{any} distribution of income shocks other than Bernoulli (for which there is no empirical evidence anyway).\footnote{If the distributions of \emph{both} shocks are Bernoulli, then there exist some parameter configurations for which matrix $\mathbf{A}$ is not full rank. Assuming away Bernoulli distributions is thus an identifying assumption.} Identification of the parameters does \emph{not} require those shocks to be non-Gaussian; the matrix is invertible even if the distributions of shocks are normal and have the same variance between them (i.e., $\sigma_{\zeta_t}^2 = \sigma_{v_t}^2$). 

In the nonlinear specification, tail shocks exhibit a different pass-through compared to average shocks, regardless of the specific form of underlying income risk. As a result, their transmission parameters cannot be identified with second-order income moments alone. Skewness and kurtosis now become essential for identification, even if the income distribution is Gaussian. This is intuitive: we want to identify the impact of tail shocks, so we must use information on said tails. In appendix \ref{SubAppendix::Identification_QuadraticConsumptionF}, we explore how varying levels of income skewness and kurtosis influence the transmission parameters.

Contrary to income, no higher-order moments of \emph{consumption} are needed; the identifying moments in \eqref{Eq::Quadratic_C_function_identification} are covariances between consumption growth and various powers of income growth. This is an advantage over quantile methods that require observing the \emph{entire} consumption distribution, e.g., ABB, because, while we may estimate higher moments of consumption growth in the PSID, there is yet no established population benchmark to assess them against. This is in contrast to higher \emph{income} moments, for which \citet{Guvenen_Karahan_Ozkan_Song2021} is a natural benchmark given their use of administrative population data. 

How biased are the parameter estimates if one assumes a linear specification when the true consumption function is quadratic? The workhorse covariance that identifies the pass-through of permanent shocks in BPP is
\begin{equation*}
    {\E}(\Delta c_{it} \times (\Delta y_{it-1} + \Delta y_{it} + \Delta y_{it+1}))/\sigma_{\zeta_t}^2 = \phi_{t}^{(1)} + \phi_{t}^{(2)}(\gamma_{\zeta_{t}}/\sigma_{\zeta_t}^{2}). 
\end{equation*}
This fails to identify the true pass-through of permanent shocks in the presence of nonlinear transmission, i.e., expression \eqref{Eq::Quadratic_C_function_Phi}. Moreover, this workhorse covariance is biased for $\phi_{t}^{(1)}$ if permanent shocks exhibit skewness. If $\gamma_{\zeta_{t}}<0$ and $\phi_{t}^{(2)}<0$, as our empirical estimates suggest, linearity biases $\phi_{t}^{(1)}$ upwards. Similarly, BPP's identifying condition for $\psi_{t}^{(1)}$ becomes 
\begin{equation*}
    -{\E}(\Delta c_{it} \times \Delta y_{it+1})/\sigma_{v_t}^2 = \psi_{t}^{(1)} + \psi_{t}^{(2)}(\gamma_{v_{t}}/\sigma_{v_t}^{2}), 
\end{equation*}
also failing to identify the true pass-through of transitory shocks in the presence of nonlinear transmission, i.e., expression \eqref{Eq::Quadratic_C_function_Psi}.
If $\gamma_{v_{t}}<0$ and $\psi_{t}^{(2)}>0$, as our baseline estimates suggest, linearity biases $\psi_{t}^{(1)}$ downwards. This may help bridge the discrepancy between covariance based methods (who find little pass-through of transitory shocks) and studies based on natural experiments (who usually find large consumption responses to said shocks).

Our estimation procedure, which we describe in the next section, is linked transparently to the above identification argument. Effectively, we bring \eqref{Eq::Quadratic_C_function_identification} to the data, along with some over-identifying restrictions, using moments estimation.

\section{Empirical implementation}\label{Sec::Empirical_Implementation}

\subsection{Data}\label{SubSec::Empirical_Implementation_Data}

As is evident from the previous section, panel data on income and consumption are needed to identify the moments of the permanent and transitory shocks and the degrees of insurance to them. The Panel Study of Income Dynamics (PSID) provides such data for a representative and continuously evolving sample of U.S. households since $1968$. For most of this period, the PSID collected expenditures only on food items. Since $1999$, however, the PSID was drastically redesigned to collect information over a broader set of consumption categories, currently comprising over 70\% of consumption in the National Income and Product Accounts \citep{Blundell_Pistaferri_Saporta-Eksten2016}. We, therefore, use the PSID over the period between $1999$ and $2019$, the last available wave when we started working on this paper; the data are available only biennially. This is in contrast to BPP, who use PSID data over $1980$--$1992$ and rely on consumption imputations from the Consumer Expenditure Survey (CEX).

Our baseline sample consists of married couples in the PSID with a male spouse aged $30$ to $65$. We focus on a sample that is representative of the U.S. population, eliminating the supplementary low-income subsample. Households are also removed in the event of a break-up or remarriage; our analysis thus focuses on income risk rather than on divorce or other household dissolution factors. Naturally, we require non-missing data on income, expenditure, and basic demographics.\footnote{We also remove observations in the bottom 0.25\% of the distributions of $\Delta y_{it} \Delta y_{it+1}$ and $\Delta c_{it} \Delta c_{it+1}$, i.e., large increases followed by large drops, likely reflecting measurement error, and observations with extreme jumps in income or consumption (\emph{jump}: $\geq$10-fold increase or $\geq$90\% reduction), which likely also reflect measurement error. Our results are not sensitive to this sample selection choice.} Our final sample consists of 20,866 household-year observations; on average, a household is observed for $5.6$ periods, which corresponds to about $10$ calendar years given the biennial nature of the data.

Appendix table \ref{AppTable::Descriptives} presents summary statistics over the sample period $1999$--$2019$. About $89\%$ of men and $79\%$ of women work for pay (but some households have intermittent employment); $68\%$ of men and $71\%$ of women have had some college education. Our measure of income is household disposable income, namely the sum of taxable income (e.g., earnings and asset income), transfers, and social security income of the couple and of other family members, minus taxes paid. Average disposable income is \$142,910 in 2018 prices.\footnote{We express all monetary figures in 2018 prices because earnings are 1-year retrospective and our latest data are from 2019. We deflate using the Consumer Price Index from the Bureau of Labor Statistics.} Our measure of consumption is real expenditure on nondurable goods and services, comprising food (at home and outside), utilities, out-of-pocket health expenses, public transport, vehicle expenses (gasoline, parking, insurance), education, and child care. Average real consumption is \$26,766, with food at home accounting for about $30\%$ of this. Housing, including rent and home insurance, is the main expenditure we exclude; we do so for comparability with BPP who also exclude it. Including housing does not meaningfully change the subsequent results.

Aside from the PSID, we use a similarly selected sample drawn from the CEX to facilitate comparison with BPP. We defer the discussion of this secondary data to section \ref{SubSec::Empirical_Implementation_Moments}.

\subsection{Estimation}\label{SubSec::Empirical_Implementation_Estimation}

Estimation proceeds as follows. In a first step, we compute income and consumption growth net of their predictable components, $\mathbf{X}_{it}$ and $\mathbf{Z}_{it}$ in model notation, by regressing real income and consumption growth on a rich set of time effects and household characteristics.\footnote{The controls include year and year-of-birth dummies, dummies for education, race, region, family size and number of children (level and change from last period), employment status (current and past), presence of outside dependents, and presence of income recipients other than the husband and wife (both current and past). The dummies for education, race, and region are interacted with the year dummies.} The residuals from these regressions are the empirical counterparts to unexplained income and consumption growth, $\Delta y_{it}$ and $\Delta c_{it}$, respectively. In a second step, we estimate the parameters of the income process -- i.e., the second and higher moments of permanent and transitory shocks -- using income data alone. In a third step, we estimate the transmission parameters of shocks and the moments of taste heterogeneity and consumption measurement error.

We estimate the second- and third-step parameters using GMM. Given the potentially poor small sample properties of optimal weighting, we use the identity matrix as our baseline weighting choice. In practice, this matters little for the parameter estimates of either stage: using the optimal weighting matrix gives almost identical point estimates, suggesting the parameters are strongly identified. We report results from equal weighting subsequently, but we also present the main estimates using optimal GMM in appendix \ref{SubAppendix::Empirics_Additonal_Results}.\footnote{Diagonal weighting also delivers similar results (those results are omitted for brevity).} 

We present estimates of the transmission parameters separately for the linear and quadratic specifications of the consumption function. For the linear function, we first present results when only second moments of income and consumption growth are targeted, exactly as in BPP. This purports to replicate BPP over the more recent sample period. We next report results for the linear function when, in addition to the second moments, third and fourth moments are also targeted. This helps assess whether and how, through the lenses of the linear specification, information in the tails of the income distribution affects the measurement of partial insurance. We then move on to the results from the quadratic specification. Our choice to estimate the income process separately in an earlier step enables the income parameters to remain unchanged regardless of the specification of the consumption function fitted. Appendix \ref{SubAppendix::Identification_Targeted_Moments} lists the moments targeted in each case. 

Unless noted otherwise, we conduct inference for the main parameters using block bootstrap. We thus account for serial correlation and heteroskedasticity of arbitrary form and for parameter uncertainty from earlier steps of the estimation process. GMM (asymptotic) standard errors clustered at the household level are very similar, suggesting that earlier-step parameter uncertainty is negligible (the income parameters are tightly estimated).\footnote{\label{Footnote::StandardErrors}We compute bootstrap standard errors as the normalized interquartile range of bootstrap replications, i.e., the difference between the $75^\text{th}$ and $25^\text{th}$ percentiles divided by the interquartile range of the standard normal cdf. This is equivalent to applying a normal approximation to the distribution of bootstrap replications, thereby shielding the standard errors from extreme resampling draws \citep{Blundell_Pistaferri_Saporta-Eksten2016,Theloudis2021}. Bootstrap standard errors without this approximation are similar. For the subsample analysis (tables \ref{Table::Consumption_function_lagY}-\ref{Table::Quadratic_Consumption_function_subsamples} and corresponding tables in the appendix), we calculate GMM (asymptotic) standard errors clustered at the household level to avoid numerical instability in bootstrap, driven by vanishing shares of certain subsamples, e.g., the wealthy non-college graduates, in many bootstrap draws. For the replication of BPP using the original BPP data (table \ref{AppTable::Replication_BPP_19801992}), we show non-clustered GMM standard errors as in BPP.} 

As we explained in section \ref{Sec::Identification}, we must use an external estimate for the variance of income measurement error. We cannot otherwise separate the moments of the transitory shock from those of the measurement error. Following \citet{MeghirPistaferri2004IncomeVariance} and \citet{Blundell_Pistaferri_Saporta-Eksten2016}, we retrieve the error variance from \citet{BoundEtAl1994PSIDerror}, a popular validation study of the PSID, setting it equal to $4\%$ of the variance of log income in any given year.\footnote{\citet{BoundEtAl1994PSIDerror} survey workers from a single manufacturing firm in 1983 and 1987. They obtain information on earnings in a way that mimics the PSID questionnaire and coding practices and compare them to administrative data from the firm. There are two caveats in this approach: first, the sample of workers in the validation study comes from two decades prior to the data in this paper; second, we use estimates of error in male earnings to correct for error in household income.}

\subsection{Empirical moments of income and consumption}\label{SubSec::Empirical_Implementation_Moments}

Table \ref{Table::Empirical_Moments_Income_Consumption} reports the empirical second, third, and fourth moments of the cross-sectional distributions of unexplained income and consumption growth in the sample.\footnote{In contrast to the identifying statements of section \ref{Sec::Identification}, the data are biennial so we report moments for income/consumption growth at a biennial frequency. We maintain the notation $\Delta x_{t}$ for the first difference of variable $x$ over time, noting that this corresponds to a difference over two years. See appendix \ref{Appendix::Identification} for details.}

\

\noindent \textbf{Income moments.} Panel A presents the moments of income growth. The variance is $0.153$, while the first-order autocovariance is $-0.049$ and highly significant. The income process of section \ref{SubSec::Model_Income_Process} implies that income growth is negatively first-order autocorrelated due to mean reversion in the transitory shock, and this is supported by the data.\footnote{The second-order autocovariance is an order of magnitude smaller and only marginally significant. This feature and the biennial nature of the data, whereby the second-order autocovariance reflects correlations 4 calendar years apart, motivates our white noise specification for the transitory shock.} 

\begin{table}[t!]  
\begin{center}
\caption{Empirical moments of income and consumption growth}\label{Table::Empirical_Moments_Income_Consumption}
\begin{tabular}{l R{2.1cm} L{2.1cm} R{2.1cm} L{2.1cm}}
\toprule								
\emph{Income data:}									& \mcl{2}{c}{PSID} 		& \mcl{2}{c}{PSID}	 			\\
\emph{Consumption data:}							& \mcl{2}{c}{PSID}  	& \mcl{2}{c}{imputed from CEX}	\\
													& \mcl{2}{c}{(1)} 		& \mcl{2}{c}{(2)} 				\\
\midrule
\mcl{5}{c}{\textbf{Panel A. Moments of income growth}} 				\\
${\Var}(\Delta y_{it})$								&0.153	&(0.005) 		&0.153		&(0.005)			\\
${\Cov}(\Delta y_{it},\Delta y_{it+1})$				&-0.049 &(0.002) 		&-0.049 	&(0.002) 			\\
${\Skew}(\Delta y_{it})$							&-0.143 &(0.098) 		&-0.143 	&(0.098) 			\\
${\Cov}((\Delta y_{it})^2,\Delta y_{it+1})$			&0.009  &(0.003) 		&0.009  	&(0.003) 			\\
${\Kurt}(\Delta y_{it})$							&10.032 &(0.509) 		&10.032 	&(0.509) 			\\
${\Cov}((\Delta y_{it})^2,(\Delta y_{it+1})^2)$		&0.080 	&(0.007) 		&0.080 		&(0.007) 			\\
\noalign{\smallskip}
\mcl{5}{c}{\textbf{Panel B. Moments of consumption growth}} 				\\
${\Var}(\Delta c_{it})$								&0.126	&(0.002)		&0.323		&(0.015)			\\
${\Cov}(\Delta c_{it},\Delta c_{it+1})$				&-0.044	&(0.001)		&-0.140		&(0.009)			\\
${\Skew}(\Delta c_{it})$							&-0.039	&(0.034)		&-0.177		&(0.083)			\\
${\Cov}((\Delta c_{it})^2,\Delta c_{it+1})$			&-0.003	&(0.001)		&0.190		&(0.036)			\\
${\Kurt}(\Delta c_{it})$							&4.445	&(0.100)		&17.109		&(2.111)			\\
${\Cov}((\Delta c_{it})^2,(\Delta c_{it+1})^2)$		&0.027	&(0.001)		&0.909		&(0.198)			\\
\noalign{\smallskip}
\mcl{5}{c}{\textbf{Panel C. Joint moments}} 				\\
${\Cov}(\Delta y_{it},\Delta c_{it})$				&0.008	&(0.001)		&0.012 		&(0.002)			\\
\bottomrule
\end{tabular}
\caption*{\fsz\emph{Notes:} The table presents the second, third, and fourth moments of income and consumption growth. Skewness and kurtosis correspond to the third and fourth standardized moments respectively. Column 1 reports moments of consumption internally in the PSID; column 2 reports moments of consumption imputed from the CEX. See appendix \ref{SubAppendix::Empirics_CEX} for details on the imputation. We maintain the notation $\Delta x_{t}$ for the first difference of variable $x$ over time, noting that, given the biennial nature of the data, this corresponds to a difference over two calendar years. Block bootstrap standard errors are in parentheses.}
\end{center}
\end{table}

The third \emph{standardized} moment of income growth is $-0.143$. Despite the small size of the cross-section and the inherent difficulties in estimating higher moments, the point estimate misses statistical significance at the 10\% level only by a small margin. ${\Cov}((\Delta y_{it})^2,\Delta y_{it+1})$, which reflects (minus) the third \emph{central} moment of the transitory shock, is significant at the 1\% level. The point estimates suggest that income growth exhibits left skewness, i.e., a long left tail. In simple terms, the left tail of the distribution of income growth (i.e., of income shocks) is longer than the right tail, so a negative shock is on average more severe, more unsettling, than a positive one. This is in line with \cite{Guvenen_Karahan_Ozkan_Song2021}, who document strong left skewness in the distribution of male earnings growth using administrative data from the U.S. Social Security Administration. 

The fourth \emph{standardized} moment of income growth is $10.032$ and, as expected, highly significant. ${\Cov}((\Delta y_{it})^2,(\Delta y_{it+1})^2)$ is also very significant; this is larger than three times the square of the variance of transitory shocks, indicating excess kurtosis in those shocks. These estimates suggest that income growth is strongly leptokurtic. In simple terms, the distribution of income growth (i.e., of income shocks) has thicker tails and less mass in the shoulders relative to a Gaussian density. Shocks are thus either close to the center of the distribution, i.e., zero, or out in the tails. This is again in line with \cite{Guvenen_Karahan_Ozkan_Song2021}, who document strong excess kurtosis in male earnings growth.

\cite{Guvenen_Karahan_Ozkan_Song2021} focus on male earnings while our model and statistics for income growth above revolve around household disposable income. To facilitate a comparison, appendix \ref{SubAppendix::Empirics_Comparison_Guvenen} reports moments of male \emph{earnings} growth in our sample. The variance is similar between the PSID and the administrative data, at $0.271$ and $0.323$ respectively. Skewness is $-0.569$ (and highly statistically significant) versus $-1.039$ in the administrative data. While the nature of skewness is similar in both cases, the population data feature substantially more left skewness than the PSID. Kurtosis, finally, is nearly identical between the two, at $13.711$ and $13.494$ respectively.

Evidently, left skewness and excess kurtosis are not only features of the growth rates in disposable household income (table \ref{Table::Empirical_Moments_Income_Consumption}), but they are also observed in alternative measures of income, such as male earnings (table \ref{AppTable::Empirical_Moments_Earnings}, panel A) and household earnings (table \ref{AppTable::Empirical_Moments_Earnings}, panel B). Together, left skewness and excess kurtosis suggest that income changes for most households are either close to their mean value of zero or far out in the left tail (large income \emph{reductions}).\footnote{Using earnings as the relevant income variable in the model does not materially change our subsequent results, in particular the asymmetry in the transmission of bad versus good (earnings) shocks.} These results corroborate (and, in fact, are close numerically with) findings by \citet{DeNardi_Fella_PazPardo2019,DeNardi_EtAl2021_Netherlands_US} over an earlier period of the PSID.

Large income reductions do not just reflect unemployment, but a multitude of events. Issues of reverse causality aside, appendix \ref{SubAppendix::Sources_Income_Shocks} documents that households with large income reductions also experience the loss of a second job, worsening health, large drops in hours worked (reflecting both demand and supply factors), and precarious housing. It is unclear how persistent those events are, but our subsequent decomposition of income to moments of permanent and transitory shocks will shed light on this.

\

\noindent \textbf{Consumption moments.} Panel B of table \ref{Table::Empirical_Moments_Income_Consumption} presents the moments of consumption growth; column 1 specifically reports moments from our baseline sample in the PSID. The variance of consumption growth is similar to that of income growth ($0.126$ versus $0.153$), reflecting either a large pass-through of income shocks or large taste heterogeneity/measurement error. The first-order autocovariance is $-0.044$ and highly significant. In the context of our model, this autocovariance reflects the variance of consumption error $\sigma_{u^c_t}^2$ (see appendix \ref{Appendix::Identification}).

The third \emph{standardized} moment is $-0.039$. Albeit statistically insignificant, it agrees with \cite{Constantinides_Ghosh2017}, who find left skewness in (quarterly) consumption growth in the CEX. Similar to income growth, the distribution of consumption growth in the PSID exhibits left skewness, i.e., a longer left tail, albeit at a lesser extent than income.\footnote{${\Cov}((\Delta c_{it})^2,\Delta c_{it+1})$ identifies, in the context of the linear model, the skewness of consumption error. The covariance is small, so measurement error does not contribute much to negative skewness in consumption.}

The fourth \emph{standardized} moment of consumption growth is $4.445$ and highly significant. There is excess kurtosis and the distribution is thus leptokurtic. In contrast to the distribution of income, however, excess kurtosis is less pronounced here.

The covariance between income and consumption growth is $0.008$. As we established in section \ref{Sec::Identification}, this moment is crucial for the identification of the pass-through of income shocks into consumption. While we cannot immediately gauge its magnitude, the strong statistical significance reflects that income shocks do transmit into consumption. 

Unlike income, there are neither administrative data on consumption, nor has there been a validation study for consumption in the PSID.\footnote{\citet{AndreskiEtAl2014_PSID_CEX_comparison} and \citet{LiEtAl_2010_PSID_CEX_consumption} compare consumption between the PSID and the CEX; \citet{MeyerSullivan2023onsumptionIncomeInequality} offer a critical discussion of the CEX data against the PSID.} As a consequence, we cannot benchmark our estimates of the consumption higher moments against the literature. The CEX, however, provides high quality consumption data for a representative sample of the U.S. population, although it lacks a long panel dimension. We can thus use the CEX to select a sample of households as close to our baseline as possible, and estimate similar cross-sectional moments across the surveys. The CEX is useful also for another reason. Since the PSID collected little consumption data prior to $1999$, the early literature on consumption insurance, notably BPP, resorted to imputing expenditure from the CEX into the PSID. To facilitate the comparison with BPP, we can use the CEX to replicate the consumption imputation into the PSID. This is not strictly needed in our case (the PSID provides comprehensive data over our sample period) but the imputation is nevertheless insightful, as we illustrate later.

\

\noindent \textbf{Consumption in the CEX.} We use CEX interview survey data over 1999-2019 and select a sample of stable married households that mimics closely our baseline selection in the PSID. We report details on the sample selection in appendix \ref{SubAppendix::Empirics_CEX}.

Appendix table \ref{AppTable::PSID_CEX_sample} offers a comparison between the PSID and CEX samples over time. The two samples are very close with respect to demographics and labor market participation. Household income is higher in the PSID; BPP noted this and argued that it is due to the more comprehensive definition of income in the PSID. Food expenditure in the PSID is also higher, in contrast to BPP who find that food expenditure over the earlier period 1980--1992 is similar across the two surveys. Consequently, overall consumption in the PSID is, on average, higher by about 25\% than a similarly selected sample in the CEX.

Appendix figure \ref{AppFigure::Comparison_PSID_CEX} plots the variance, skewness, and kurtosis of \emph{log} consumption in the PSID and CEX. The patterns for the variance and kurtosis are similar in the two samples over time, though the variance in the PSID is lower and the kurtosis is slightly higher. Skewness of log consumption, however, is markedly different: it is strongly positive in the PSID but mostly negative in the CEX; the patterns also diverge over time. 

This comparison is not informative about the moments of consumption \emph{growth}, which is the relevant variable in our case. The CEX lacks the panel dimension needed for our analysis, so it cannot be used to obtain measures of consumption growth. This is why BPP imputed consumption from the CEX into the PSID: the CEX lacked the necessary panel dimension while the PSID lacked comprehensive consumption data before 1999. We repeat BPP's imputation over the \emph{recent} years, i.e., our sample period 1999--2019. The imputation involves the estimation of food demand in the CEX (food expenditure expressed as a function of prices and nondurable consumption). The demand equation is then inverted (nondurable consumption expressed as a function of prices and food) and used to impute nondurables in the PSID based on food expenditure and prices that are available in both surveys. In this way, we obtain an alternative, external measure of consumption in the PSID, and of consumption growth. Appendix \ref{SubAppendix::Empirics_CEX} provides details on the imputation. 

The moments of imputed consumption growth are in column 2 of table \ref{Table::Empirical_Moments_Income_Consumption}. The variance at $0.323$ is two-and-a-half times larger than the variance in the true series (column 1). In fact, the variance of imputed consumption growth is more than double the variance of income growth. The magnitude of the higher-order moments is also multiple times larger in the imputed versus the true series. Skewness in imputed consumption growth is $-0.177$ (versus $-0.039$ in the true series), while kurtosis is $17.109$, almost four times higher than kurtosis in the true series. The imputation thus seems to impart substantial measurement/imputation error to consumption.

Since the imputation is not strictly needed over our sample period, we use consumption internally available in the PSID as our \emph{baseline} measure of consumption. The present analysis also suggests that we may not want to rely on higher-order moments of consumption, given the discrepancies in skewness of \emph{log} consumption between the PSID and the CEX, which are independent of the imputation. As we established in section \ref{SubSec::Identification_Quadratic_Consumption_Function}, an advantage of our approach is precisely its non-reliance on higher-order consumption moments.

\section{Results}\label{Sec::Results}

We first present baseline results imposing stationarity in the distributions of income shocks, taste heterogeneity, consumption error, and a constant pass-through of income shocks over the lifecycle. We subsequently relax stationarity/lifecycle-invariance in section \ref{SubSec::Results_Quadratic_Consumption_Function}.\footnote{Stationarity is not an identifying assumption but helps with precision, especially of higher moments.}

\subsection{Income process}\label{SubSec::Results_Income_Process}

We present estimates of the income process in three specifications. First, we target only the second moments of income growth, which enables us to estimate the variance of income shocks. This is similar to what BPP do over the earlier period 1980--1992. Second, we target second \emph{and} third moments of income growth, which enables us to estimate also the skewness of shocks. Finally, we target \emph{up to} the fourth moments, which allows us to estimate all the parameters of the income process as presented in section \ref{SubSec::Model_Income_Process}. The results are in table \ref{Table::Income_process}, while the full set of moments targeted in each case is shown in appendix \ref{SubSubAppendix::Identification_Targeted_Moments_Income}.

When only the second moments of income growth are targeted, we estimate the variance of permanent and transitory shocks at $0.027$ and $0.033$, respectively. These are very close to BPP and \citet{MeghirPistaferri2004IncomeVariance}, who estimate these parameters over an earlier period. As the variance of income growth is driven by the variance of the contemporaneous permanent shock and the sum of the variances of transitory shocks in the current and previous periods, the contribution of the transitory shock to the variance of income growth is almost $2.5$-fold that of the permanent shock \citep[as in][]{GottschalkMoffitt2009_RisingInstability}.

\begin{table}[t!]  
\begin{center}
\caption{Estimates of the income process}\label{Table::Income_process}
\begin{tabular}{ll L{2.5cm} R{1.3cm} L{1.3cm} R{1.3cm} L{1.3cm} R{1.5cm} L{1.5cm}}
\toprule
\mcl{3}{l}{\emph{Income moments:}}                      & \mcl{2}{c}{2\textsuperscript{nd}} & \mcl{2}{c}{2\textsuperscript{nd}, 3\textsuperscript{rd}}  & \mcl{2}{c}{2\textsuperscript{nd}, 3\textsuperscript{rd}, 4\textsuperscript{th}}   \\
\mcl{3}{l}{\emph{Income data:}}                         & \mcl{2}{c}{PSID}                  & \mcl{2}{c}{PSID}                                          & \mcl{2}{c}{PSID} \\
&&                                                      & \mcl{2}{c}{(1)}                   & \mcl{2}{c}{(2)}                                           & \mcl{2}{c}{(3)}  \\
\midrule
\mcl{9}{c}{\textbf{Panel A. Permanent shocks}}\\
Variance    & $\sigma_{\zeta}^2$        &               &0.027      &(0.002)                &0.027          &(0.002)                                    &0.030      &(0.002)    \\
Skewness    & $\gamma_{\zeta}$          & central       &           &                       &-0.004         &(0.003)                                    &-0.004     &(0.003)    \\
            &                           & standardized  &           &                       &-0.947         &(0.649)                                    &-0.831     &(0.571)    \\
Kurtosis    & $\kappa_{\zeta}$          & central       &           &                       &               &                                           &0.039      &(0.007)    \\
            &                           & standardized  &           &                       &               &                                           &44.393     &(11.080)   \\
\noalign{\smallskip}
\mcl{9}{c}{\textbf{Panel B. Transitory shocks}}\\
Variance    & $\sigma_{v}^2$            &               &0.033      &(0.002)                &0.033          &(0.002)                                    &0.031      &(0.003)    \\
Skewness    & $\gamma_{v}$              & central       &           &                       &-0.008         &(0.003)                                    &-0.008     &(0.003)    \\
            &                           & standardized  &           &                       &-1.310         &(0.510)                                    &-1.431     &(0.566)    \\
Kurtosis    & $\kappa_{v}$              & central       &           &                       &               &                                           &0.048      &(0.005)    \\
            &                           & standardized  &           &                       &               &                                           &49.226     &(10.119)   \\
\bottomrule
\end{tabular}
\caption*{\fsz\emph{Notes:} The table presents the estimates of the parameters of the income process, imposing stationarity over time. Column 1 targets only the second moments of income growth; column 2 targets also third moments; column 3 targets all moments up to fourth-order. Estimation is via equally weighted GMM; block bootstrap standard errors are in parentheses. Results from optimal GMM are similar (appendix table \ref{AppTable::Income_process_OptimalMD}).}
\end{center}
\end{table}

Targeting the third moments of income growth (in addition to the second moments) leaves the variances unchanged. We estimate the skewness (third standardized moment) at $-0.947$ and $-1.31$ for permanent and transitory shocks, respectively. Both shocks thus exhibit substantial negative skewness, i.e., their distribution has a longer left than right tail. This is consistent with \citet{Geweke_Keane2000} who find left skewness of earnings shocks in the early PSID, and \citet{Theloudis2021} who finds left skewness of wage shocks in the recent period. These results vary with the level of income, but we postpone that discussion for later. Assuming for now that good and bad shocks transmit similarly into consumption, the long left tail implies that bad shocks are more damaging than good ones because they are farther away from the zero mean -- they are more extreme. This is true for both permanent and transitory shocks. Skewness of income growth, however, is driven mostly by permanent shocks (fully so under stationarity). Therefore, if targeting the third moments of income growth subsequently changes the degree of consumption insurance in the household, this must reflect the implications of permanent rather than of transitory shocks. 

Targeting all moments of income growth up to fourth-order leaves the third central moments unchanged and modifies the variances by little. We estimate the kurtosis (fourth standardized moment) at $44.393$ and $49.226$ respectively. In both cases, the distributions are highly leptokurtic: they are sharp-pointed around the mean, have relatively little mass in the shoulders and very fat tails. This is consistent with \citet{Busch_Ludwig2024}, who use the PSID until 2012 and estimate numerically similar kurtosis coefficients for both shocks. 

Seen together, left skewness and excess kurtosis imply that most households face either negligible shocks (close to zero) or shocks that are quite bad. Ignoring these stark features may lead to erroneous inference of the degree of partial insurance, to which we now turn.

\subsection{Linear consumption function}\label{SubSec::Results_Linear_Consumption_Function}

We present estimates of the linear consumption function, first, targeting only second moments of the joint income-consumption distribution and, subsequently, targeting second as well as higher-order moments. The full set of targeted moments is shown in appendix \ref{SubSubAppendix::Identification_Targeted_Moments_LinearFunction}.

\

\noindent \textbf{Targeting second-order moments only.} The estimation of the linear consumption function targeting second moments alone is effectively a replication of BPP over the recent years (1999-2019) while using consumption data that are internally available in the PSID.\footnote{We use the pre-estimated second moments of income shocks from column 1 of table \ref{Table::Income_process}.} This replication is important in its own right. The recent years offer superior consumption data (in the sense that we do not need to rely on an imputation from the CEX), thus reducing the scope of measurement/imputation error. The data, however, come at a biennial rather than annual frequency. Moreover, there have been many changes in taxation and redistribution policies since 1980--1992, the original period studied in BPP, which may matter for the extent of partial insurance available to households more recently. 

The results are in column 1 of table \ref{Table::Consumption_function}. We estimate the transmission parameter of permanent shocks at $\phi^{(1)}=0.152$. A $10\%$ permanent change in income thus results in only $1.52\%$ shift in consumption. This reflects a very low pass-through of permanent shocks and a substantial degree of consumption insurance. While highly significant, the estimate for $\phi^{(1)}$ is strikingly different from BPP, whose baseline estimate of $0.64$, four times bigger, indicates a much larger pass-through of permanent shocks, hence a much lower degree of insurance. By contrast, we find evidence of full insurance to transitory shocks, similarly to BPP. We estimate the pass-through of said shocks at $\psi^{(1)}=-0.006$, statistically indistinguishable from zero. Therefore, transitory shocks do not pass through into consumption.\footnote{\cite{Commault2022} generalizes BPP's linear consumption function by allowing \emph{past} transitory shocks to enter the specification in \eqref{Eq::Consumption_Function_Linear}. She then finds a strong consumption response to transitory income, similar to the response estimated from quasi-experimental data. We abstract from this generalization, as our focus here is on higher-order income moments and on non-linearities in the consumption response.}

What explains the discrepancy in $\phi^{(1)}$ with BPP? As suggested earlier, there are a few important differences between our replication and the original exercise in BPP: new consumption data and a different time period. To help decipher their role, we impute consumption from the CEX into the modern PSID in the same way BPP did for the earlier period, and we repeat the estimation. The results are in column 2 of table \ref{Table::Consumption_function}. The pass-through of permanent shocks doubles to $\phi^{(1)}=0.288$, while the pass-through of transitory shocks remains statistically zero (but at a larger magnitude). Using imputed consumption data thus inflates the transmission of permanent shocks and lowers the degree of insurance. This is likely the byproduct of measurement/imputation error, evident in the variance $\sigma^2_{u_c}$ of consumption error that more than triples between the true and the imputed data. By contrast, the variance of taste heterogeneity $\sigma^2_{\xi}$ remains similar in both cases, and very close to BPP. The imputation, therefore, seems to bring in large amounts of measurement error, explaining about 1/3 of the difference between our lower $\phi^{(1)}$ and BPP's higher estimate. Appendix \ref{SubAppendix::Empirics_CEX} discusses the details of the imputation and a mechanism through which in particular it inflates $\phi^{(1)}$: the imputation involves the \emph{inversion} of a demand system, so certain types of measurement error can inflate the imputed income-consumption covariance -- evident also in table \ref{Table::Empirical_Moments_Income_Consumption}.

\begin{table}[t!]  
\begin{center}
\caption{Estimates of the consumption function}\label{Table::Consumption_function}
\begin{tabular}{l C{1.8cm} C{1.8cm} C{1.9cm} C{2.1cm} c C{2.4cm}}
\toprule
\emph{Consumption fn.:}				& \mcl{4}{c}{\textbf{Linear}} 							&& \textbf{Quadratic} \\
\cmidrule{2-5}\cmidrule{7-7}
\emph{Income moments:} 					& 2\textsuperscript{nd}     & 2\textsuperscript{nd}     & 2\textsuperscript{nd}, 3\textsuperscript{rd}      
                                        & 2\textsuperscript{nd}, 3\textsuperscript{rd}, 4\textsuperscript{th} 
                                        && 2\textsuperscript{nd}, 3\textsuperscript{rd}, 4\textsuperscript{th}   	\\                                        
\emph{Consumption data:}     			& PSID          & CEX           & PSID          & PSID          && PSID     \\
                                       	& (1)         	& (2)         	& (3)           & (4)           && (5)      \\
\midrule                                       	
$\phi^{(1)}$							&0.152			&0.288			&0.160			&0.112			&&0.134\\
										&(0.028)		&(0.044)		&(0.029)		&(0.041)		&&(0.026)\\
$\psi^{(1)}$							&-0.006			&-0.109			&-0.012			&0.007			&&-0.003\\
										&(0.045)		&(0.078)		&(0.038)		&(0.056)		&&(0.052)\\
$\phi^{(2)}$							&				&				&				&				&&-0.040\\
										&				&				&				&				&&(0.026)\\
$\psi^{(2)}$							&				&				&				&				&&0.015\\
										&				&				&				&				&&(0.034)\\
$\omega^{(22)}$							&				&				&				&				&&0.320\\
										&				&				&				&				&&(0.770)\\         	
\midrule
$\sigma^2_{\xi}$						&0.019			&0.019			&0.019			&0.020			&&0.019\\
										&(0.001)		&(0.004)		&(0.001)		&(0.001)		&&(0.001)\\
$\gamma_{\xi}$							&				&				&-0.001			&-0.001			&&\\
										&				&				&(0.001)		&(0.001)		&&\\
$\kappa_{\xi}$							&				&				&				&0.005			&&\\
										&				&				&				&(0.001)		&&\\
\midrule
$\sigma^2_{u_c}$						&0.044			&0.140			&0.044			&0.044			&&0.044\\
										&(0.002)		&(0.009)		&(0.002)		&(0.001)		&&(0.002)\\
$\gamma_{u_c}$							&				&				&0.002			&0.002			&&\\
										&				&				&(0.001)		&(0.001)		&&\\
$\kappa_{u_c}$							&				&				&				&0.013			&&\\
										&				&				&				&(0.001)		&&\\					
\bottomrule
\end{tabular}
\caption*{\fsz\emph{Notes:} The table presents the estimates of the parameters of the consumption function, imposing homogeneous transmission parameters over the lifecycle/in the cross-section and stationarity of taste heterogeneity and consumption measurement error. Columns 1-4 present parameter estimates in the linear function, while column 5 presents estimates in the quadratic case; the order of moments targeted in each case is shown at the top of the table. Except for column 2 where we use imputed data from the CEX, all other columns use consumption data internally available in the PSID. Estimation is via equally weighted GMM; block bootstrap standard errors are in parentheses. Results from optimal GMM are similar (appendix table \ref{AppTable::Consumption_function_OptimalMD}).}
\end{center}
\end{table}

Although the imputation inflates $\phi^{(1)}$, the inflated value in column 2 remains substantially lower than BPP. The PSID switched in 1997 from annual to biennial frequency. It is thus possible that, at the lower frequency, some higher-frequency income shocks are smoothed out and their transmission into consumption muted/averaged out. Ideally, we would assess this by re-estimating our model using annual growth rates, but this is not feasible after 1997. However, we can use the original BPP data from 1980--1992 to estimate the model \emph{annually} versus \emph{biennially}. We do this in appendix \ref{SubAppendix::Empirics_ComparisonBPP}. While we obtain $\phi^{(1)}=0.64$ using annual growth rates (as in BPP), this drops to $\phi^{(1)}=0.4$--0.5 using biennial rates. The decline is sharper when we target time-invariant moments, which is recently also shown in \citet{ChatterjeeMorleySingh2021_BPP}. The biennial nature of the modern data thus explains another 1/3 of the gap between our lower $\phi^{(1)}$ and BPP's higher estimate.\footnote{$\phi^{(1)}$ is the pass-through of permanent shocks at an \emph{annual} rate in both annual and biennial cases. This becomes clear in appendix \ref{Appendix::Identification}. The difference in the estimates of $\phi^{(1)}$ between the two cases arises because with biennial data we measure an ‘aggregate’ pass-through within a given biennial period. Some higher-frequency shocks are inevitably smoothed out at the observed lower frequency and their transmission muted.} We attribute the remaining gap to calendar time effects, e.g., changes in taxes/benefits and other insurance channels between 1980--1992 and 1999--2019. \citet{BorellaDeNardiEtAl2022_Taxes} offer a discussion of such changes. It is worth noting that we are not the first to find a low pass-through of permanent shocks in the recent years. ABB and \citet{Arellano_Blundell_Bonhomme_Light2021} use recent PSID data and find an average marginal propensity to consume at about $0.3$ (specification with heterogeneity closest to ours). The implied pass-through rates in ABB are similar to ours, as we further show below.

\

\noindent \textbf{Introducing higher-order moments.} The estimation of the linear consumption function targeting higher-order moments is our first attempt to uncover how the tails of income risk affect the pass-through of shocks into consumption. In other words, we want to measure the degree of partial insurance targeting moments beyond the workhorse income-consumption covariance in order to account for the non-Gaussian features of modern income dynamics. We introduce the higher-order moments incrementally.\footnote{We use the corresponding pre-estimated income parameters from columns 2 and 3 of table \ref{Table::Income_process} respectively.}

Column 3 of table \ref{Table::Consumption_function} presents the results from targeting third- in addition to second-order moments of consumption and income in the PSID. The transmission parameters remain largely unchanged. We estimate $\phi^{(1)}=0.160$ (as opposed to $0.152$ with only second moments) and $\psi^{(1)}=-0.012$ (again indistinguishable from zero). We thus again find large partial insurance to permanent shocks and full insurance to transitory shocks. Skewness in taste heterogeneity and consumption error is small (much smaller than skewness in income shocks), while their variances remain unchanged from the specification with only second moments. Therefore, left skewness in consumption is driven by negatively skewed income shocks rather than by taste heterogeneity or measurement error. 

Column 4 of table \ref{Table::Consumption_function} presents the results from targeting all income and consumption moments up to fourth-order. The earlier findings hold up. We estimate $\phi^{(1)}=0.112$ (versus $0.152$ and $0.160$ in the lower-order cases) and $\psi^{(1)}=0.007$ (indistinguishable from zero). A $10\%$ permanent change in income thus results in $1.12\%$ shift in consumption. Therefore, if anything, we find evidence for slightly \emph{more} insurance to permanent shocks upon inclusion of the fourth moments. The dispersion and skewness of taste heterogeneity and consumption error do not change much from the lower-order cases; we find evidence for excess kurtosis in both cases, albeit at a much lesser extent than for income shocks.\footnote{Appendix table \ref{AppTable::Consumption_function_CEX} uses imputed consumption from the CEX to target higher-order moments as in columns 3-4 of table \ref{Table::Consumption_function}. We do see slightly higher pass-through of permanent shocks upon inclusion of the higher-order moments, but we also find a strong negative response to transitory shocks, which is hard to explain. We do not think those results are very credible; the imputation brings in substantial measurement and imputation error, which likely becomes worse the higher the order of the target moments is.}

Clearly, targeting higher-order moments within the linear consumption function has not altered much the degree of partial insurance vis-\`a-vis targeting second-order moments alone. This is counterintuitive given the extent of tail income risk in the data and ample evidence in the literature to the contrary.\footnote{\citet{Ai_Bhandari2021} show that tail risk is uninsured in realistically parameterized labor markets. However, \citet{DeNardi_Fella_PazPardo2019} and \citet{Busch_Ludwig2024} find higher insurance when shocks are non-normal, due to stronger precautionary motives. The latter caution against BPP's linear framework, showing evidence of bias in the presence of tail risk. Our quadratic specification addresses their concern.} One would expect that tail risk \emph{increases} the pass-through of shocks, as there must be a limit beyond which households cannot further insure consumption against extreme events, even if such risk triggers a stronger precautionary motive. Yet, the linear consumption function leaves no room to such events, because it reflects a linearization of the consumption rule around a deterministic lifecycle profile. Tail events, however, move the agent far from that profile: if utility is sufficiently concave, a large shock will command a large consumption response. A linearized marginal utility fails to pick up such large response, which in turn explains why we do not see much action upon including higher-order moments. In other words, tail risk has, by design, no role in the linear consumption function, which in turn motivates our higher-order specification subsequently.

\

\noindent To summarize our findings from the linear consumption function: 1.) we replicate BPP over the recent years and find much lower pass-through of permanent shocks (larger degree of partial insurance); the imputation of consumption from the CEX and the biennial nature of the modern PSID explain the bulk of the difference from BPP; 2.) targeting higher-order moments does not alter the overall degree of consumption partial insurance, which is not entirely surprising as tail risk has, by design, no role in the linear framework.

\subsection{Quadratic consumption function}\label{SubSec::Results_Quadratic_Consumption_Function}

We now turn to the estimation of the quadratic consumption function. The results appear in column 5 of table \ref{Table::Consumption_function}; the full set of targeted moments is shown in appendix \ref{SubSubAppendix::Identification_Targeted_Moments_QuadrFunction}.\footnote{Estimation of the quadratic consumption function requires up to 4\textsuperscript{th}-order moments of the income shocks, so we use the pre-estimated income parameters from column 3, table \ref{Table::Income_process}.}

\subsubsection{Main results}\label{SubSubSec::Results_Quadratic_Consumption_Function_Baseline}

We estimate the transmission parameter of the average \emph{permanent} shock (the coefficient on the linear part of the consumption polynomial) at $\phi^{(1)}=0.134$, close to the values $0.11$--$0.16$ obtained from the linear specification. We estimate the transmission parameter of the average \emph{transitory} shock at $\psi^{(1)}=-0.003$, statistically indistinguishable from zero. As $\phi^{(1)}$ and $\psi^{(1)}$ remain largely unaltered relative to the linear case, interest lies in the transmission parameters of the typical shocks \emph{away} from the middle of the distribution, i.e., the coefficients on the quadratic part of the polynomial. 

We estimate a negative $\phi^{(2)}=-0.04$ for permanent shocks but a positive $\psi^{(2)}=0.015$ for transitory ones. $\phi^{(2)}$ is statistically insignificant at the margin for now but turns highly significant subsequently. By contrast, $\psi^{(2)}$ is almost always insignificant. The coefficient $\omega^{(22)}$ on the product of the two shocks is large but always imprecise. The variance of taste heterogeneity $\sigma_{\xi}^2$ and consumption measurement error $\sigma_{u_c}^2$ remain effectively unchanged from the linear case.\footnote{Skewness and kurtosis of taste heterogeneity and consumption error are not identified because they require higher-order moments of consumption, which depend on higher than fourth-order moments of income.}\textsuperscript{,}\footnote{The consumption-income moments targeted in the quadratic case differ from the moments targeted in the most general linear case in column 4, table \ref{Table::Consumption_function}. This is because in the quadratic case certain 3\textsuperscript{rd}- and 4\textsuperscript{th}-order consumption moments require up to 8\textsuperscript{th} moments of income shocks, which we do not model. This raises an issue of comparability between the linear and quadratic specifications, as the target moments differ (appendix table \ref{AppTable::List_of_Moments_C}). However, we can estimate the linear function targeting the exact same moments that we use in the quadratic case, as the linear function is identified by a smaller set of lower-order moments anyway. Our estimates of the linear function in that case do not differ from those in column 4, table \ref{Table::Consumption_function}, so estimation is robust to the choice of moments. These results are available upon request.}

At face value, $\phi^{(2)}$ has important implications for the pass-through of permanent shocks into consumption. To help interpret the results, recall that the pass-through is given by \eqref{Eq::Quadratic_C_function_Phi} and depends on $\phi^{(1)}$, $\phi^{(2)}$, $\omega^{(22)}$, and the magnitudes and signs of both income shocks. If, for simplicity, we set the transitory shock to its mean value ($v_{it}=0$) and plug in the parameter estimates from table \ref{Table::Consumption_function}, we get
\begin{equation*}
	\partial \Delta c_{it}/\partial \zeta_{it} = 0.134 - 0.08 \times \zeta_{it}.
\end{equation*}
Two main results follow. 

First, bad (negative) permanent shocks are more unsettling than good (positive) ones because they are associated with a larger transmission parameter into consumption. Consider, for example, a large permanent shock to household income, $|\zeta_{it}|=0.5$, about three standard deviations away from the mean. If the shock is negative, $\partial \Delta c_{it}/\partial \zeta_{it} = 0.174$ (\emph{s.e.}$=0.036$, estimated by the delta method); by contrast, if the shock is positive, then $\partial \Delta c_{it}/\partial \zeta_{it} = 0.094$ (\emph{s.e.}$=0.036$). We can generalize this to bad and good shocks of any magnitude and conclude that
\begin{equation*} 
	\left.\frac{\partial \Delta c_{it}}{\partial \zeta_{it}}\right|_{\zeta<0} ~>~ \left.\frac{\partial \Delta c_{it}}{\partial \zeta_{it}}\right|_{\zeta>0}. 
\end{equation*}
Therefore, bad permanent shocks are not just more extreme than good shocks, i.e., farther away from the zero mean -- see left skewness in table \ref{Table::Income_process}, but they also have larger pass-through rates into consumption. Bad permanent shocks are thus \emph{double hurtful}. 

Second, severe bad shocks are more hurtful than moderate ones because they too are associated with larger transmission parameters into consumption. Compare, for example, a very large bad permanent shock, $\zeta_{it}=-0.7$, about four standard deviations away from the mean, with a moderate one, $\zeta_{it}=-0.07$, half standard deviation from the mean. In the first case, $\partial \Delta c_{it}/\partial \zeta_{it} = 0.19$ (\emph{s.e.}$=0.044$) while in the latter case $\partial \Delta c_{it}/\partial \zeta_{it} = 0.139$ (\emph{s.e.}$=0.026$). We can generalize this to bad shocks of any magnitude and conclude that
\begin{equation*} 
    \left.\frac{\partial \Delta c_{it}}{\partial \zeta_{it}}\right|_{\zeta<<0} ~>~ \left.\frac{\partial \Delta c_{it}}{\partial \zeta_{it}}\right|_{\zeta<0}. 
\end{equation*}

Why would a positive or small shock be better insured? When a good shock hits, households save part of it and consumption does not respond too much. When a bad shock hits, however, households insure themselves by running down savings or by using other formal or informal insurance arrangements. Yet there is clearly a limit to how much they can dissave or borrow. As the magnitude of the bad shock worsens, these insurance arrangements will be eventually exhausted, resulting in a larger fraction of the shock transmitting into consumption. A large negative shock thus affects consumption by relatively more than a small or positive shock does.\footnote{The results also imply that the pass-through of a \emph{good} shock decreases with its magnitude. The reason behind this counterintuitive result is that we lack sufficient degrees of freedom to separate among all combinations of large/small, positive/negative shocks. However, large left skewness and excess kurtosis imply that \emph{large} good shocks are rare. We should thus not extrapolate the implications of the parameter estimates for the case of \emph{large} good shocks, as such shocks are effectively not present in our sample.}

The asymmetric response to bad versus good income news has recently been documented in survey questions in which subjects directly report how they would adjust consumption in different hypothetical scenarios. For example, \cite{Bunn_LeRoux_Reinold_Surico2018} find that the marginal propensity to consume from negative income shocks is larger than from positive shocks, while \citet{ChristelisEtAl_2019_AsymmetricConsumptionEffects} and \cite{Fuster_Kaplan_Zafar2021} find that responses to losses are much larger than responses to gains, and the magnitude of the response increases with the magnitude of the bad shock. This literature, however, elicits responses to \emph{transitory} shocks. By contrast, we establish these asymmetries in response to \emph{permanent} shocks in a context in which the shocks are not directly observed (namely in survey data). 

Moving on to transitory shocks, their pass-through is given by \eqref{Eq::Quadratic_C_function_Psi}; this depends on $\psi^{(1)}$, $\psi^{(2)}$, $\omega^{(22)}$, and on the magnitudes and signs of both income shocks. If we set the permanent shock to its mean value ($\zeta_{it}=0$) and plug in the parameter estimates from table \ref{Table::Consumption_function}, we get
\begin{equation*}
    \partial \Delta c_{it}/\partial v_{it} = -0.003 + 0.03 \times v_{it}.\\[-3pt]
\end{equation*}
Unlike the transmission parameters of permanent shocks, neither $\psi^{(1)}$ nor $\psi^{(2)}$ are statistically significant. This implies that transitory shocks, be they good or bad, moderate or extreme, do not pass through into consumption.

Despite their statistical insignificance, the transmission parameters of transitory shocks have, at face value, an interesting implication: good and large shocks have a numerically larger pass-through than smaller ones, whose pass-through is economically zero. While we should be cautious in drawing conclusions about \emph{large} good transitory shocks (given left skewness and excess kurtosis, such shocks are rare in our sample), this implication helps reconcile the large pass-through found in quasi-experimental studies with the negligible pass-through found in BPP and similar studies that employ covariance restrictions on survey data. The former typically exploit large infrequent transitory gains \citep[e.g., tax rebates or lotteries as in][]{ParkerSoulelesJohnsonMcClelland2013_StimulusPayments,MisraSurico2014_EvidenceFiscalStimulus,FagerengEtAl_2021LotteryWins}, which are unlikely to be present in the pooled survey data that the latter studies exploit \citep[e.g., BPP,][]{Blundell_Pistaferri_Saporta-Eksten2016,Theloudis2021}. In an environment where the transmission of shocks depends on their magnitude, these two types of settings will lead to opposite conclusions. Our framework can potentially reconcile the two by permitting transitory shocks to pass through to consumption at a varying rate that depends on the shocks' magnitude and sign.

To summarize our findings from the quadratic function up to this point: 1.) bad permanent shocks reduce consumption by more than equal-sized good permanent shocks increase it; 2.) severe bad permanent shocks reduce consumption by more than moderate ones do. For a given value of $\phi^{(1)}$, the econometrician will overestimate the true degree of partial insurance against bad permanent shocks if she assumes a linear consumption function instead of the nonlinear specification herein. By contrast, transitory shocks, be they good or bad, small or large, do not affect consumption, at least not in a statistically significant way.\footnote{Similar conclusions are obtained if we use imputed consumption from the CEX. In particular, $\phi^{(2)}<0$, with a magnitude similar to that obtained from the PSID, and $\psi^{(2)}\approx>0$ (appendix table \ref{AppTable::Consumption_function_CEX}).}

\subsubsection{Results by age and position in the income distribution}\label{SubSubSec::Results_Quadratic_Consumption_Function_AgeIncome}

We now consider how partial insurance changes with age, by repeating the estimation over three, almost equally sized, subsamples of households with the male spouse between 30--40, 41--50, and 51--65 years old respectively.\footnote{We target the same moments as earlier, broken down by the age of the male spouse. The moments of income shocks are also age-specific (appendix table \ref{AppTable::Income_process_age_lagY}).} Intuitively, the transmission parameters of shocks should exhibit some age dependence since wealth accumulates over the lifecycle, which in turn influences the household's ability to insure against income fluctuations. 

Table \ref{Table::Consumption_function_age} reports results from the quadratic specification (see table \ref{AppTable::Linear_Consumption_function_age} for the linear specification). $\phi^{(1)}$ decreases with age, so the average permanent shock affects consumption less as households age. In line with theory, this pattern reflects the increasing self-insurance role of financial wealth over the lifecycle. $\phi^{(2)}$, the transmission parameter of the typical shock away from the mean, also decreases with age. It drops from $0.006$ to $-0.008$ as we move from ages 30--40 to 41--50 (neither statistically significant) and becomes very negative and significant at $\phi^{(2)} = -0.104$ in the 51--65 bracket. As a result, large negative permanent shocks impact consumption more severely at older ages. Consider, again, $\zeta_{it}=-0.5$. At age 45, its pass-through is $\partial \Delta c_{it}/\partial \zeta_{it} = 0.123$ (\emph{s.e.}$=0.080$), while its pass-through at age 60 is almost double, $\partial \Delta c_{it}/\partial \zeta_{it} = 0.218$ (\emph{s.e.}$=0.077$). 

\begin{table}[]  
\begin{center}
\caption{Estimates of the consumption function, by age}\label{Table::Consumption_function_age}
\begin{tabular}{L{4cm} R{1.4cm} L{1.4cm} c R{1.4cm} L{1.4cm} c R{1.4cm} L{1.4cm}}
\toprule
\emph{Consumption fn.:}                 & \mcl{8}{c}{\textbf{Quadratic}} \\
\cmidrule{2-9}
\emph{Age household head:}              & \mcl{2}{c}{30--40}      && \mcl{2}{c}{41--50}     && \mcl{2}{c}{51--65} \\
\cmidrule{2-3}\cmidrule{5-6}\cmidrule{8-9}
\emph{Income moments:}                  &  \mcl{2}{c}{2\textsuperscript{nd}, 3\textsuperscript{rd}, 4\textsuperscript{th}}  
                                        && \mcl{2}{c}{2\textsuperscript{nd}, 3\textsuperscript{rd}, 4\textsuperscript{th}}  
                                        && \mcl{2}{c}{2\textsuperscript{nd}, 3\textsuperscript{rd}, 4\textsuperscript{th}}  \\
\emph{Consumption data:}                & \mcl{2}{c}{PSID}              && \mcl{2}{c}{PSID}         && \mcl{2}{c}{PSID}     \\
                                        & \mcl{2}{c}{(1)}               && \mcl{2}{c}{(2)}          && \mcl{2}{c}{(3)}      \\
\midrule                                        
$\phi^{(1)}$                            &0.149      &(0.050)            &&0.115     &(0.049)        &&0.113     &(0.057)    \\
$\psi^{(1)}$                            &0.105      &(0.090)            &&0.019     &(0.090)        &&-0.063    &(0.085)    \\
$\phi^{(2)}$                            &0.006      &(0.038)            &&-0.008    &(0.063)        &&-0.104    &(0.052)    \\
$\psi^{(2)}$                            &0.036      &(0.061)            &&0.048     &(0.046)        &&-0.052    &(0.073)    \\
$\omega^{(22)}$                         &-1.089     &(1.465)            &&-0.294    &(1.195)        &&2.984     &(0.791)    \\
\midrule
$\sigma^2_{\xi}$                        &0.013      &(0.004)            &&0.022     &(0.002)        &&0.010     &(0.005)    \\
\midrule
$\sigma^2_{u_c}$                        &0.046      &(0.003)            &&0.040     &(0.002)        &&0.043     &(0.002)    \\
\bottomrule
\end{tabular}
\caption*{\fsz\emph{Notes:} The table presents the estimates of the parameters of the quadratic consumption function, allowing them to vary over ages 30--40, 41--50, and 51--65 of the household head (male spouse). The underlying income moments also vary over these age brackets (appendix table \ref{AppTable::Income_process_age_lagY}). All columns use consumption data internally available in the PSID. Estimation is via equally weighted GMM; block bootstrap standard errors are in parentheses. Linear specification estimates appear in appendix table \ref{AppTable::Linear_Consumption_function_age}.}
\end{center}
\end{table}

In our data, skewness of permanent shocks becomes more severe as households age, increasing in magnitude from $-0.65$ to $-2.39$ between ages 30--40 and 51--65 (table \ref{AppTable::Income_process_age_lagY}). This is consistent with \citet{Guvenen_Karahan_Ozkan_Song2021}, who document that skewness of earnings growth becomes more negative with the level of earnings and, consequently, with the earner's age. There is more room for income to fall sharply in this case than among the young. The results in table \ref{Table::Consumption_function_age}, however, reveal that large bad shocks affect consumption more at older ages not only because those shocks become more extreme \emph{but also} because the consumption response itself changes ($\phi^{(2)}$ becomes more negative). The type of events that underlie large income downturns at older ages, such as unemployment or declining health -- see table \ref{AppTable::Sources_Income_Shocks} for sources of bad shocks, also trigger permanent cuts of certain types of expenditure (food away from home, offspring education, or work-related expenses), which translate into a consumption response that differs between the old and the young. This does not mean that \emph{moderate} bad shocks are worse insured among the old; for small or moderate shocks, the decreasing pattern in $\phi^{(1)}$ offsets the increasing magnitude of $|\phi^{(2)}|$. 

Finally, the transmission parameters of transitory shocks remain statistically indistinguishable from zero over the entire lifecycle.

Next, we consider how partial insurance varies over the household's position in the income distribution prior to the realization of shocks, i.e., the distribution of $y_{it-1}$. Our linear income process and first-difference approach does not directly link the transmission parameters to the levels of income, therefore we take an empirical approach and estimate the model over different quantiles of $y_{it-1}$. Sample size considerations prevent us from being overly granular; we tried estimation over deciles, quartiles, and tertiles, and the latter was the most robust. For brevity we report results in the bottom (`poor') and top (`rich') tertiles.\footnote{We target the same moments as earlier, broken down by tertiles of the distribution of past income $y_{it-1}$. The moments of shocks are also income level-specific; we report them in appendix table \ref{AppTable::Income_process_age_lagY}.}   

Table \ref{Table::Consumption_function_lagY} reports results from the quadratic specification (see table \ref{AppTable::Linear_Consumption_function_lagY} for the linear specification). Three points emerge. First, $\phi^{(1)}$ among the rich is half the value of the poor, so the average permanent shock affects consumption less strongly among the rich. This reflects again the self-insurance role of wealth given that income-richer households are also wealthier. Second, $\phi^{(2)}$ is negative, confirming the baseline finding that negative permanent shocks have larger pass-through than equal-sized positive shocks. Third, the transmission of transitory shocks shows mixed patterns but it is never statistically significant.

At face value, $\phi^{(2)}$ is more negative among the poor, and highly significant, suggesting that a given bad shock transmits more strongly among them. However, this provides only a partial picture of the underlying dynamics because the nature of income risk is vastly different between poor and rich. Appendix table \ref{AppTable::Income_process_age_lagY} shows that skewness of permanent shocks is \emph{positive} among the poor, with a coefficient of $1.34$ (marginally insignificant), but negative among the rich, with a coefficient of $-4.86$ (multiple times larger than the baseline, and highly significant). One may thus study how large bad/good shocks from \emph{a given position} in the underlying distributions, e.g., shocks in the 3\textsuperscript{rd} or 97\textsuperscript{th} percentiles, affect consumption in each case. Simulations are needed to infer the value of the specific shocks.\footnote{We simulate a mixture of normal distributions separately for the poor and the rich, given the estimated income parameters of appendix table \ref{AppTable::Income_process_age_lagY}. We provide details on similar simulations in appendix \ref{Appendix::WelfareCosts}.} 

Adopting this approach, the lower-tail shock shifts $c_{it}$ by $-0.017$ among the poor ($\Delta c_{it}=-0.017$, associated pass-through $0.129$) and by $-0.035$, twice as much, among the rich (pass-through $0.084$). By contrast, the upper-tail shock increases $c_{it}$ by $0.022$ among the poor (pass-through $0.075$) but by only $0.006$ among the rich (pass-through $0.057$). In other words, households in the top of the income distribution (the rich) exhibit a stronger consumption change in response to \emph{bad} news, while households in the bottom (the poor) exhibit a stronger consumption change in response to \emph{good} news. 

\begin{table}[t!]  
\begin{center}
\caption{Estimates of the consumption function, over position in income distribution}\label{Table::Consumption_function_lagY}
\begin{tabular}{L{4cm} R{1.8cm} L{1.8cm} c R{1.8cm} L{1.8cm}}
\toprule
\emph{Consumption fn.:}                 & \mcl{5}{c}{\textbf{Quadratic}} \\
\cmidrule{2-6}
\emph{Lag income $y_{it-1}$:}           & \mcl{2}{c}{Bottom tertile}      && \mcl{2}{c}{Top tertile}     \\
\cmidrule{2-3}\cmidrule{5-6}
\emph{Income moments:}                  &  \mcl{2}{c}{2\textsuperscript{nd}, 3\textsuperscript{rd}, 4\textsuperscript{th}}  
                                        && \mcl{2}{c}{2\textsuperscript{nd}, 3\textsuperscript{rd}, 4\textsuperscript{th}}  \\
\emph{Consumption data:}                & \mcl{2}{c}{PSID}              && \mcl{2}{c}{PSID}             \\
                                        & \mcl{2}{c}{(1)}               && \mcl{2}{c}{(2)}          \\
\midrule                                        
$\phi^{(1)}$                            &0.113          &(0.029)        &&0.063     &(0.100)    \\
$\psi^{(1)}$                            &0.186          &(0.199)        &&0.275     &(0.200)    \\
$\phi^{(2)}$                            &-0.064         &(0.029)        &&-0.025    &(0.101)    \\
$\psi^{(2)}$                            &0.210          &(0.220)        &&-0.268    &(0.197)    \\
$\omega^{(22)}$                         &1.965          &(1.513)        &&1.072     &(2.144)    \\
\midrule
$\sigma^2_{\xi}$                        &0.011          &(0.005)        &&0.025     &(0.004)    \\
\midrule
$\sigma^2_{u_c}$                        &0.044          &(0.002)        &&0.041     &(0.002)    \\
\bottomrule
\end{tabular}
\caption*{\fsz\emph{Notes:} The table presents the estimates of the parameters of the quadratic consumption function in the bottom and top tertiles of the distribution of lag income $y_{it-1}$. The underlying income moments also vary over the income distribution (appendix table \ref{AppTable::Income_process_age_lagY}). All columns use consumption data internally available in the PSID. Estimation is via equally weighted GMM; GMM standard errors clustered at the household level are in parentheses. Linear specification estimates appear in appendix table \ref{AppTable::Linear_Consumption_function_lagY}.}
\end{center}
\end{table}

These findings agree with ABB, who simulate the transmission of shocks in the 10\textsuperscript{th} and 90\textsuperscript{th} percentiles. The reported $\Delta c_{it}$ -- impulse responses with heterogeneity in their appendix figure S30 -- are in the same ballpark, especially for bad shocks. While this is reassuring, our findings further suggest that ABB's nonlinear persistence is not crucial for the main consumption results, i.e., the asymmetric impact of different shocks or at different income levels. Our approach also highlights two distinct forces at play: the underlying nature of income risk (i.e., income parameters) and the nature of the consumption response (i.e., consumption parameters). The larger consumption change among the rich after a lower-tail shock is not because their consumption response differs materially from the poor ($\phi^{(2)}$'s smaller magnitude among the rich in fact mitigates said shocks), but mostly due to stronger downside risk that they face (the magnitude of the bad shock is simply larger in this case). 

To distinguish between these two forces, one must evaluate the marginal distribution of income risk. While ABB's quantile method can flexibly estimate consumption changes over different income quantiles -- and compare changes after, say, 10- and 90-percentile shocks -- it does so without explicitly accounting for the higher-order features of the distribution these shocks come from. Technically, in the context of a linear income process, and assuming transitory shocks away and no interaction between the income level and its shock, ABB estimate  
\begin{equation*}
    c_{it} = g_{t}(\underbrace{y_{it-1} + F_{t}^{-1}(u_{it})}_{y_{it}},~\dots),\\[-4pt]
\end{equation*}
where $\zeta_{it} = F_{t}^{-1}(u_{it})$ is the persistent shock and $u_{it}$ denotes the quantile of the shocks' distribution $F_{t}$. They then simulate the consumption pass-through over different values of $y_{it-1}$ and $u_{it}$, without accounting for $F_{t}$. This may lead inadvertently to comparisons between consumption responses to very different income shocks, even if those shocks come from the same position in their respective distributions (the nature of income risk varies over the distribution of income).\footnote{Apparently, this may also lead inadvertently to comparisons between consumption responses to seemingly different shocks, when these shocks may in fact be similar to one another. For instance, our estimates for skewness and kurtosis reveal that the 10- and 90-percentile shocks are \emph{both} negative and have almost the same size among the poor; this is why we looked farther out at the 3\textsuperscript{rd} and 97\textsuperscript{th} percentiles.} By contrast, our approach allows us to evaluate the second and higher moments of $F_{t}$ and estimate pass-through rates that explicitly shift with the economically relevant shock, i.e., with $\zeta_{it} = F_{t}^{-1}(u_{it})$. It can thus describe more transparently how income shocks of different signs/sizes shift consumption.

How does a less negative $\phi^{(2)}$ among the rich (smaller response to larger bad shocks) reconcile with a more negative $\phi^{(2)}$ among the old (stronger response), who are presumably also rich? The answer lies in the composition of the old (they are in fact evenly present in each income tertile) and in the interplay between income risk and consumption response. Skewness of permanent shocks is twice as large among the rich ($-4.86$) as among the old ($-2.39$). The consumption distribution is similar for both, so the rich must have overall \emph{more} insurance to a given bad shock. This explains why $\phi^{(2)}$'s magnitude is four times smaller among them. In both cases (old, rich), bad permanent shocks from a given position in the distribution reduce consumption more than among their counterparts (young, poor).

\subsubsection{Results by wealth and education}\label{SubSubSec::Results_Quadratic_Consumption_Function_WealthEduc}

We next consider four subsamples formed on the basis of household wealth and education. We use conventional definitions for `low' versus `high' wealth (the cutoff is median wealth in the sample) and `no' versus `some' college (the cutoff is 12 years of education). Appendix table \ref{AppTable::Descriptives} provides summary statistics for these subsamples.\footnote{Wealth is defined as the sum of the value of housing, vehicles, business, investment, pension, savings, and other assets, net of mortgage and medical, student, and credit card debt.} 

\begin{table}[t!]  
\begin{center}
\caption{Estimates of the consumption function, by wealth and education}\label{Table::Quadratic_Consumption_function_subsamples}
\begin{tabular}{l C{2.3cm} C{2.3cm} c C{2.3cm} C{2.3cm}}
\toprule
\emph{Consumption fn.:}                 & \mcl{5}{c}{\textbf{Quadratic}}                                \\
\cmidrule{2-6}
                                        & \mcl{2}{c}{Low wealth}        && \mcl{2}{c}{High wealth}      \\
                                        & No college    & Some college  && No college   & Some college  \\
\cmidrule{2-3}\cmidrule{5-6}
\emph{Income moments:}                  &  2\textsuperscript{nd}, 3\textsuperscript{rd}, 4\textsuperscript{th}     
                                        &  2\textsuperscript{nd}, 3\textsuperscript{rd}, 4\textsuperscript{th}     
                                        && 2\textsuperscript{nd}, 3\textsuperscript{rd}, 4\textsuperscript{th} 
                                        &  2\textsuperscript{nd}, 3\textsuperscript{rd}, 4\textsuperscript{th} \\                                        
\emph{Consumption data:}                & PSID          & PSID          && PSID         & PSID          \\
                                        & (1)           & (2)           && (3)          & (4)           \\
\midrule                                        
$\phi^{(1)}$                            &0.242          &0.192          &&0.164         &0.079          \\
                                        &(0.051)        &(0.052)        &&(0.067)       &(0.034)        \\
$\psi^{(1)}$                            &-0.251         &0.015          &&0.006         &0.004          \\
                                        &(0.131)        &(0.108)        &&(0.085)       &(0.073)        \\
$\phi^{(2)}$                            &-0.094         &0.005          &&-0.076        &-0.032         \\
                                        &(0.062)        &(0.052)        &&(0.057)       &(0.041)        \\
$\psi^{(2)}$                            &-0.162         &-0.001         &&0.177         &0.001          \\
                                        &(0.087)        &(0.074)        &&(0.089)       &(0.048)        \\
$\omega^{(22)}$                         &2.998          &-0.660         &&-2.228        &0.441          \\
                                        &(1.423)        &(1.530)        &&(1.568)       &(0.995)        \\
\midrule            
$\sigma ^2_{\xi}$                       &0.006          &0.013          &&0.006         &0.025          \\
                                        &(0.008)        &(0.002)        &&(0.010)       &(0.002)        \\
\midrule            
$\sigma ^2_{u_c}$                       &0.046          &0.045          &&0.038         &0.043          \\
                                        &(0.003)        &(0.003)        &&(0.004)       &(0.002)        \\              
\bottomrule
\end{tabular}
\caption*{\fsz\emph{Notes:} The table presents the estimates of the parameters of the quadratic consumption function, allowing them to vary across wealth and education groups. The underlying income moments also vary by wealth and education (results not shown for brevity). Low (high) wealth is defined on the basis of household wealth being less  (more) than median real wealth over $1999$--$2019$. No versus some college is defined on the basis of the highest level of education attained by the male spouse. All columns use consumption data internally available in the PSID. Estimation is via equally weighted GMM; GMM standard errors clustered at the household level are in parentheses. Linear specification estimates appear in appendix table \ref{AppTable::Linear_Consumption_function_subsamples}.}
\end{center}
\end{table}

Table \ref{Table::Quadratic_Consumption_function_subsamples} reports results from the quadratic model (see table \ref{AppTable::Linear_Consumption_function_subsamples} for the linear model). Three points stand out. First, $\phi^{(1)}$ decreases monotonically from group 1 (low wealth, no college) to group 4 (high wealth, some college). $\phi^{(1)}$ measures the transmission of the \emph{average} permanent shock, so these groups possess gradually more insurance to average shocks. This reflects the insurance role of wealth, which increases monotonically across groups. 

Second, $\phi^{(2)}$ is negative but increases (i.e., less negative) as we move rightwards across groups, with the exception of group 2 (low wealth, some college) where it equals zero. Negative shocks thus transmit weakly more strongly than positive shocks among everyone, but wealthier and more educated agents have more insurance to bad shocks of a \emph{given} size. They also face, however, much stronger downside income risk, a pattern consistent with the gradient in income risk between poor and rich in the earlier discussion. 

Third, the transmission parameters of transitory shocks are mostly insignificant, with two exceptions. In group 1 (low wealth, no college), $\psi^{(1)}$ and $\psi^{(2)}$ are both significant and negative, leading to the odd prediction that larger bad shocks may increase consumption slightly while smaller good shocks reduce it. In group 3 (high wealth, no college; the smallest group in our sample), $\psi^{(2)}$ is positive and significant, suggesting that large good shocks have a larger pass-through than smaller ones (thus opposite from group 1). Both cases are at odds with the bulk of estimates for the transmission of transitory shocks throughout the paper that suggest that such shocks are fully insured, so these results remain a puzzle.

\subsubsection{Link with structural model and some estimates of welfare costs}\label{SubSubSec::Results_Quadratic_Consumption_Function_QuantModel}

Our quasi reduced-form approach so far has enabled us to estimate an asymmetric transmission of income shocks without relying on a particular parameterization of preferences, expectations, or the income distribution. Estimating a fully specified model is outside the scope of this paper, but it is useful to evaluate our findings against the few structural lifecycle models that have recently incorporated higher-order income risk. 

\citet{Madera2019_ConsumptionResponseTailShocks} finds that earnings and nondurables are correlated mainly in the middle of the earnings distribution, with weaker correlations in the tails -- especially when compared to earnings and durables. This suggests that tail shocks pass through to durables (which we do not model) more than to nondurables. Negative shocks exhibit larger pass-through rates than equal-sized positive shocks, albeit at twice the magnitude that we estimate here.

\citet{DeNardi_Fella_PazPardo2019} find that tail risk has large welfare costs because, among other things, it increases precautionary savings enabling \emph{more} insurance to a given shock. They show that the BPP method relying on the workhorse consumption-income covariance implies a substantially higher pass-through rate than the true rate in the model.

\citet{Busch_Ludwig2024} simulate a model with borrowing constraints. While we do not explicitly model such constraints, a strength of our approach is that it measures insurance without relying on specific economic mechanisms. They find that the marginal propensity to consume rises with the size of a negative permanent shock as agents approach the constraint, but falls with the size of a positive shock. Their pass-through rates are slightly higher than ours, though their model excludes other form of insurance beyond self-insurance. 

\citet{GuvenenOzkanMadera2024NonGaussianRisk} simulate a model with retirement, stochastic lifetimes, and altruism. The consumption response varies with the size and sign of shocks; the pass-through of the \emph{average negative} permanent shock is between 0.29 and 0.46, depending on age, which is higher than ours but lower than BPP. They show that measuring pass-through as in BPP overestimates the true transmission rate by 1.5 times. 

We read these results as not too different from ours. Structural models seem to produce similar asymmetries in the transmission of income shocks, with some evidence that, as bad shocks hits, there is a limit to how much one can dissave or borrow. The magnitude of our pass-through rate of permanent shocks is uniformly lower than its structural counterpart (consumption transmission in the data likely reflects \emph{multiple} insurance sources, beyond self-insurance and taxes examined in these models), yet other empirical studies using the updated PSID, e.g., ABB and \citet{Arellano_Blundell_Bonhomme_Light2021}, find similar rates to ours.

Finally, we obtain back-of-the-envelope estimates of the welfare costs of tail income risk. Specifically, we simulate permanent income shocks from mixtures of normal distributions, we use those shocks to simulate consumption responses based on the quasi-reduced form consumption function in this paper, and we use power utility to calculate the discounted present value of a lifetime consumption stream associated with a given profile of income risk. Our baseline choice is tail risk similar to what we estimate in the data, while our counterfactual is Gaussian risk with similar variance. We then estimate the welfare costs of tail risk as the fraction of lifetime consumption a household facing Gaussian risk (counterfactual) would give up to avoid tail risk (baseline). We provide full details in appendix \ref{Appendix::WelfareCosts}. 

Households facing large downside risk (ranging from a permanent shock 3 standard deviations below the mean at least once in their lifetime -- 40\% of households experience this -- to a shock 5 standard deviations below the mean -- a year's long unemployment, 17\% of households experience this at least once) would give up 4\%-8\% of lifetime consumption when income risk transmits linearly. This rises to 6\%-12\%, almost up to 1/8 of lifetime consumption, in the presence of nonlinear transmission. These estimates vary across the income and wealth distributions (appendix \ref{Appendix::WelfareCosts}), but the results remain in the same ballpark. 

Although we calculate the welfare costs outside of a full structural model (among other things, taking wealth as given; hence `back-of-the-envelope'), our estimates seem to fall in the same ballpark as estimates from the structural papers described above. As a comparison, households in \citet{Busch_Ludwig2024} sacrifice 0.5\%-14\% of lifetime consumption to live in a world with Gaussian risk, while households in \citet{GuvenenOzkanMadera2024NonGaussianRisk} sacrifice almost 22\% of consumption to live in a world without any risk.

\section{An alternative income process: persistent shocks}\label{Sec::Income_Persistence}

Our results so far rely on a permanent-transitory income process that leaves no room for shocks to have persistent, yet not fully permanent, effects on income. This unit root restriction in income may be perceived as unappealing in the context of biennial data and non-Gaussian shocks, especially if the various moments of shocks, on which the partial insurance estimates depend, are sensitive to it. Moreover, \citet{Kaplan_Violante2010} show that BPP's method of measuring insurance to permanent shocks is biased in the presence of borrowing limits, but the bias lessens if the framework is adapted to measure insurance to persistent shocks. This is unlikely to explain the low magnitude of $\phi^{(1)}$ that we find here (the bias tends to \emph{increase} the linear pass-through rate), but it is unclear how it might affect $\phi^{(2)}$. For both these reasons, we relax the unit root restriction in income in this section.

Let persistent income $P_{it}$ follow an AR(1) process rather than a unit root. We may write this as $P_{it} = \rho P_{it-1} + \zeta_{it}$, where $0<\rho\leq 1$ measures income persistence. All other assumptions (including the distributions of income shocks) remain similar to the baseline. This modeling choice gives rise to the \emph{persistent}-transitory formulation $\Delta y_{it} = (\rho-1)P_{it-1} + \zeta_{it} + \Delta v_{it}$ and the quasi-difference
\begin{equation}\label{Eq::Income_Process_rho}
    \Delta_\rho y_{it} = \zeta_{it} + \Delta_\rho v_{it},
\end{equation}
where $\Delta_\rho x_{t} = x_t - \rho x_{t-1}$ and $\zeta$ reflects a persistent rather than a permanent shock. Clearly, the permanent-transitory process is a special case of this for $\rho=1$.

Despite the modification in the income process, the specification of the linear and quadratic consumption functions remains unchanged from the baseline processes \eqref{Eq::Consumption_Function_Linear} and \eqref{Eq::Consumption_Function_Quadratic}, respectively. However, the structure that underlies the transmission parameters $\phi^{(1)}$, $\phi^{(2)}$, etc, explicitly involves $\rho$ in this case. Appendix \ref{Appendix::Derivation_Consumption_Function} details the derivation of the consumption function with income persistence and some further adjustments when the data are observed biennially.

The identification strategy for the income process must now be adjusted to account for the AR(1) component and the presence of an additional parameter, $\rho$. Formally, the variance of shocks and $\rho$ are jointly identified through
\begin{align*}
    \sigma_{\zeta_t}^{2}  &=  {\E}(\Delta_\rho y_{it} \times (\rho\Delta_\rho y_{it-1}+\Delta_\rho y_{it}+\rho^{-1}\Delta_\rho y_{it+1})), \\
    \sigma_{v_t}^2        &= -\rho^{-1}{\E}(\Delta_\rho y_{it} \times \Delta_\rho y_{it+1}),\\
    {\E}(\Delta_\rho y_{it} \times \Delta y_{it-2}) &= {\E}((y_{it}-\rho y_{it-1}) \times (y_{it-2}-y_{it-3})) = 0.
\end{align*}
The first two moments appear slightly modified also in the baseline; the third moment is new and pins down $\rho$. An additional time period of data is needed ($t-3$, for a minimum of 5 periods in total) to separately identify income persistence from the variance of persistent shocks. The skewness and kurtosis of shocks are then identified from similarly modified higher-order income moments that mimic the baseline. Identification of the consumption transmission parameters is analogous: the quasi-difference $\Delta_\rho y_{it}$ replaces $\Delta y_{it}$, and $\sum_{\tau=-1}^1\rho^{-\tau} \Delta_\rho y_{it+\tau}$ replaces the long sum $\sum_{\tau=-1}^1 \Delta y_{it+\tau}$ in all previous identifying statements, but otherwise the strategy is similar to the baseline. Appendix \ref{Appendix::Identification} shows the identifying statements in detail, including the adjustments needed for biennial data and measurement error; identification in the quadratic case again obtains for \emph{any} distribution of income shocks except Bernoulli.

\begin{table}[t!]  
\begin{center}
\caption{Estimates of the consumption function, income persistence}\label{Table::Consumption_function_rho}
\begin{tabular}{L{4cm} R{2cm} L{2cm}}
\toprule
\emph{Consumption fn.:}                 & \mcl{2}{c}{\textbf{Quadratic}} \\
\cmidrule{2-3}
\emph{Income moments:}                  & \mcl{2}{c}{2\textsuperscript{nd}, 3\textsuperscript{rd}, 4\textsuperscript{th}} \\
\emph{Income persistence:}              & \mcl{2}{c}{$\rho = 0.931$ (\emph{s.e.}$=0.028$)}              \\
\emph{Consumption data:}                & \mcl{2}{c}{PSID}              \\
                                        & \mcl{2}{c}{(1)}               \\
\midrule                                        
$\phi^{(1)}$                            &0.112          &(0.024)        \\
$\psi^{(1)}$                            &-0.010         &(0.066)        \\
$\phi^{(2)}$                            &-0.049         &(0.026)        \\
$\psi^{(2)}$                            &0.019          &(0.037)        \\
$\omega^{(22)}$                         &0.430          &(0.847)        \\
\midrule
$\sigma^2_{\xi}$                        &0.019          &(0.001)        \\
\midrule
$\sigma^2_{u_c}$                        &0.044          &(0.002)        \\
\bottomrule
\end{tabular}
\caption*{\fsz\emph{Notes:} The table presents the estimates of the parameters of the quadratic consumption function, when income features linear persistence. Estimation is via equally weighted GMM; block bootstrap standard errors are in parentheses. The estimates of the income process appear in appendix table \ref{AppTable::Income_process_rho}; linear consumption specification estimates appear in appendix table \ref{AppTable::Consumption_function_rho}.}
\end{center}
\end{table}

Appendix table \ref{AppTable::Income_process_rho} reports the estimates of the income process from this specification. Three points emerge. First, when targeting only the second moments of income, we estimate $\rho = 0.846$ (\emph{s.e.}$=0.038$), which lies at the lower end of persistence estimates in the literature but within ABB's nonlinear persistence range of 0.5--1 over a similar period. The biennial nature of the data likely influences this, even though $\rho$ reflects \emph{annual} persistence. As $\rho$ is below its baseline value, the variance of persistent shocks is higher than the baseline to help match the variance of income growth in the data (yet the transitory variance is smaller). Second, adding third-order moments yields negative skewness for both shocks but leaves the estimates of $\rho$ and the variances unchanged. Third, targeting all moments up to fourth-order, we estimate $\rho = 0.931$ (\emph{s.e.}$=0.028$). The estimates of the variances align with the baseline values of table \ref{Table::Income_process}, as also do the skewness coefficients at -0.91 and -1.178 for the two types of shocks, respectively. We estimate kurtosis at 28.891 and 52.091; for persistent shocks, this is lower than the baseline, yet both distributions remain highly leptokurtic and left-skewed despite the AR(1) component.

Table \ref{Table::Consumption_function_rho} reports the estimates of the quadratic consumption function (see table \ref{AppTable::Consumption_function_rho} for the linear specification). We estimate the transmission parameter of the average persistent shock at $\phi^{(1)} = 0.112$, somewhat lower than the value 0.134 when shocks are permanent. The average persistent shock is therefore slightly better insured than the average permanent shock. \citet{Kaplan_Violante2010} similarly observe a drop in the pass-through rate (an increase in partial insurance) when $\rho$ falls and shocks become less permanent. We further estimate the transmission parameter of persistent shocks \emph{away} from the middle of the distribution at $\phi^{(2)}=-0.049$, negative and larger in magnitude than the baseline value of -0.04. Negative persistent shocks remain more unsettling than positive ones: they are associated with a larger pass-through rate into consumption, one that increases with the severity of the shock. The asymmetric response to bad versus good income news is therefore qualitatively and numerically robust to the earlier unit root assumption. Finally, transitory shocks have economically small and statistically zero pass-through rates.\footnote{In the context of the linear consumption specification (appendix table table \ref{AppTable::Consumption_function_rho}), we estimate $\phi^{(1)} = 0.089$ when we target only second moments, which gradually increases to match its quadratic specification counterpart as we introduce higher-order moments. This is driven by the increase in the value of $\rho$ as we add higher moments. \citet{Kaplan_Violante2010} also find lower pass-through rates (more insurance) at lower values of $\rho$. Transitory shocks remain fully insured, statistically.}\textsuperscript{,}\footnote{It is straightforward to further condition the results on age, position in the income distribution, wealth and education. We do not report these results for brevity.}

\section{Conclusion}\label{Sec::Conclusion}

We introduce a simple analytical framework to measure the degree of consumption partial insurance to income shocks, accounting for higher-order moments of the income distribution. Such moments, in particular left skewness and excess kurtosis, are important features of unexplained income growth, documented recently in the U.S. and other countries. 

We use recent income and consumption data from the PSID and establish several findings.

First, permanent and transitory shocks to disposable household income exhibit negative skewness and very large excess kurtosis.

Second, in estimating a linear consumption function over the recent years, we find a low pass-through of permanent shocks (large insurance) and full insurance to transitory ones. We attribute one third of the difference from BPP to the consumption imputation from the CEX, another third to the biennial nature of our data, and the remainder to time effects.

Third, in introducing third- and fourth-order income and consumption moments into the linear consumption model does not alter the above results.

Fourth, in estimating a nonlinear consumption function, we find that bad permanent shocks transmit more strongly than good ones, and the pass-through increases with the severity of the shock. By contrast, transitory shocks are mostly fully insured. The nature of income risk and consumption transmission vary substantially by age and the household's position in the income distribution. 

In summary, the paper offers a straightforward analytical framework to measure the transmission of income shocks into consumption, allowing such transmission to vary with the sign and size of the shock. To do so, we exploit higher-order income moments but we do not otherwise need to observe the entire consumption-income joint distribution. Our results consistently point to asymmetries in the pass-through of good and bad shocks, or by age and income level, suggesting that tail income risk matters substantially for consumption.


\begin{singlespace}
\bibliographystyle{chicago}
\bibliography{GT_Partial_Insurance.bib}

\begin{thebibliography}{}

\bibitem[\protect\citeauthoryear{Ai and Bhandari}{Ai and
  Bhandari}{2021}]{Ai_Bhandari2021}
Ai, H. and A.~Bhandari (2021).
\newblock Asset pricing with endogenously uninsurable tail risk.
\newblock {\em Econometrica\/}~{\em 89\/}(3), 1471--1505.

\bibitem[\protect\citeauthoryear{Andreski, Li, Samancioglu, and
  Schoeni}{Andreski et~al.}{2014}]{AndreskiEtAl2014_PSID_CEX_comparison}
Andreski, P., G.~Li, M.~Z. Samancioglu, and R.~Schoeni (2014).
\newblock {Estimates of annual consumption expenditures and its major
  components in the PSID in comparison to the CE}.
\newblock {\em American Economic Review\/}~{\em 104\/}(5), 132--35.

\bibitem[\protect\citeauthoryear{Arellano, Blundell, and Bonhomme}{Arellano
  et~al.}{2017}]{Arellano_Blundell_Bonhomme2017}
Arellano, M., R.~Blundell, and S.~Bonhomme (2017).
\newblock Earnings and consumption dynamics: A nonlinear panel data framework.
\newblock {\em Econometrica\/}~{\em 85\/}(3), 693--734.

\bibitem[\protect\citeauthoryear{Arellano, Blundell, Bonhomme, and
  Light}{Arellano et~al.}{2023}]{Arellano_Blundell_Bonhomme_Light2021}
Arellano, M., R.~Blundell, S.~Bonhomme, and J.~Light (2023).
\newblock Heterogeneity of consumption responses to income shocks in the
  presence of nonlinear persistence.
\newblock {\em Journal of Econometrics\/}.

\bibitem[\protect\citeauthoryear{Attanasio and Davis}{Attanasio and
  Davis}{1996}]{AttanasioDavis1996_RelativeWageMovements}
Attanasio, O. and S.~J. Davis (1996).
\newblock Relative wage movements and the distribution of consumption.
\newblock {\em Journal of Political Economy\/}~{\em 104\/}(6), 1227--1262.

\bibitem[\protect\citeauthoryear{Backus, Chernov, and Martin}{Backus
  et~al.}{2011}]{Backus_Chernov_Martin2011}
Backus, D., M.~Chernov, and I.~Martin (2011).
\newblock Disasters implied by equity index options.
\newblock {\em Journal of Finance\/}~{\em 66\/}(6), 1969--2012.

\bibitem[\protect\citeauthoryear{Barro}{Barro}{2009}]{Barro2009}
Barro, R.~J. (2009).
\newblock Rare disasters, asset prices, and welfare costs.
\newblock {\em American Economic Review\/}~{\em 99\/}(1), 243--264.

\bibitem[\protect\citeauthoryear{Blundell, Borella, Commault, and
  Nardi}{Blundell et~al.}{2020}]{Blundell_Borella_Commault_DeNardi2020}
Blundell, R., M.~Borella, J.~Commault, and M.~D. Nardi (2020).
\newblock Why does consumption fluctuate in old age and how should the
  government insure it?
\newblock NBER WP 27348.

\bibitem[\protect\citeauthoryear{Blundell, Low, and Preston}{Blundell
  et~al.}{2013}]{BlundellPrestonLow2013decomposing}
Blundell, R., H.~Low, and I.~Preston (2013).
\newblock Decomposing changes in income risk using consumption data.
\newblock {\em Quantitative Economics\/}~{\em 4\/}(1), 1--37.

\bibitem[\protect\citeauthoryear{Blundell, Pistaferri, and Preston}{Blundell
  et~al.}{2004}]{BlundellPistaferriPreston_2004_Imputation}
Blundell, R., L.~Pistaferri, and I.~Preston (2004).
\newblock {Imputing consumption in the PSID using food demand estimates from
  the CEX}.
\newblock Technical report, IFS Working Papers.

\bibitem[\protect\citeauthoryear{Blundell, Pistaferri, and Preston}{Blundell
  et~al.}{2008}]{Blundell_Pistaferri_Preston2008}
Blundell, R., L.~Pistaferri, and I.~Preston (2008).
\newblock Consumption inequality and partial insurance.
\newblock {\em American Economic Review\/}~{\em 98\/}(5), 1887--1921.

\bibitem[\protect\citeauthoryear{Blundell, Pistaferri, and
  Saporta-Eksten}{Blundell
  et~al.}{2016}]{Blundell_Pistaferri_Saporta-Eksten2016}
Blundell, R., L.~Pistaferri, and I.~Saporta-Eksten (2016).
\newblock Consumption inequality and family labor supply.
\newblock {\em {American Economic Review}\/}~{\em 106\/}(2), 387--435.

\bibitem[\protect\citeauthoryear{Blundell and Preston}{Blundell and
  Preston}{1998}]{Blundell_Preston1998}
Blundell, R. and I.~Preston (1998).
\newblock Consumption inequality and income uncertainty.
\newblock {\em Quarterly Journal of Economics\/}~{\em 113\/}(2), 603--640.

\bibitem[\protect\citeauthoryear{Bonhomme and Robin}{Bonhomme and
  Robin}{2010}]{Bonhomme_Robin2010}
Bonhomme, S. and J.-M. Robin (2010).
\newblock Generalized non-parametric deconvolution with an application to
  earnings dynamics.
\newblock {\em Review of Economic Studies\/}~{\em 77\/}(2), 491--533.

\bibitem[\protect\citeauthoryear{Borella, {De Nardi}, Pak, Russo, and
  Yang}{Borella et~al.}{2022}]{BorellaDeNardiEtAl2022_Taxes}
Borella, M., M.~{De Nardi}, M.~Pak, N.~Russo, and F.~Yang (2022).
\newblock {The importance of modeling income taxes over time. U.S. reforms and
  outcomes}.
\newblock NBER WP 30725.

\bibitem[\protect\citeauthoryear{Bound, Brown, Duncan, and Rodgers}{Bound
  et~al.}{1994}]{BoundEtAl1994PSIDerror}
Bound, J., C.~Brown, G.~J. Duncan, and W.~L. Rodgers (1994).
\newblock Evidence on the validity of cross-sectional and longitudinal labor
  market data.
\newblock {\em Journal of Labor Economics\/}~{\em 12\/}(3), 345--368.

\bibitem[\protect\citeauthoryear{Brav, Constantinides, and Geczy}{Brav
  et~al.}{2002}]{BravConstantinidesGeczy_2002_AssetPricing}
Brav, A., G.~M. Constantinides, and C.~C. Geczy (2002).
\newblock Asset pricing with heterogeneous consumers and limited participation:
  Empirical evidence.
\newblock {\em Journal of Political Economy\/}~{\em 110\/}(4), 793--824.

\bibitem[\protect\citeauthoryear{Bunn, {Le Roux}, Reinold, and Surico}{Bunn
  et~al.}{2018}]{Bunn_LeRoux_Reinold_Surico2018}
Bunn, P., J.~{Le Roux}, K.~Reinold, and P.~Surico (2018).
\newblock The consumption response to positive and negative income shocks.
\newblock {\em Journal of Monetary Economics\/}~{\em 96}, 1--15.

\bibitem[\protect\citeauthoryear{Busch, Domeij, Guvenen, and Madera}{Busch
  et~al.}{2018}]{Busch_Domeij_Guvenen_Madera2018}
Busch, C., D.~Domeij, F.~Guvenen, and R.~Madera (2018).
\newblock Asymmetric business-cycle risk and social insurance.
\newblock NBER WP 24569.

\bibitem[\protect\citeauthoryear{Busch and Ludwig}{Busch and
  Ludwig}{2024}]{Busch_Ludwig2024}
Busch, C. and A.~Ludwig (2024).
\newblock Higher-order income risk over the business cycle.
\newblock {\em International Economic Review\/}~{\em 65\/}(3), 1105--1131.

\bibitem[\protect\citeauthoryear{Campbell and Hercowitz}{Campbell and
  Hercowitz}{2019}]{CampbellHercowitz2019LiquidityConstraints}
Campbell, J.~R. and Z.~Hercowitz (2019).
\newblock Liquidity constraints of the middle class.
\newblock {\em American Economic Journal: Economic Policy\/}~{\em 11\/}(3),
  130–55.

\bibitem[\protect\citeauthoryear{Carroll}{Carroll}{2001}]{Carroll2001_Death}
Carroll, C.~D. (2001).
\newblock {Death to the log-linearized consumption Euler equation! (And very
  poor health to the second-order approximation)}.
\newblock {\em Advances in Macroeconomics\/}~{\em 1\/}(1).

\bibitem[\protect\citeauthoryear{Chatterjee, Morley, and Singh}{Chatterjee
  et~al.}{2021}]{ChatterjeeMorleySingh2021_BPP}
Chatterjee, A., J.~Morley, and A.~Singh (2021).
\newblock Estimating household consumption insurance.
\newblock {\em Journal of Applied Econometrics\/}~{\em 36\/}(5), 628--635.

\bibitem[\protect\citeauthoryear{Christelis, Georgarakos, Jappelli, Pistaferri,
  and {van Rooij}}{Christelis
  et~al.}{2019}]{ChristelisEtAl_2019_AsymmetricConsumptionEffects}
Christelis, D., D.~Georgarakos, T.~Jappelli, L.~Pistaferri, and M.~{van Rooij}
  (2019).
\newblock Asymmetric consumption effects of transitory income shocks.
\newblock {\em Economic Journal\/}~{\em 129\/}(622), 2322--2341.

\bibitem[\protect\citeauthoryear{Cochrane}{Cochrane}{1991}]{Cochrane1991_TestConsumptionInsurance}
Cochrane, J.~H. (1991).
\newblock A simple test of consumption insurance.
\newblock {\em Journal of Political Economy\/}~{\em 99\/}(5), 957--976.

\bibitem[\protect\citeauthoryear{Commault}{Commault}{2022}]{Commault2022}
Commault, J. (2022).
\newblock {Does consumption respond to transitory shocks? Reconciling natural
  experiments and semistructural methods}.
\newblock {\em American Economic Journal: Macroeconomics\/}~{\em 14\/}(2),
  96--122.

\bibitem[\protect\citeauthoryear{Constantinides and Ghosh}{Constantinides and
  Ghosh}{2017}]{Constantinides_Ghosh2017}
Constantinides, G.~M. and A.~Ghosh (2017).
\newblock Asset pricing with countercyclical household consumption risk.
\newblock {\em Journal of Finance\/}~{\em 72\/}(1), 415--460.

\bibitem[\protect\citeauthoryear{Crawley and Kuchler}{Crawley and
  Kuchler}{2023}]{Crawley_Kuchler2023}
Crawley, E. and A.~Kuchler (2023).
\newblock Consumption heterogeneity: Micro drivers and macro implications.
\newblock {\em American Economic Journal: Macroeconomics\/}~{\em 15\/}(1),
  314--41.

\bibitem[\protect\citeauthoryear{Crawley and Theloudis}{Crawley and
  Theloudis}{2024}]{CrawleyTheloudis2024_IncomeShocks}
Crawley, E. and A.~Theloudis (2024).
\newblock Income shocks and their transmission into consumption.
\newblock {\em Encyclopedia of Consumption\/}.
\newblock Forthcoming.

\bibitem[\protect\citeauthoryear{{De Nardi}, Fella, Knoef, Paz-Pardo, and {Van
  Ooijen}}{{De Nardi} et~al.}{2021}]{DeNardi_EtAl2021_Netherlands_US}
{De Nardi}, M., G.~Fella, M.~Knoef, G.~Paz-Pardo, and R.~{Van Ooijen} (2021).
\newblock {Family and government insurance: Wage, earnings, and income risks in
  the Netherlands and the U.S.}
\newblock {\em Journal of Public Economics\/}~{\em 193}, 104327.

\bibitem[\protect\citeauthoryear{{De Nardi}, Fella, and Paz-Pardo}{{De Nardi}
  et~al.}{2019}]{DeNardi_Fella_PazPardo2019}
{De Nardi}, M., G.~Fella, and G.~Paz-Pardo (2019).
\newblock Nonlinear household earnings dynamics, self-insurance, and welfare.
\newblock {\em Journal of the European Economic Association\/}~{\em 18\/}(2),
  890--926.

\bibitem[\protect\citeauthoryear{Deaton and Paxson}{Deaton and
  Paxson}{1994}]{Deaton_Paxson1994}
Deaton, A. and C.~Paxson (1994).
\newblock Intertemporal choice and inequality.
\newblock {\em Journal of Political Economy\/}~{\em 102\/}(3), 437--467.

\bibitem[\protect\citeauthoryear{Fagereng, Holm, and Natvik}{Fagereng
  et~al.}{2021}]{FagerengEtAl_2021LotteryWins}
Fagereng, A., M.~B. Holm, and G.~J. Natvik (2021).
\newblock {MPC heterogeneity and household balance sheets}.
\newblock {\em American Economic Journal: Macroeconomics\/}~{\em 13\/}(4),
  1--54.

\bibitem[\protect\citeauthoryear{Fuster, Kaplan, and Zafar}{Fuster
  et~al.}{2020}]{Fuster_Kaplan_Zafar2021}
Fuster, A., G.~Kaplan, and B.~Zafar (2020).
\newblock {What would you do with \$500? Spending responses to gains, losses,
  news, loans}.
\newblock {\em Review of Economic Studies\/}~{\em 88\/}(4), 1760--1795.

\bibitem[\protect\citeauthoryear{Georgarakos, Kim, Coibion, Shim, Lee,
  Gorodnichenko, Kenny, Han, and Weber}{Georgarakos
  et~al.}{2025}]{Georgarakos2025HowCostlyBusinessCycle}
Georgarakos, D., K.~H. Kim, O.~Coibion, M.~Shim, M.~A. Lee, Y.~Gorodnichenko,
  G.~Kenny, S.~Han, and M.~Weber (2025).
\newblock {How costly are business cycle volatility and inflation? A vox populi
  approach}.
\newblock Working Paper 33476, National Bureau of Economic Research.

\bibitem[\protect\citeauthoryear{Geweke and Keane}{Geweke and
  Keane}{2000}]{Geweke_Keane2000}
Geweke, J. and M.~Keane (2000).
\newblock {An empirical analysis of earnings dynamics among men in the PSID:
  1968–1989}.
\newblock {\em Journal of Econometrics\/}~{\em 96\/}(2), 293--356.

\bibitem[\protect\citeauthoryear{Ghosh and Julliard}{Ghosh and
  Julliard}{2012}]{Julliard_Ghosh2012}
Ghosh, A. and C.~Julliard (2012).
\newblock Can rare events explain the equity premium puzzle?
\newblock {\em Review of Financial Studies\/}~{\em 25\/}(10), 3037--3076.

\bibitem[\protect\citeauthoryear{Gottschalk and Moffitt}{Gottschalk and
  Moffitt}{2009}]{GottschalkMoffitt2009_RisingInstability}
Gottschalk, P. and R.~Moffitt (2009).
\newblock {The rising instability of U.S. earnings}.
\newblock {\em Journal of Economic Perspectives\/}~{\em 23\/}(4), 3--24.

\bibitem[\protect\citeauthoryear{Gourinchas and Parker}{Gourinchas and
  Parker}{2002}]{Gourinchas_Parker2002}
Gourinchas, P.-O. and J.~A. Parker (2002).
\newblock Consumption over the life cycle.
\newblock {\em Econometrica\/}~{\em 70\/}(1), 47--89.

\bibitem[\protect\citeauthoryear{Guvenen, Karahan, Ozkan, and Song}{Guvenen
  et~al.}{2021}]{Guvenen_Karahan_Ozkan_Song2021}
Guvenen, F., F.~Karahan, S.~Ozkan, and J.~Song (2021).
\newblock {What do data on millions of U.S. workers reveal about lifecycle
  earnings dynamics?}
\newblock {\em Econometrica\/}~{\em 89\/}(5), 2303--2339.

\bibitem[\protect\citeauthoryear{Guvenen, Ozkan, and Madera}{Guvenen
  et~al.}{2024}]{GuvenenOzkanMadera2024NonGaussianRisk}
Guvenen, F., S.~Ozkan, and R.~Madera (2024).
\newblock {Consumption dynamics and welfare under non-Gaussian earnings risk}.
\newblock NBER WP 32298.

\bibitem[\protect\citeauthoryear{Guvenen, Ozkan, and Song}{Guvenen
  et~al.}{2014}]{Guvenen_Ozkan_Song2014}
Guvenen, F., S.~Ozkan, and J.~Song (2014).
\newblock The nature of countercyclical income risk.
\newblock {\em Journal of Political Economy\/}~{\em 122\/}(3), 621--660.

\bibitem[\protect\citeauthoryear{Guvenen and Smith}{Guvenen and
  Smith}{2014}]{Guvenen_Smith2014}
Guvenen, F. and A.~A. Smith (2014).
\newblock Inferring labor income risk and partial insurance from economic
  choices.
\newblock {\em Econometrica\/}~{\em 82\/}(6), 2085--2129.

\bibitem[\protect\citeauthoryear{Hall and Mishkin}{Hall and
  Mishkin}{1982}]{Hall_Mishkin1982}
Hall, R.~E. and F.~S. Mishkin (1982).
\newblock The sensitivity of consumption to transitory income: Estimates from
  panel data on households.
\newblock {\em Econometrica\/}~{\em 50\/}(2), 461--481.

\bibitem[\protect\citeauthoryear{Hayashi, Altonji, and Kotlikoff}{Hayashi
  et~al.}{1996}]{Hayashi_Altonji_Kotlikoff1996}
Hayashi, F., J.~Altonji, and L.~Kotlikoff (1996).
\newblock Risk-sharing between and within families.
\newblock {\em Econometrica\/}~{\em 64\/}(2), 261--294.

\bibitem[\protect\citeauthoryear{Heathcote, Storesletten, and
  Violante}{Heathcote et~al.}{2014}]{Heathcote_Storesletten_Violante2014}
Heathcote, J., K.~Storesletten, and G.~L. Violante (2014).
\newblock Consumption and labor supply with partial insurance: An analytical
  framework.
\newblock {\em American Economic Review\/}~{\em 104\/}(7), 2075--2126.

\bibitem[\protect\citeauthoryear{Hryshko and Manovskii}{Hryshko and
  Manovskii}{2022}]{Hryshko_Manovskii2022}
Hryshko, D. and I.~Manovskii (2022).
\newblock {How much consumption insurance in the U.S.?}
\newblock {\em Journal of Monetary Economics\/}~{\em 130}, 17--33.

\bibitem[\protect\citeauthoryear{Jappelli and Pistaferri}{Jappelli and
  Pistaferri}{2010}]{JappelliPistaferri2010Review}
Jappelli, T. and L.~Pistaferri (2010).
\newblock The consumption response to income changes.
\newblock {\em Annual Review of Economics\/}~{\em 2\/}(1), 479--506.

\bibitem[\protect\citeauthoryear{Kaplan and Violante}{Kaplan and
  Violante}{2010}]{Kaplan_Violante2010}
Kaplan, G. and G.~L. Violante (2010).
\newblock How much consumption insurance beyond self-insurance?
\newblock {\em American Economic Journal: Macroeconomics\/}~{\em 2\/}(4),
  53--87.

\bibitem[\protect\citeauthoryear{Kaplan and Violante}{Kaplan and
  Violante}{2014}]{KaplanViolante2014ConsumptionResponseFiscalStimulus}
Kaplan, G. and G.~L. Violante (2014).
\newblock A model of the consumption response to fiscal stimulus payments.
\newblock {\em Econometrica\/}~{\em 82\/}(4), 1199--1239.

\bibitem[\protect\citeauthoryear{Krueger and Perri}{Krueger and
  Perri}{2006}]{Krueger_Perri2006}
Krueger, D. and F.~Perri (2006).
\newblock {Does income inequality lead to consumption inequality? Evidence and
  theory}.
\newblock {\em Review of Economic Studies\/}~{\em 73\/}(1), 163--193.

\bibitem[\protect\citeauthoryear{Lewis, Melcangi, and Pilossoph}{Lewis
  et~al.}{2022}]{Lewis_Melcangi_Pilossoph2022}
Lewis, D., D.~Melcangi, and L.~Pilossoph (2022).
\newblock Latent heterogeneity in the marginal propensity to consume.
\newblock Unpublished manuscript.

\bibitem[\protect\citeauthoryear{Li, Schoeni, Danziger, and Charles}{Li
  et~al.}{2010}]{LiEtAl_2010_PSID_CEX_consumption}
Li, G., R.~Schoeni, S.~Danziger, and K.~Charles (2010).
\newblock {New expenditure data in the PSID: Comparisons with the CE}.
\newblock {\em Monthly Labor Review\/}~{\em 133}.

\bibitem[\protect\citeauthoryear{Madera}{Madera}{2019}]{Madera2019_ConsumptionResponseTailShocks}
Madera, R. (2019).
\newblock The consumption response to tail earnings shocks.
\newblock Unpublished manuscript.

\bibitem[\protect\citeauthoryear{McKay}{McKay}{2017}]{McKay2017}
McKay, A. (2017).
\newblock Time-varying idiosyncratic risk and aggregate consumption dynamics.
\newblock {\em Journal of Monetary Economics\/}~{\em 88}, 1--14.

\bibitem[\protect\citeauthoryear{Meghir and Pistaferri}{Meghir and
  Pistaferri}{2004}]{MeghirPistaferri2004IncomeVariance}
Meghir, C. and L.~Pistaferri (2004).
\newblock Income variance dynamics and heterogeneity.
\newblock {\em Econometrica\/}~{\em 72\/}(1), 1--32.

\bibitem[\protect\citeauthoryear{Meghir and Pistaferri}{Meghir and
  Pistaferri}{2011}]{MeghirPistaferri2011}
Meghir, C. and L.~Pistaferri (2011).
\newblock Earnings, consumption and life cycle choices.
\newblock Volume~4 of {\em Handbook of Labor Economics}, pp.\  773--854.

\bibitem[\protect\citeauthoryear{Meyer and Sullivan}{Meyer and
  Sullivan}{2023}]{MeyerSullivan2023onsumptionIncomeInequality}
Meyer, B.~D. and J.~X. Sullivan (2023).
\newblock {Consumption and income inequality in the United States since the
  1960s}.
\newblock {\em Journal of Political Economy\/}~{\em 131\/}(2), 247--284.

\bibitem[\protect\citeauthoryear{Misra and Surico}{Misra and
  Surico}{2014}]{MisraSurico2014_EvidenceFiscalStimulus}
Misra, K. and P.~Surico (2014).
\newblock Consumption, income changes, and heterogeneity: Evidence from two
  fiscal stimulus programs.
\newblock {\em American Economic Journal: Macroeconomics\/}~{\em 6\/}(4),
  84--106.

\bibitem[\protect\citeauthoryear{Parker, Souleles, Johnson, and
  McClelland}{Parker
  et~al.}{2013}]{ParkerSoulelesJohnsonMcClelland2013_StimulusPayments}
Parker, J.~A., N.~S. Souleles, D.~S. Johnson, and R.~McClelland (2013).
\newblock Consumer spending and the economic stimulus payments of 2008.
\newblock {\em American Economic Review\/}~{\em 103\/}(6), 2530--53.

\bibitem[\protect\citeauthoryear{Theloudis}{Theloudis}{2021}]{Theloudis2021}
Theloudis, A. (2021).
\newblock Consumption inequality across heterogeneous families.
\newblock {\em European Economic Review\/}~{\em 136}.

\bibitem[\protect\citeauthoryear{Wu and Krueger}{Wu and
  Krueger}{2021}]{WuKrueger2021_ConsumptionInsurance}
Wu, C. and D.~Krueger (2021).
\newblock Consumption insurance against wage risk: Family labor supply and
  optimal progressive income taxation.
\newblock {\em American Economic Journal: Macroeconomics\/}~{\em 13\/}(1),
  79--113.

\end{thebibliography}
\addcontentsline{toc}{section}{Bibliography} 
\end{singlespace}


\pagebreak
\let\origappendix\appendix 
\renewcommand\appendix{\clearpage\pagenumbering{arabic}\origappendix}
\appendix
\renewcommand\appendixtocname{Online Appendix}
\renewcommand\appendixpagename{Online Appendix}
\appendixpage           
\addappheadtotoc        
\numberwithin{equation}{section} 
\numberwithin{table}{section}   
\numberwithin{figure}{section}
\setcounter{footnote}{0} 


\section{Derivation of the consumption function}\label{Appendix::Derivation_Consumption_Function}

This appendix shows how to derive the consumption function under the general AR(1) specification for log income in section \ref{Sec::Income_Persistence}; the consumption function based on the baseline permanent-transitory income process is a special case of this. We follow similar derivations in \citet{Blundell_Pistaferri_Preston2008}, BPP hereafter, though our approach is a refinement of theirs given the second-order approximation we carry out herein and the relaxation of the unit root assumption for income. We show below how to derive the most general (quadratic) consumption function; the linear function is a special case of this. We let a time unit in the model correspond to one calendar year. However, data in the PSID in recent years are available only every second year, so we adapt the derivations to reflect this.

Let the utility function take the form $U(C_{it};\mathbf{Z}_{it}) = \widetilde{U}(C_{it}\exp(-\mathbf{Z}_{it}^\prime \boldsymbol{\alpha}))$, which simplifies the subsequent statements. The first-order conditions of the household problem \eqref{Eq::Household_Problem} \emph{s.t.} \eqref{Eq::Budget_Constraint} in a generic period $t+s$ are
\begin{alignat*}{2}
[C_{it+s}]    &:{}& \quad \widetilde{U}_{C} (\widetilde{C}_{it+s}) \exp(-\mathbf{Z}_{it+s}^{\prime} \boldsymbol{\alpha})   &= \lambda_{it+s} \\
[A_{it+s+1}]  &:{}& \quad \beta (1+r) {\E}_{t+s} \lambda_{it+s+1} &= \lambda_{it+s},
\end{alignat*}
where $\widetilde{U}_{C}$ is the marginal utility of consumption and $\widetilde{C}_{it} = C_{it} \exp(-\mathbf{Z}_{it}^{\prime} \boldsymbol{\alpha})$. $\lambda_{it}$ is the Lagrange multiplier on the sequential budget constraint (the marginal utility of wealth).

\subsection{Taylor approximation to optimality conditions}\label{SubAppendix::Derivation_Consumption_Function_ApproximationFOCs}

\textbf{Approximation to static optimality condition.} Applying logs to the static condition and taking a first difference over time yields $\Delta \ln \widetilde{U}_{C} (\widetilde{C}_{it+s}) - \Delta (\mathbf{Z}_{it+s}^{\prime} \boldsymbol{\alpha}) = \Delta \ln \lambda_{it+s}$. A first-order Taylor expansion of $\ln \widetilde{U}_{C} (\widetilde{C}_{it+s})$ around $\widetilde{C}_{it+s-1}$ yields 
\begin{align*}
	\Delta \ln \widetilde{U}_{C} (\widetilde{C}_{it+s}) 
	&\approx \frac{\widetilde{U}_{CC} (\widetilde{C}_{it+s-1})}{\widetilde{U}_{C} (\widetilde{C}_{it+s-1})} \exp(-\mathbf{Z}_{it+s-1}^{\prime} \boldsymbol{\alpha}) (C_{it+s}-C_{it+s-1}) \\
	&\approx \frac{\widetilde{U}_{CC} (\widetilde{C}_{it+s-1})}{\widetilde{U}_{C} (\widetilde{C}_{it+s-1})} C_{it+s-1}\exp(-\mathbf{Z}_{it+s-1}^{\prime} \boldsymbol{\alpha}) \Delta \ln C_{it+s} 
= \theta^{-1}_{it+s-1} \Delta \ln C_{it+s},
\end{align*}
where $\widetilde{U}_{CC}$ is the second derivative of the utility function and $\theta_{it} = \widetilde{U}_{CC}^{-1} (\widetilde{C}_{it}) \widetilde{C}^{-1}_{it} \widetilde{U}_{C} (\widetilde{C}_{it})$ is the inverse of the consumption substitution elasticity. Plugging the last expression in the log-linearized first-order condition yields $\Delta \ln C_{it+s} - \theta_{it+s-1} \Delta (\mathbf{Z}_{it+s}^{\prime} \boldsymbol{\alpha}) = \theta_{it+s-1} \Delta \ln \lambda_{it+s}$. 


\noindent \textbf{Approximation to Euler equation.} The approximation to the intertemporal condition involves future expectations. Let $\exp(\Gamma) = \beta^{-1} (1+r)^{-1}$ for some $\Gamma$. A second-order Taylor expansion of $\exp(\ln \lambda_{it+s+1})$ around $\ln \lambda_{it+s} + \Gamma$ yields
\begin{equation*}
	\lambda_{it+s+1} \approx \frac{\lambda_{it+s}}{\beta (1+r)} \left( 1 + (\Delta \ln \lambda_{it+s+1} - \Gamma) + \frac{1}{2}(\Delta \ln \lambda_{it+s+1} - \Gamma)^{2}\right).
\end{equation*}
Taking expectations at time $t+s$ and noting that ${\E}_{t+s} \lambda_{it+s+1} = \lambda_{it+s} \beta^{-1} (1+r)^{-1}$, we obtain
\begin{align*}
    {\E}_{t+s} \Delta \ln \lambda_{it+s+1} \approx \Gamma - \frac{1}{2}{\E}_{t+s}(\Delta \ln \lambda_{it+s+1} - \Gamma)^{2} \Rightarrow
    \Delta \ln \lambda_{it+s+1}  \approx \omega_{it+s+1} + \epsilon_{it+s+1}.
\end{align*}
The first term $\omega_{it+s+1} = \Gamma - \frac{1}{2} {\E}_{t} (\Delta \ln \lambda_{it+s+1} - \Gamma)^{2}$ captures the impact of the interest rate, impatience, and precautionary motives on the growth of the marginal utility of wealth. To maintain tractability, we assume that $\omega$ is non-stochastic but possibly heterogeneous across households. The second term is an expectation error with ${\E}_{t}(\epsilon_{it+1})=0$; it captures idiosyncratic revisions to the marginal utility of wealth upon arrival of income shocks.

Combining the approximations to the optimality conditions, we obtain
\begin{align}\label{AppEq:Linearized_FOCs}
	\notag
	\Delta \ln C_{it+s} - \theta_{it+s-1} \Delta (\mathbf{Z}_{it+s}^{\prime} \boldsymbol{\alpha}) 
		&= \theta_{it+s-1} (\omega_{it+s} + \epsilon_{it+s}) \\
	\notag
	\Delta \ln C_{it+s} - \theta_{it+s-1} \Delta (\mathbf{Z}_{it+s}^{\prime} \boldsymbol{\alpha}) - \Xi_{it+s} 
		&= \xi_{it+s} + \theta_{it+s-1} \epsilon_{it+s} \\
	\Delta c_{it+s} &= \xi_{it+s} + \theta_{it+s-1} \epsilon_{it+s}, 
\end{align}
where $\Delta c_{it+s} = \Delta \ln C_{it+s} - \theta_{it+s-1} \Delta (\mathbf{Z}_{it+s}^{\prime} \boldsymbol{\alpha}) - \Xi_{it+s}$. We have split $\theta_{it+s-1}\omega_{it+s}$ into $\Xi_{it+s}$ and $\xi_{it+s}$; the first term reflects the gradient of consumption due to impatience (discounting) and the interest rate, the second term reflects unobserved consumption taste heterogeneity. We let demographics and time fixed effects pick up the effect of taste shifters and the gradient in the consumption path. Therefore $\Delta c_{it+s}$ can be obtained empirically from a regression of consumption growth on appropriate observables (demographics, time fixed effects).

\subsection{Taylor approximation to lifetime budget constraint}\label{SubAppendix::Derivation_Consumption_Function_ApproximationIBC}

Let $G(\boldsymbol{\xi}) = \ln \sum_{s=0}^{N} \exp \xi_s$ for $\boldsymbol{\xi} = (\xi_0, \xi_1, \dots, \xi_{N})^\prime$. Applying a second-order Taylor expansion of $G(\boldsymbol{\xi})$ around a deterministic $\boldsymbol{\xi}^0$, and taking expectations given information ${\cal I}$, yields
\begin{align}\label{AppEq:Approximation_general_rule}
\begin{split}
	{\E}_{\cal I} G(\boldsymbol{\xi}) \approx G(\boldsymbol{\xi}^0) 
	&+ \sum_{s=0}^{N} \frac{\exp \xi^0_s}{\sum_{\kappa=0}^{N} \exp \xi^0_\kappa} {\E}_{\cal I}(\xi_s - \xi^0_s) \\
	&+ \frac{1}{2} \sum_{s=0}^{N} \sum_{\ell=0}^{N} \frac{\exp \xi^0_s}{\sum_{\kappa=0}^{N} \exp \xi^0_\kappa} \left(\delta_{s\ell} - \frac{\exp \xi^0_\ell}{\sum_{\kappa=0}^{N} \exp \xi^0_\kappa}\right){\E}_{\cal I}(\xi_s - \xi^0_s)(\xi_\ell - \xi^0_\ell),
\end{split}
\end{align}
where $\delta_{s\ell}$ is the Kronecker delta ($\delta_{s\ell}=1$ if $s=\ell$, $\delta_{s\ell}=0$ otherwise). We apply this approximation to both sides of the intertemporal budget constraint.

Assuming expectations away, the logarithm of the \emph{left} hand side of the budget constraint \eqref{Eq::Budget_Constraint} in a generic period $t$ is $\ln \left( A_{it} + \sum_{s=0}^{T-t} \exp \left( \ln Y_{it+s} -s\ln(1+r) \right)\right)$. Letting
\begin{align*}
	\xi_s &=
		\begin{cases}
		    \ln A_{it+s}  										&\text{ for } s=0 \\
		    \ln Y_{it+s-1} - (s-1)\ln(1+r) \qquad \qquad		&\text{ for } s=1,\dots,T-t+1  \\
		\end{cases}\\
	\xi^0_s &=
		\begin{cases}
		    {\E}_{t-2} \ln A_{it+s} 							&\text{ for } s=0 \\
		    {\E}_{t-2} \ln Y_{it+s-1} - (s-1)\ln(1+r) 			&\text{ for } s=1,\dots,T-t+1,
		\end{cases}
\end{align*}
and declaring the information set to contain time $t$ information, we follow \eqref{AppEq:Approximation_general_rule} to write
\begin{align*}
	&{\E}_{t} \ln \left( A_{it} + \sum_{s=0}^{T-t} \frac{Y_{it+s}}{(1+r)^s} \right) 
	\approx \ln \left( \exp({\E}_{t-2} \ln A_{it}) + \sum_{s=0}^{T-t} \exp \left( {\E}_{t-2} \ln Y_{it+s} - s\ln(1+r) \right)\right) \\
	&+ \pi_{it} {\E}_{t} (\ln A_{it} - {\E}_{t-2} \ln A_{it}) \\
	&+ (1-\pi_{it})\sum_{s=0}^{T-t} \vartheta_{it+s}^{Y} {\E}_{t} (\ln Y_{it+s} - {\E}_{t-2} \ln Y_{it+s})\\
	&+ \frac{1}{2} \pi_{it} (1-\pi_{it}) {\E}_{t} (\ln A_{it} - {\E}_{t-2} \ln A_{it})^2 \\
	&+ \frac{1}{2} (1-\pi_{it}) \sum_{s=0}^{T-t} \sum_{\ell=0}^{T-t} \vartheta_{it+s}^{Y} (\delta_{s\ell} - (1-\pi_{it})\vartheta_{it+\ell}^{Y}) {\E}_{t} (\ln Y_{it+s} - {\E}_{t-2} \ln Y_{it+s}) (\ln Y_{it+\ell} - {\E}_{t-2} \ln Y_{it+\ell}) \\
	&-  \pi_{it} (1-\pi_{it}) \sum_{s=0}^{T-t} \vartheta_{it+s}^{Y} {\E}_{t} (\ln A_{it} - {\E}_{t-2} \ln A_{it})(\ln Y_{it+s} - {\E}_{t-2} \ln Y_{it+s}).
\end{align*}
The notation is: $\pi_{it}=\frac{Q_{1t}}{Q_{1t}+Q_{2t}}$ is equal to the expected share of financial wealth $Q_{1t}=\exp({\E}_{t-2} \ln A_{it})$ in the household's total financial and human wealth at $t$, the latter defined as $Q_{2t}=\sum_{\kappa=0}^{T-t} \exp \left( {\E}_{t-2} \ln Y_{it+\kappa} -\kappa\ln(1+r) \right)$, i.e., the household's expected lifetime income for the remaining of life; $\vartheta_{it+s}^Y = \exp \left(\E_{t-2} \ln Y_{it+s} - s\ln(1+r)\right) / Q_{2t}$ is an annuitization factor equal to the expected share of time $t+s$ household income in total lifetime income. $\pi_{it}$ and $\vartheta_{it+s}^Y$ pertain to expectations at $t-2$ so they are both known at $t-2$ and later.

Using the income process and assuming the deterministic observables away (they cancel out in a first difference in expectations at a later stage), we reach $\ln Y_{it+s} = \rho^{s+2}\ln Y_{it-2} + \sum_{\tau=-1}^s \rho^{s-\tau}\zeta_{it+\tau} + v_{it+s} - \rho^{s+2}v_{it-2}$, therefore $\ln Y_{it+s} - {\E}_{t-2}\ln Y_{it+s} = \sum_{\tau=-1}^s \rho^{s-\tau}\zeta_{it+\tau} + v_{it+s}$, given that $s\geq -1$. It follows that $\sum_{s=0}^{T-t} \vartheta_{it+s}^{Y} {\E}_{t} (\ln Y_{it+s} - {\E}_{t-2} \ln Y_{it+s}) = \Theta_{it}^Y (\zeta_{it} + \rho\zeta_{it-1}) + \vartheta_{it}^{Y} v_{it}$, where $\Theta_{it}^Y = \sum_{\kappa=0}^{T-t} \vartheta_{it+\kappa}^{Y} \rho^\kappa$, because ${\E}_{t}\zeta_{it+\kappa}={\E}_{t}v_{it+\kappa}=0$ for $\kappa>0$. The second-order term in income is more involved. This is given by
\begin{align*}
    &\sum_{s=0}^{T-t} \sum_{\ell=0}^{T-t} \vartheta_{it+s}^{Y} (\delta_{s\ell} - (1-\pi_{it})\vartheta_{it+\ell}^{Y}) 
    {\E}_{t} (\ln Y_{it+s} - {\E}_{t-2} \ln Y_{it+s}) (\ln Y_{it+\ell} - {\E}_{t-2} \ln Y_{it+\ell}) \\
    &=\sum_{s=0}^{T-t} \sum_{\ell=0}^{T-t} \vartheta_{it+s}^{Y} (\delta_{s\ell} - (1-\pi_{it})\vartheta_{it+\ell}^{Y}) 
    {\E}_{t}\left( \rho^s \rho^\ell (\zeta_{it} + \rho\zeta_{it-1})^2 + (\zeta_{it} + \rho\zeta_{it-1})(\rho^s v_{it+\ell} + \rho^\ell v_{it+s}) + v_{it+s}v_{it+\ell} \right)\\
    &= \widetilde{\pi}_{it}(\zeta_{it} + \rho\zeta_{it-1})^2 
    + 2\widetilde{\widetilde{\pi}}_{it} \vartheta_{it}^{Y}(\zeta_{it} + \rho\zeta_{it-1})v_{it} 
    + \vartheta_{it}^{Y}(1-(1-\pi_{it})\vartheta_{it}^{Y})v_{it}^2,
\end{align*}
where $\widetilde{\pi}_{it} = \widetilde{\Theta}_{it}^Y - (1-\pi_{it}) (\Theta_{it}^Y)^2$, $\widetilde{\Theta}_{it}^Y = \sum_{\kappa=0}^{T-t} \vartheta_{it+\kappa}^{Y} \rho^{2\kappa}$, and $\widetilde{\widetilde{\pi}}_{it} = 1 - (1-\pi_{it})\Theta_{it}^Y$. We derive this noting that the expectation of future shocks is zero by construction.

Plugging these results in the approximation of the previous page, noting that assets at $t$ are `beginning-of-period' and thus independent of the time $t$ income shocks, and taking a first difference in expectations between $t-2$ and $t$ yields
\begin{align*}
    {\E}_{t} \ln \left( A_{it} + \sum_{s=0}^{T-t} \frac{Y_{it+s}}{(1+r)^s} \right) - {\E}_{t-2} \ln \left( A_{it} + \sum_{s=0}^{T-t} \frac{Y_{it+s}}{(1+r)^s} \right) 
    &\approx (1-\pi_{it})\Theta_{it}^Y(\zeta_{it}+\rho\zeta_{it-1}) \\
    &+ (1-\pi_{it})\vartheta_{it}^{Y} v_{it} \\
    &+\frac{1}{2} (1 - \pi_{it})\widetilde{\pi}_{it} (\zeta_{it}+\rho\zeta_{it-1})^2 \\
    &+\frac{1}{2} (1 - \pi_{it})\vartheta_{it}^{Y}(1-(1 - \pi_{it})\vartheta_{it}^{Y}) v_{it}^2 \\
    &+(1 - \pi_{it})\widetilde{\widetilde{\pi}}_{it}\vartheta_{it}^{Y}(\zeta_{it}+\rho\zeta_{it-1})v_{it}.
\end{align*}
The linear terms in the first two lines are the first-order approximation of BPP (adapted here for biennial data and income persistence); the quadratic terms in the next lines are the refinement from the second-order approximation.

Assuming expectations away, the logarithm of the \emph{right} hand side of the budget constraint \eqref{Eq::Budget_Constraint} is $\ln \sum_{s=0}^{T-t} \exp ( \ln C_{it+s} -s\ln(1+r) )$. Adopting the notation of \eqref{AppEq:Approximation_general_rule}, we let $\xi_s = \ln C_{it+s} - s \ln(1+r)$ and $\xi_s^0 = {\E}_{t-2}\ln C_{it+s} - s \ln(1+r)$. Letting the information set contain time $t$ information, it follows that
\begin{align*}
	{\E}_{t} \ln \sum_{s=0}^{T-t}\frac{C_{it+s}}{(1+r)^s} 
	&\approx \ln \sum_{s=0}^{T-t} \exp ( {\E}_{t-2} \ln C_{it+s} -s\ln(1+r) ) \\[-1pt]
	&+ \sum_{s=0}^{T-t} \vartheta_{it+s}^{C} {\E}_{t}(\ln C_{it+s} - {\E}_{t-2} \ln C_{it+s})\\[-1pt]
	&+ \frac{1}{2}\sum_{s=0}^{T-t} \sum_{\ell=0}^{T-t} \vartheta_{it+s}^{C} (\delta_{s\ell}-\vartheta_{it+\ell}^{C}){\E}_{t}(\ln C_{it+s} - {\E}_{t-2} \ln C_{it+s})(\ln C_{it+\ell} - {\E}_{t-2} \ln C_{it+\ell}),
\end{align*}
where $\vartheta_{it+s}^C = \exp \left(\E_{t-2} \ln C_{it+s} - s\ln(1+r)\right)/\sum_{\kappa=0}^{T-t} \exp \left(\E_{t-2} \ln C_{it+\kappa} - \kappa\ln(1+r)\right)$ is an annuitization factor equal to the expected share of time $t+s$ consumption in total lifetime consumption. $\vartheta_{it+s}^C$  pertains to expectations at $t-2$, so it is known at $t-2$ and later.

Using the linearized consumption function in \eqref{AppEq:Linearized_FOCs} to replace $\ln C_{it+s}$ recursively, we reach $\ln C_{it+s} = \ln C_{it-2} + \sum_{\tau=-1}^{s} \Psi_{it+\tau} + \sum_{\tau=-1}^{s}\xi_{it+\tau} + \sum_{\tau=-1}^{s} \theta_{it+\tau-1} \epsilon_{it+\tau}$, where $\Psi_{it+\tau} = \theta_{it+\tau-1} \Delta (\mathbf{Z}_{it+\tau}^{\prime} \boldsymbol{\alpha}) + \Xi_{it+\tau}$, therefore $\ln C_{it+s} - {\E}_{t-2} \ln C_{it+s} = \sum_{\tau=-1}^{s} \theta_{it+\tau-1} \epsilon_{it+\tau}$ because $\Psi_{it}$ and $\xi_{it}$ are non-stochastic by assumption and they cancel out in the first difference. It follows that $\sum_{s=0}^{T-t} \vartheta_{it+s}^{C} {\E}_{t} (\ln C_{it+s} - {\E}_{t-2} \ln C_{it+s}) = \theta_{it-1}\epsilon_{it} + \theta_{it-2}\epsilon_{it-1}$ because ${\E}_{t}\epsilon_{it+\kappa}=0$ for $\kappa>0$ and $\sum_{s=0}^{T-t} \vartheta_{it+s}^{C}=1$ by construction. The second-order term is more involved; it is given by
\begin{align*}
	&\sum_{s=0}^{T-t} \sum_{\ell=0}^{T-t} \vartheta_{it+s}^{C} (\delta_{s\ell}-\vartheta_{it+\ell}^{C}){\E}_{t}(\ln C_{it+s} - {\E}_{t-2} \ln C_{it+s})(\ln C_{it+\ell} - {\E}_{t-2} \ln C_{it+\ell})\\
	&=\sum_{s=0}^{T-t} \sum_{\ell=0}^{T-t} \vartheta_{it+s}^{C} (\delta_{s\ell}-\vartheta_{it+\ell}^{C}) (\theta_{it-1}\epsilon_{it} + \theta_{it-2}\epsilon_{it-1})^2 = 0.
\end{align*}
We derive this by noting that future expectations errors are zero by assumption, and that $\sum_{s=0}^{T-t} \sum_{\ell=0}^{T-t} \vartheta_{it+s}^{C} (\delta_{s\ell}-\vartheta_{it+\ell}^{C}) = \sum_{s=0}^{T-t} \vartheta_{it+s}^{C}\delta_{ss} + \sum_{s=0}^{T-t} \sum_{\ell \neq s} \vartheta_{it+s}^{C} \delta_{s\ell} - (\sum_{s=0}^{T-t} \vartheta_{it+s}^{C})(\sum_{\ell=0}^{T-t} \vartheta_{it+\ell}^{C})=0$. Plugging these results in the approximation above and taking a first difference in expectations between $t-2$ and $t$ yields
\begin{equation*}
	{\E}_{t} \ln \sum_{s=0}^{T-t}\frac{C_{it+s}}{(1+r)^s} - {\E}_{t-2} \ln \sum_{s=0}^{T-t}\frac{C_{it+s}}{(1+r)^s} \approx \theta_{it-1}\epsilon_{it} + \theta_{it-2}\epsilon_{it-1}.
\end{equation*}

\subsection{Biennial consumption growth}\label{SubAppendix::Derivation_Consumption_Function_CombiningSteps}

Given the biennial nature of our data, it follows from \eqref{AppEq:Linearized_FOCs} that \emph{observed} consumption growth is $\Delta^2 c_{it+s} = \xi_{it+s} + \xi_{it+s-1} + \theta_{it+s-1} \epsilon_{it+s} + \theta_{it+s-2} \epsilon_{it+s-1}$. The notation $\Delta^2 x_{t} = \Delta x_{t} + \Delta x_{t-1} = x_{t}-x_{t-2}$ indicates a difference between $t$ and $t-2$. Dropping the index $s$ and replacing the expectations errors with the quadratic expression in income shocks (the budget must balance so we bring the two sides of the budget constraint together), we obtain the final expression for consumption growth with biennial survey data and income persistence, given by 
\begin{align}\label{AppEq::Consumption_Function_Approximation_Final}
\begin{split}
    \Delta^2 c_{it} 
    &\approx \xi_{it} + \xi_{it-1} + (1-\pi_{it})\Theta_{it}^Y(\zeta_{it}+\rho\zeta_{it-1})
    + (1-\pi_{it})\vartheta_{it}^{Y} v_{it} \\
    &+\frac{1}{2} (1 - \pi_{it})\widetilde{\pi}_{it} (\zeta_{it}+\rho\zeta_{it-1})^2
    +\frac{1}{2} (1 - \pi_{it})\vartheta_{it}^{Y}(1-(1 - \pi_{it})\vartheta_{it}^{Y}) v_{it}^2 \\
    &+(1 - \pi_{it})\widetilde{\widetilde{\pi}}_{it}\vartheta_{it}^{Y}(\zeta_{it}+\rho\zeta_{it-1})v_{it}.
\end{split}
\end{align}

The first line of \eqref{AppEq::Consumption_Function_Approximation_Final} corresponds to the linear consumption function, adapted here for income persistence and the biennial nature of the data, and given compactly by  
\begin{equation*}
	\Delta^2 c_{it} = \xi_{it} +  \xi_{it-1} + \phi_{it}^{(1)}\left(\zeta_{it} + \rho\zeta_{it-1}\right) + \psi_{it}^{(1)}v_{it}.
\end{equation*}
The transmission parameters of income shocks are given by $\phi_{it}^{(1)} = (1-\pi_{it})\Theta_{it}^Y$ and $\psi_{it}^{(1)} = (1 - \pi_{it})\vartheta_{it}^{Y}$. In the baseline, the permanent-transitory income process has $\rho=1$ and further restricts $\phi_{it}^{(1)} = 1 - \pi_{it}$ because $\Theta_{it}^Y = \sum_{s=0}^{T-t} \vartheta_{it+s}^{Y}=1$.

The quadratic terms in the next two lines of \eqref{AppEq::Consumption_Function_Approximation_Final} reflect the refinement from the higher-order approximation. This leads to the quadratic consumption function, adapted here for income persistence and the biennial nature of the data, and given compactly by  
\begin{equation*}
    \Delta^2 c_{it} = \xi_{it} + \xi_{it-1} 
    + \phi_{it}^{(1)}\left(\zeta_{it} + \rho\zeta_{it-1}\right) + \psi_{it}^{(1)}v_{it} \\
    + \phi_{it}^{(2)}\left(\zeta_{it} + \rho\zeta_{it-1}\right)^2 + \psi_{it}^{(2)}v_{it}^2 + \omega_{it}^{(22)}\left(\zeta_{it} + \rho\zeta_{it-1}\right)v_{it}.
\end{equation*}
The additional transmission parameters of income shocks are given by $\phi_{it}^{(2)} = \frac{1}{2}(1 - \pi_{it})\widetilde{\pi}_{it}$, $\psi_{it}^{(2)} = \frac{1}{2}(1 - \pi_{it})\vartheta_{it}^{Y}(1-(1 - \pi_{it})\vartheta_{it}^{Y})$, and $\omega_{it}^{(22)} = (1 - \pi_{it})\widetilde{\widetilde{\pi}}_{it}\vartheta_{it}^{Y}$. The baseline permanent-transitory income process has $\rho=1$ and further restricts $\phi_{it}^{(2)} = \frac{1}{2}(1 - \pi_{it})\pi_{it}$ and $\omega_{it}^{(22)} = (1 - \pi_{it})\pi_{it}\vartheta_{it}^{Y}$, because $\widetilde{\Theta}_{it}^Y = \Theta_{it}^Y = \sum_{s=0}^{T-t} \vartheta_{it+s}^{Y}=1$.

\section{Identification details}\label{Appendix::Identification}

This appendix provides detailed identification statements based on our use of the recent biennial PSID data. There are three points of departure from section \ref{Sec::Identification} in the paper. 

First, we use the most general AR(1) specification for income, shown in section \ref{Sec::Income_Persistence}. The baseline permanent-transitory specification is a special case with $\rho=1$. Second, we allow for classical measurement error. We let observed income $y_{it}^{*}$ and consumption $c_{it}^{*}$ be the sum of their corresponding true value and measurement error, namely
\begin{equation*}
    y_{it}^{*} = y_{it} + u_{it}^{y} \qquad \text{and} \qquad c_{it}^{*} = c_{it} + u_{it}^{c}.
\end{equation*}
The moments of income measurement error are not separately identifiable from the moments of the transitory shock. This is why we assume the income error to be Gaussian (but let the consumption error unrestricted), so we specify
\begin{equation*}
    {\E}((u^y_{it})^{m}) =
    \begin{cases}
        0                       & \text{for } m=1 \\[-1pt]
        \sigma_{u^y_t}^2        & \text{for } m=2 \\[-1pt]
        0                       & \text{for } m=3 \\[-1pt]
        3(\sigma_{u^y_t}^2)^2   & \text{for } m=4 \\[-1pt]
    \end{cases} \qquad \text{and} \qquad
    {\E}((u^c_{it})^{m}) =
    \begin{cases}
        0                       & \text{for } m=1 \\[-1pt]
        \sigma_{u^c_t}^2        & \text{for } m=2 \\[-1pt]
        \gamma_{u^c_t}          & \text{for } m=3 \\[-1pt]
        \kappa_{u^c_t}          & \text{for } m=4.\\[-1pt]
    \end{cases}
\end{equation*}
We retrieve the variance of income measurement error from the validation study of the PSID in \citet{BoundEtAl1994PSIDerror}, as we explain in section \ref{Sec::Empirical_Implementation}. We do not need similar assumptions for the consumption measurement error, whose moments we identify below. We further assume that income and consumption errors are mutually independent (but note the discussion of consumption imputation from the CEX in appendix \ref{SubAppendix::Empirics_CEX}), independent over time, and independent of the income shocks and taste heterogeneity. Third, we recast the income and consumption processes to reflect the biennial frequency of the modern PSID. 

These choices lead to the theoretical expressions that we bring to the data, namely the process for income growth, the quasi-difference in income, consumption growth in the linear specification, and consumption growth in the quadratic specification, given respectively by
\begin{alignat*}{3}
    \Delta^2        y_{it}  &= (\rho^2-1)P_{it-2}~ &&+ &&~\zeta_{it}+\rho \zeta_{it-1} + \Delta^2 v_{it} + \Delta^2 u^y_{it},\\
    \Delta^2_\rho   y_{it}  &= &&&&~\zeta_{it}+\rho \zeta_{it-1} + \Delta^2_\rho v_{it} + \Delta^2_\rho u^y_{it}, \\
	\Delta^2        c_{it}  &= \xi_{it} + \xi_{it-1} &&+ &&~\phi_{it}^{(1)}(\zeta_{it}+\rho\zeta_{it-1}) + \psi_{it}^{(1)}v_{it} + \Delta^2 u^c_{it}, \\
    \Delta^2        c_{it}  &= \xi_{it} + \xi_{it-1} &&+ &&~\phi_{it}^{(1)}(\zeta_{it}+\rho\zeta_{it-1}) + \psi_{it}^{(1)}v_{it} \\
                            & &&+&&~\phi_{it}^{(2)}(\zeta_{it}+\rho\zeta_{it-1})^2 + \psi_{it}^{(2)}v_{it}^2 +\omega_{it}^{(22)}(\zeta_{it}+\rho\zeta_{it-1})v_{it} + \Delta^2 u^c_{it}.
\end{alignat*}
The notation $\Delta^2 x_{t} = \Delta x_{t} + \Delta x_{t-1} = x_{t}-x_{t-2}$ indicates a first difference in $x$ between $t$ and $t-2$, i.e., the observed frequency. The quasi-difference is defined as $\Delta^2_\rho x_{t} = \Delta_\rho x_{t} + \rho \Delta_\rho x_{t-1} = x_{t}-\rho^2 x_{t-2}$, where $\Delta_\rho x_{t} = x_{t}-\rho x_{t-1}$ is the quasi-difference between $t$ and $t-1$.

Due to the biennial nature of the data, we cannot separately identify the moments of the \emph{yearly} permanent shocks within a two-year period. We thus assume $\sigma_{\zeta_{t}}^2 \approx \sigma_{\zeta_{t-1}}^2$, $\gamma_{\zeta_{t}} \approx \gamma_{\zeta_{t-1}}$, and $\kappa_{\zeta_{t}} \approx \kappa_{\zeta_{t-1}}$, which are exactly true if the underlying distributions are stationary. Similarly, we assume that $\sigma_{\xi_{t}}^2 \approx \sigma_{\xi_{t-1}}^2$. For simplicity of the next statements, we also assume $\sigma_{\zeta_{t}}^2 \approx \sigma_{\zeta_{t+1}}^2 \approx \sigma_{\zeta_{t+2}}^2$, $\sigma_{v_{t}}^2 \approx \sigma_{v_{t-2}}^2 \approx \sigma_{v_{t+2}}^2$, $\sigma_{u^y_{t}}^2 \approx \sigma_{u^y_{t-2}}^2 \approx \sigma_{u^y_{t+2}}^2$, and $\kappa_{v_t} \approx \kappa_{v_{t-2}}$. These last restrictions are not needed for identification, but they simplify the illustration.

\subsection{Income process parameters}\label{SubAppendix::Identification_IncomeProcess}

 The second and third moments of shocks are given by
\begin{alignat*}{3}
	\sigma_{\zeta_{t}}^2 	&= \frac{1}{1+\rho^2} {\E}(\Delta^2_\rho y_{it} \times \sum_{\mathclap{\kappa=\{-2,0,2\}}} \rho^{-\kappa} \Delta^2_\rho y_{it+\kappa}) 
    \qquad &\text{and} \qquad
	\sigma_{v_{t}}^2 		&= - \frac{1}{\rho^2} {\E}(\Delta^2_\rho y_{it} \times \Delta^2_\rho y_{it+2}) - \sigma_{u^y_t}^2, \\
	\gamma_{\zeta_{t}} 		&= \frac{1}{1+\rho^3} {\E}((\Delta^2_\rho y_{it})^2 \times \sum_{\mathclap{\kappa=\{-2,0,2\}}} \rho^{-\kappa} \Delta^2_\rho y_{it+\kappa})  
    \qquad &\text{and} \qquad
	\gamma_{v_{t}} 			&= - \frac{1}{\rho^2} {\E}((\Delta^2_\rho y_{it})^2 \times \Delta^2_\rho y_{it+2}). 
\end{alignat*}
The fourth moments are more involved and are given by
\begin{align*}
	\kappa_{\zeta_t} 
	&= \frac{1}{1+\rho^4} {\E}((\Delta^2_\rho y_{it})^4) - \frac{1+\rho^8}{1+\rho^4} \kappa_{v_t} - 6\frac{\rho^2}{1+\rho^4}(\sigma_{\zeta_{t}}^2)^2 - 6\frac{\rho^4}{1+\rho^4}(\sigma_{v_{t}}^2)^2 - 3(1+\rho^4)(\sigma_{u^y_t}^2)^2\\
	&- 6(1+\rho^2)\sigma_{\zeta_{t}}^2\sigma_{v_{t}}^2 - 6(1+\rho^2)\sigma_{\zeta_{t}}^2\sigma_{u^y_t}^2 - 6(1+\rho^4)\sigma_{v_{t}}^2\sigma_{u^y_t}^2, \\[5pt]
	\kappa_{v_t} 
	&= \frac{1}{\rho^4} {\E}((\Delta^2_\rho y_{it})^2 \times (\Delta^2_\rho y_{it+2})^2) - \frac{(1+\rho^2)^2}{\rho^4}(\sigma_{\zeta_{t}}^2)^2 - \frac{1+\rho^4+\rho^8}{\rho^4}(\sigma_{v_{t}}^2)^2 - \frac{1+4\rho^4+\rho^8}{\rho^4}(\sigma_{u^y_t}^2)^2\\
	&- 2(1+\rho^2)\frac{1+\rho^4}{\rho^4}\sigma_{\zeta_{t}}^2\sigma_{v_{t}}^2 - 2(1+\rho^2)\frac{1+\rho^4}{\rho^4}\sigma_{\zeta_{t}}^2\sigma_{u^y_t}^2 - \left(2\frac{(1+\rho^4)^2}{\rho^4}+4\right)\sigma_{v_{t}}^2\sigma_{u^y_t}^2. 
\end{align*}

In all cases, either we need an external estimate for $\rho$ (e.g., $\rho=1$ as in the permanent-transitory process) or treat $\rho$ as a free parameter to be jointly estimated with the moments of income shocks. In the latter case, $\rho$ is pinned down by 
\begin{equation*}
    {\E}(\Delta^2_\rho y_{it} \times \Delta^2 y_{it-4}) = {\E}((y_{it}-\rho^2 y_{it-2}) \times (y_{it-4}-y_{it-6})) = 0,
\end{equation*}
namely a ``long'' intertemporal covariance. Note that when $\rho$ is a free parameter, five time period of data are needed at a minimum (i.e., $t-6$ through $t+2$ biennially) to identify both the variance of persistent shocks and $\rho$ itself, versus four periods when $\rho$ is set externally.

\subsection{Linear consumption function parameters}\label{SubAppendix::Identification_LinearConsumptionF}

Given the general income process, the linear consumption function, and the properties of shocks, the transmission parameters are given by
\begin{align*}
	\phi^{(1)}_t &= \frac{1}{(1+\rho^2)\sigma_{\zeta_{t}}^2} {\E}(\Delta^2 c_{it} \times \sum_{\mathclap{\kappa=\{-2,0,2\}}} \rho^{-\kappa} \Delta^2_\rho y_{it+\kappa}), \\
	\psi^{(1)}_t &= -\frac{1}{\rho^2 \sigma_{v_{t}}^2}{\E}(\Delta^2 c_{it} \times \Delta^2_\rho y_{it+2}).
\end{align*}
The variance of taste heterogeneity and consumption measurement error are identified as
\begin{align*}
    \sigma_{\xi_t}^2    &= \frac{1}{2}{\E}(\Delta^2 c_{it} \times \sum_{\mathclap{\kappa=\{-2,0,2\}}} \Delta^2 c_{it+\kappa}) - (\phi^{(1)}_t)^2 \frac{1+\rho^2}{2}\sigma_{\zeta_{t}}^2 - \frac{1}{2}(\psi^{(1)}_t)^2 \sigma_{v_{t}}^2, \\
	\sigma_{u^c_t}^2 	&= -{\E}(\Delta^2 c_{it} \times \Delta^2 c_{it+2}).
\end{align*}
The third and fourth moments of taste heterogeneity and consumption error are identified by higher-order consumption moments through expressions analogous to those above.

\subsection{Quadratic consumption function parameters}\label{SubAppendix::Identification_QuadraticConsumptionF}

Given the general income process, the quadratic consumption function, and the properties of shocks, the transmission parameters are given by
\begin{equation}\label{AppEq::IdentifyingEqQuadratic}
\left(\begin{array}{l}
\phi^{(1)}_t \\
\psi^{(1)}_t \\
\phi^{(2)}_t \\
\psi^{(2)}_t \\
\omega^{(22)}_t 
\end{array}\right)
= \mathbf{A}^{-1}
\left(\begin{array}{l}
    {\E}(\Delta^2 c_{it} \times \Delta^2_\rho y_{it}) \\
    {\E}(\Delta^2 c_{it} \times (\Delta^2_\rho y_{it})^2) \\
    {\E}(\Delta^2 c_{it} \times \Delta^2_\rho y_{it+2}) \\
    {\E}(\Delta^2 c_{it} \times (\Delta^2_\rho y_{it+2})^2) \\
    {\E}(\Delta^2 c_{it} \times \Delta^2_\rho y_{it} \times \Delta^2_\rho y_{it+2}) \\
\end{array}\right),
\end{equation}
where the matrix of coefficients $\mathbf{A}$ depends exclusively on the second, third, and fourth moments of income shocks, the AR(1) persistence parameter $\rho$, and the variance of income measurement error.\footnote{While the paper shows matrix $\mathbf{A}$ for the case of annual data, unit root in income, and no measurement error, the general form of $\mathbf{A} = [a_{ij}]_{5\times5}$ with \emph{biennial} data, income persistence, and measurement error has $a_{15}=a_{31}=a_{33}=a_{35}=a_{41}=a_{45}=a_{51}=0$, and
    \begin{align*}
        a_{11} &= (1+\rho^2)\sigma_{\zeta_{t}}^2,        \\
        a_{12} &= \sigma_{v_{t}}^2,                      \\
        a_{13} &= (1+\rho^3)\gamma_{\zeta_{t}},          \\
        a_{14} &= \gamma_{v_{t}},                        \\
        a_{21} &= (1+\rho^3)\gamma_{\zeta_{t}},          \\
        a_{22} &= \gamma_{v_{t}},                        \\  
        a_{23} &= (1+\rho^4)\kappa_{\zeta_{t}} + 6\rho^2(\sigma_{\zeta_{t}}^2)^2 + (1+\rho^2)(1+\rho^4)\sigma_{\zeta_{t}}^2(\sigma_{v_{t}}^2+\sigma_{u^y_{t}}^2),        \\      
        a_{24} &= \kappa_{v_{t}} + \rho^4(\sigma_{v_{t}}^2)^2 + (1+\rho^2)\sigma_{\zeta_{t}}^2\sigma_{v_{t}}^2 +(1+\rho^4)\sigma_{v_{t}}^2\sigma_{u^y_{t}}^2,            \\   
        a_{25} &= 2(1+\rho^2)\sigma_{\zeta_{t}}^2\sigma_{v_{t}}^2,       \\
        a_{32} &= -\rho^2\sigma_{v_{t}}^2,          \\  
        a_{34} &= -\rho^2\gamma_{v_{t}},            \\   
        a_{42} &= \rho^4\gamma_{v_{t}},          \\
        a_{43} &= (1+\rho^2)^2 (\sigma_{\zeta_{t}}^2)^2 + (1+\rho^2)(1+\rho^4) \sigma_{\zeta_{t}}^2 (\sigma_{v_{t}}^2+\sigma_{u^y_{t}}^2),                               \\ 
        a_{44} &= \rho^4\kappa_{v_{t}} + (\sigma_{v_{t}}^2)^2 + (1+\rho^2)\sigma_{\zeta_{t}}^2\sigma_{v_{t}}^2 + (1+\rho^4)\sigma_{v_{t}}^2\sigma_{u^y_{t}}^2,           \\
        a_{52} &= -\rho^2\gamma_{v_{t}},               \\
        a_{53} &= -\rho^2(1+\rho^2)\sigma_{\zeta_{t}}^2 (\sigma_{v_{t}}^2+\sigma_{u^y_{t}}^2),       \\
        a_{54} &= -\rho^2(\kappa_{v_{t}} + \sigma_{v_{t}}^2 \sigma_{u^y_t}^2),                       \\
        a_{55} &= -\rho^2(1+\rho^2)\sigma_{\zeta_{t}}^2\sigma_{v_{t}}^2.
    \end{align*}} 
Its determinant\footnote{The simpler derivation for $\rho=1$ appears in older versions of the paper, see \href{https://arxiv.org/abs/2306.13208}{arxiv.org/abs/2306.13208}.} is given by
\begin{align*}
    \text{det}(\mathbf{A}) 
    &= -\rho^4(1+\rho^2)\sigma_{\zeta_{t}}^4\sigma_{v_{t}}^4 \times \\
    \bigg\{ 
    &\bigg( (1+\rho^2)(1+\rho^4)(k_{\zeta_{t}}-\frac{\gamma_{\zeta_{t}}^2}{\sigma_{\zeta_{t}}^2}-\sigma_{\zeta_t}^4) + (\rho^2-\rho)^2\frac{\gamma_{\zeta_{t}}^2}{\sigma_{\zeta_{t}}^2} \\
    &+ \frac{1+\rho^2}{\rho^4}(1+2\rho^2+2\rho^4+6\rho^6+\rho^8)\sigma_{\zeta_{t}}^4 + \frac{(1+\rho^2)^2}{\rho^4}(1+\rho^8)\sigma_{\zeta_{t}}^2(\sigma_{v_{t}}^2 + \sigma_{u^y_{t}}^2) \bigg) \times \\ 
    &\bigg( \rho^4(k_{v_{t}} -\frac{\gamma_{v_{t}}^2}{\sigma_{v_{t}}^2} - \sigma_{v_{t}}^4) + (1+\rho^4)\sigma_{v_{t}}^4 + (1+\rho^2) \sigma_{\zeta_{t}}^2 \sigma_{v_{t}}^2  + (1+\rho^4) \sigma_{v_{t}}^2 \sigma_{u^y_{t}}^2 \bigg)\\
    &- \frac{(1+\rho^2)^4}{\rho^4}(1+\rho^4)\sigma_{\zeta_{t}}^6 \sigma_{v_{t}}^2 - \frac{(1+\rho^2)^2}{\rho^4}(1+\rho^4)(1+\rho^8)\sigma_{\zeta_{t}}^2 \sigma_{v_{t}}^6 \\
    &- 2\frac{(1+\rho^2)^3}{\rho^4}(1+\rho^4+\rho^8)\sigma_{\zeta_{t}}^4 \sigma_{v_{t}}^4 - \frac{(1+\rho^2)^2}{\rho^4}(1+\rho^4)(1+\rho^8)\sigma_{\zeta_{t}}^2 \sigma_{v_{t}}^2 \sigma_{u^y_{t}}^4 \\
    &- 2\frac{(1+\rho^2)^2}{\rho^4}(1+\rho^4)(1+\rho^8)\sigma_{\zeta_{t}}^2 \sigma_{v_{t}}^4 \sigma_{u^y_{t}}^2 - 2\frac{(1+\rho^2)^3}{\rho^4}(1+\rho^4+\rho^8)\sigma_{\zeta_{t}}^4 \sigma_{v_{t}}^2 \sigma_{u^y_{t}}^2 \bigg\}.
\end{align*}

Assuming that the shocks are \emph{not} drawn from a Bernoulli distribution, kurtosis is bounded below by $\gamma^2/\sigma^2 + \sigma^4$. Given that $0<\rho\leq 1$, it follows that
\begin{align}
\label{AppEq::KurtosisBound_zeta}
    (1+\rho^2)(1+\rho^4)(k_{\zeta_{t}} - \gamma_{\zeta_{t}}^2/\sigma_{\zeta_{t}}^2 - \sigma_{\zeta_{t}}^4) &>0, \\
\label{AppEq::KurtosisBound_v}
    \rho^4(k_{v_{t}} - \gamma_{v_{t}}^2/\sigma_{v_{t}}^2 - \sigma_{v_{t}}^4) &>0.
\end{align}
Note also that 
\begin{align}
\label{AppEq::QuadraticIdentification_Ineq1A}
    (\rho^2-\rho)^2\frac{\gamma_{\zeta_{t}}^2}{\sigma_{\zeta_{t}}^2} &\geq 0, \\
\label{AppEq::QuadraticIdentification_Ineq1B}
    \frac{1+\rho^2}{\rho^4}(1+2\rho^2+2\rho^4+6\rho^6+\rho^8)\sigma_{\zeta_{t}}^4 + \frac{(1+\rho^2)^2}{\rho^4}(1+\rho^8)\sigma_{\zeta_{t}}^2(\sigma_{v_{t}}^2 + \sigma_{u^y_{t}}^2) &> 0, \\
\label{AppEq::QuadraticIdentification_Ineq2}
    (1+\rho^4)\sigma_{v_{t}}^4 + (1+\rho^2) \sigma_{\zeta_{t}}^2 \sigma_{v_{t}}^2  + (1+\rho^4) \sigma_{v_{t}}^2 \sigma_{u^y_{t}}^2 &> 0,
\end{align}
as the sum of mostly strictly positive terms. Adding the inequalities \eqref{AppEq::KurtosisBound_zeta}+\eqref{AppEq::QuadraticIdentification_Ineq1A}+\eqref{AppEq::QuadraticIdentification_Ineq1B} and \eqref{AppEq::KurtosisBound_v}+\eqref{AppEq::QuadraticIdentification_Ineq2} by parts, it follows that the two long bracketed terms in the first lines of the determinant are strictly positive, so their product is also strictly positive. Given the lower bound on the kurtosis, the term inside the curly brackets is then strictly greater than 
\begin{align*}
    \bigg( (\rho^2-\rho)^2\frac{\gamma_{\zeta_{t}}^2}{\sigma_{\zeta_{t}}^2} + \frac{1+\rho^2}{\rho^4}(1+2\rho^2+2\rho^4+6\rho^6+\rho^8)\sigma_{\zeta_{t}}^4 + \frac{(1+\rho^2)^2}{\rho^4}(1+\rho^8)\sigma_{\zeta_{t}}^2(\sigma_{v_{t}}^2 + \sigma_{u^y_{t}}^2) \bigg) &\times \\ 
    \bigg( (1+\rho^4)\sigma_{v_{t}}^4 + (1+\rho^2) \sigma_{\zeta_{t}}^2 \sigma_{v_{t}}^2  + (1+\rho^4) \sigma_{v_{t}}^2 \sigma_{u^y_{t}}^2 \bigg)&\\
    - \frac{(1+\rho^2)^4}{\rho^4}(1+\rho^4)\sigma_{\zeta_{t}}^6 \sigma_{v_{t}}^2 - \frac{(1+\rho^2)^2}{\rho^4}(1+\rho^4)(1+\rho^8)\sigma_{\zeta_{t}}^2 \sigma_{v_{t}}^6 &\\
    - 2\frac{(1+\rho^2)^3}{\rho^4}(1+\rho^4+\rho^8)\sigma_{\zeta_{t}}^4 \sigma_{v_{t}}^4 - \frac{(1+\rho^2)^2}{\rho^4}(1+\rho^4)(1+\rho^8)\sigma_{\zeta_{t}}^2 \sigma_{v_{t}}^2 \sigma_{u^y_{t}}^4 &\\
    - 2\frac{(1+\rho^2)^2}{\rho^4}(1+\rho^4)(1+\rho^8)\sigma_{\zeta_{t}}^2 \sigma_{v_{t}}^4 \sigma_{u^y_{t}}^2 - 2\frac{(1+\rho^2)^3}{\rho^4}(1+\rho^4+\rho^8)\sigma_{\zeta_{t}}^4 \sigma_{v_{t}}^2 \sigma_{u^y_{t}}^2 &=\\
    (\rho^2-\rho)^2\frac{\gamma_{\zeta_{t}}^2}{\sigma_{\zeta_{t}}^2}\bigg( (1+\rho^4)\sigma_{v_{t}}^4 + (1+\rho^2) \sigma_{\zeta_{t}}^2 \sigma_{v_{t}}^2  + (1+\rho^4) \sigma_{v_{t}}^2 \sigma_{u^y_{t}}^2 \bigg)&\\
    + 4\rho^2(1+\rho^2)^2\sigma_{\zeta_{t}}^6 \sigma_{v_{t}}^2 + 4\rho^2(1+\rho^2)(1+\rho^4)\sigma_{\zeta_{t}}^4 \sigma_{v_{t}}^4 + 4\rho^2(1+\rho^2)(1+\rho^4)\sigma_{\zeta_{t}}^4 \sigma_{v_{t}}^2 \sigma_{u^y_{t}}^2 &>0.
\end{align*}
The last inequality follows from $0<\rho\leq 1$ and the fact that we are adding mostly strictly positive terms (and one weakly positive, the term involving $\gamma_{\zeta_{t}}^2$). The expression inside the curly brackets in the determinant is thus strictly positive, so $\text{det}(\mathbf{A}) < 0$. $\mathbf{A}$ is thus full rank and invertible for \emph{any} distribution of permanent \& transitory shocks except Bernoulli.\footnote{If the distributions of both permanent and transitory shocks were Bernoulli, kurtosis would equal its lower bound $\gamma^2/\sigma^2 + \sigma^4$ and the determinant may not always be strictly away from zero.} 

How do different levels of skewness and kurtosis influence the transmission parameters? Consider the transmission of permanent shocks, abstracting from transitory shocks, i.e., $\psi^{(1)}_t=\psi^{(2)}_t=\omega^{(22)}_t=0$, and let $\rho=1$. $\phi^{(1)}_t$ and $\phi^{(2)}_t$ are pinned down by the first equation in \eqref{AppEq::IdentifyingEqQuadratic}, namely $2\sigma_{\zeta_{t}}^2 \phi^{(1)}_t + 2\gamma_{\zeta_{t}} \phi^{(2)}_t = {\E}(\Delta^2 c_{it} \times \Delta^2_\rho y_{it})$. Absent transmission of transitory shocks, the right hand side reflects the strength of the covariance between consumption growth and permanent shocks. Suppose $\phi^{(1)}_t$ is fixed. We highlight three cases: 
\begin{itemize}
    \item ${\E}(\Delta^2 c_{it} \times \Delta^2_\rho y_{it}) - 2\sigma_{\zeta_{t}}^2 \phi^{(1)}_t = 0$. Letting $\gamma_{\zeta_{t}}<0$ as in the data, this equation restricts $\phi^{(2)}_t=0$. Tail shocks thus transmit similarly to non-tail ones, i.e., this configuration identifies a \emph{linear} consumption process. 
    \item ${\E}(\Delta^2 c_{it} \times \Delta^2_\rho y_{it}) - 2\sigma_{\zeta_{t}}^2 \phi^{(1)}_t > 0$. Consumption growth and permanent shocks co-vary more strongly than the linear consumption process suggests, i.e., there is excess co-movement. For $\gamma_{\zeta_{t}}<0$, this configuration identifies $\phi^{(2)}_t<0$ and implies that negative shocks transmit more strongly than other shocks. Negative shocks are often the \emph{largest} shocks (due to left skewness), which explains the excess co-movement. Ceteris paribus, a \emph{less skewed} distribution ($\gamma_{\zeta_{t}}$ drops in magnitude) implies a more negative $\phi^{(2)}_t$, because a more symmetric distribution associated with fewer large shocks requires a stronger transmission of said shocks to maintain a given excess co-movement. If $\gamma_{\zeta_{t}}>0$, this configuration identifies $\phi^{(2)}_t>0$ and analogous arguments apply to positive shocks.
    \item ${\E}(\Delta^2 c_{it} \times \Delta^2_\rho y_{it}) - 2\sigma_{\zeta_{t}}^2 \phi^{(1)}_t < 0$. Consumption growth and permanent shocks co-move less strongly than the linear process suggests, i.e., there is excess smoothness. For $\gamma_{\zeta_{t}}<0$, this identifies $\phi^{(2)}_t>0$ and suggests a muted pass-through of negative shocks. These are the largest shocks to income, so muting their pass-through suppresses the covariance between consumption and income growth. The opposite is true if $\gamma_{\zeta_{t}}>0$.
\end{itemize}

Another equation that pins down $\phi^{(1)}_t$ and $\phi^{(2)}_t$ is $2\gamma_{\zeta_{t}} \phi^{(1)}_t + 2(\kappa_{\zeta_{t}} + \sigma_{\zeta_{t}}^2(3\sigma_{\zeta_{t}}^2+2\sigma_{v_{t}}^2)) \phi^{(2)}_t = {\E}(\Delta^2 c_{it} \times (\Delta^2_\rho y_{it})^2)$, which, abstracting from income measurement error, corresponds to the second line in \eqref{AppEq::IdentifyingEqQuadratic}. Absent transmission of transitory shocks, the right hand side reflects the strength of the covariance between consumption growth and squared shocks so it captures the pass-through of shocks of large magnitude. We highlight again three cases (for $\phi^{(1)}_t$ fixed): 
\begin{itemize}
    \item ${\E}(\Delta^2 c_{it} \times (\Delta^2_\rho y_{it})^2) - 2\gamma_{\zeta_{t}} \phi^{(1)}_t = 0$. This identifies $\phi^{(2)}_t=0$, so tail shocks transmit similarly to non-tail ones. In fact, ${\E}(\Delta^2 c_{it} \times (\Delta^2_\rho y_{it})^2) = 2\gamma_{\zeta_{t}} \phi^{(1)}_t$ is implied by $\Delta^2 c_{it} = \phi_{t}^{(1)}(\zeta_{it}+\rho\zeta_{it-1})$, so this configuration identifies the \emph{linear} consumption process.
    \item ${\E}(\Delta^2 c_{it} \times (\Delta^2_\rho y_{it})^2) - 2\gamma_{\zeta_{t}} \phi^{(1)}_t < 0$, which, for $\gamma_{\zeta_{t}}<0$, is true when ${\E}(\Delta^2 c_{it} \times (\Delta^2_\rho y_{it})^2)$ is negative and large. This suggests that large shocks (of which negative shocks are the most likely) transmit strongly with a negative coefficient. It thus identifies $\phi^{(2)}_t<0$. Ceteris paribus, a \emph{less leptokurtic} income distribution (lower $\kappa_{\zeta_{t}}$) turns $\phi^{(2)}_t$ more negative; a bell-shaped distribution is associated with fewer large shocks, thus necessitating a stronger transmission of said shocks to maintain a given excess co-movement. 
    \item ${\E}(\Delta^2 c_{it} \times (\Delta^2_\rho y_{it})^2) - 2\gamma_{\zeta_{t}} \phi^{(1)}_t > 0$. This identifies $\phi^{(2)}_t>0$ for analogous reasons. For given excess co-movement, the smaller the kurtosis of permanent shocks, the larger the value of $\phi^{(2)}_t$ must be to achieve said excess co-movement.
\end{itemize}

In practice, $\phi^{(1)}_t$ and $\phi^{(2)}_t$ are jointly pinned down by these and other equations of system \eqref{AppEq::IdentifyingEqQuadratic}, and they are also determined jointly with the transmission parameters of transitory shocks, to which analogous statements apply -- skipped here for brevity. The sensitivity analysis is thus more nuanced, so simulations can prove useful. Starting from the baseline estimates of the income process (table \ref{Table::Income_process}), reducing the left skewness of permanent shocks by half increases the magnitude of $\phi^{(2)}_t$ by almost 10\% (more negative). Independently reducing the kurtosis of permanent shocks by half increases the magnitude of $\phi^{(2)}_t$ by 60\% (more negative). The response of $\phi^{(1)}_t$ is much smaller, confirming that the values of skewness and kurtosis are crucial for identifying the pass-through of \emph{tail} shocks.

An interesting special case arises when $\rho=1$, the distributions of permanent and transitory shocks are normal and have the same variance, and there is no measurement error. In this case, $\text{det}(\mathbf{A}) = -288 \sigma_t^{16}$, $\phi_t^{(1)} = (1/(2\sigma_t^2)) {\E}\left(\Delta^2 c_{it} \times \Delta^2_\rho y_{it}\right) + (1/(2\sigma_t^2)){\E}\left(\Delta^2 c_{it} \times \Delta^2_\rho y_{it+2}\right)$ and $\phi_t^{(2)} = (1/(12\sigma_t^4)) {\E}\left(\Delta^2 c_{it} \times (\Delta^2_\rho y_{it})^2\right) + (1/(6\sigma_t^4)){\E}\left(\Delta^2 c_{it} \times \Delta^2_\rho y_{it} \times \Delta^2_\rho y_{it+2}\right)$, where $\sigma_t$ is the dispersion of the hypothetical distribution. Even in such a special environment, $\phi_t^{(2)}$, the transmission parameter of tail shocks, is separately identified from $\phi_t^{(1)}$. Identification does \emph{not} require shocks to be non-Gaussian, but skewness and kurtosis (both defined also in the Gaussian case) are still crucial for identifying the pass-through of tail shocks. Despite normality, shocks in the (normal) tails are associated with a different pass-through from that of average shocks. Analogous expressions separately identify $\psi_t^{(1)}$ and $\psi_t^{(2)}$.

Upon identification of the transmission parameters, ${\E}(\Delta^2 c_{it} \times \Delta^2 c_{it+2})$ identifies $\sigma_{u^{c}_t}^{2}$, while ${\E}((\Delta^2 c_{it})^2)$ identifies $\sigma_{\xi_t}^{2}$. Higher moments of the consumption error or of the taste shock involve higher than fourth-order moments of income, which we do not model.

\subsection{Targeted moments}\label{SubAppendix::Identification_Targeted_Moments}

The estimation targets several over-identifying moments. Table \ref{AppTable::List_of_Moments_Y} lists the target income moments, while table \ref{AppTable::List_of_Moments_C} lists the moments targeted in the various specifications of the consumption model. For simplicity, we illustrate these moments assuming stationarity in the distributions of income shocks, taste heterogeneity, and measurement error, and time-invariance of the partial insurance parameters. These are \emph{not} identifying assumptions but they make the illustration more compact by allowing us to pool common terms together.

\begin{table}[] 
\begin{center}
\caption{Targeted moments of income}\label{AppTable::List_of_Moments_Y}
\begin{tabular}{l L{4.2cm} L{4.3cm} L{4.3cm}}
\toprule
\emph{Moments:}         
    & \mcl{1}{c}{2\textsuperscript{nd}}         &  \mcl{1}{c}{2\textsuperscript{nd}, 3\textsuperscript{rd}} &  \mcl{1}{c}{2\textsuperscript{nd}, 3\textsuperscript{rd}, 4\textsuperscript{th}} \\
\midrule
1   & ${\E}((\Delta^{2}_\rho y_{it})^2)$                         
    & ${\E}((\Delta^{2}_\rho y_{it})^2)$                             
    & ${\E}((\Delta^{2}_\rho y_{it})^2)$         \\
2   & ${\E}(\Delta^{2}_\rho y_{it} \times \Delta^{2}_\rho y_{it-2})$  
    & ${\E}(\Delta^{2}_\rho y_{it} \times \Delta^{2}_\rho y_{it-2})$      
    & ${\E}(\Delta^{2}_\rho y_{it} \times \Delta^{2}_\rho y_{it-2})$\\
3   & ${\E}(\Delta^{2}_\rho y_{it}\times \Delta^2 y_{it-4})$                                                      
    & ${\E}(\Delta^{2}_\rho y_{it}\times \Delta^2 y_{it-4})$                             
    & ${\E}(\Delta^{2}_\rho y_{it}\times \Delta^2 y_{it-4})$  \\
4   &                                                       
    & ${\E}((\Delta^{2}_\rho y_{it})^3)$                             
    & ${\E}((\Delta^{2}_\rho y_{it})^3)$         \\
5   &                                                       
    & ${\E}((\Delta^{2}_\rho y_{it})^2 \times \Delta^{2}_\rho y_{it-2})$  
    & ${\E}((\Delta^{2}_\rho y_{it})^2 \times \Delta^{2}_\rho y_{it-2})$ \\
6   &                                                       
    & ${\E}(\Delta^{2}_\rho y_{it} \times (\Delta^{2}_\rho y_{it-2})^2)$  
    & ${\E}(\Delta^{2}_\rho y_{it} \times (\Delta^{2}_\rho y_{it-2})^2)$ \\
7   &                                                       
    &                                                           
    & ${\E}((\Delta^{2}_\rho y_{it})^4)$         \\
8   &                                                       
    &                                                           
    & ${\E}((\Delta^{2}_\rho y_{it})^2 \times (\Delta^{2}_\rho y_{it-2})^2)$\\
9   &                                                       
    &                                                           
    & ${\E}((\Delta^{2}_\rho y_{it})^3 \times \Delta^{2}_\rho y_{it-2})$\\
10  &                                                       
    &                                                           
    & ${\E}(\Delta^{2}_\rho y_{it} \times (\Delta^{2}_\rho y_{it-2})^3)$\\
\bottomrule
\end{tabular}
\caption*{\fsz\emph{Notes:} The table lists the moments targeted in the estimation of the various specifications of the income process. $\Delta^2 y_{t} = y_{t} - y_{t-2}$ indicates a first difference between periods $t$ and $t-2$; $\Delta^2_\rho y_{t} = y_{t} - \rho^2 y_{t-2}$ indicates the analogous quasi-difference. Note that $\Delta^2 y_{t} = \Delta^2_\rho y_{t}$ if $\rho=1$ (baseline permanent-transitory process).}
\end{center}
\end{table}

\subsubsection{Moments of income}\label{SubSubAppendix::Identification_Targeted_Moments_Income}

\begin{align*}
    {\E}((\Delta^{2}_\rho y_{it})^2) 
        &= (1+\rho^2)\sigma_{\zeta}^{2} + (1+\rho^4)\sigma_{v}^{2} + (1+\rho^4)\sigma_{{u}^{y}}^{2} 
        \\
    {\E}(\Delta^{2}_\rho y_{it}\times\Delta^{2}_\rho y_{it-2}) 
        &= - \rho^2\sigma_{v}^{2} - \rho^2\sigma_{{u}^{y}}^{2} 
        \\
    {\E}(\Delta^{2}_\rho y_{it}\times \Delta^2 y_{it-4}) 
        &= 0 
        \\
    {\E}((\Delta^{2}_\rho y_{it})^3)
        &= (1+\rho^3)\gamma_{\zeta} + (1-\rho^6)\gamma_{v}
        \\
    {\E}((\Delta^{2}_\rho y_{it})^2 \times \Delta^{2}_\rho y_{it-2}) 
        &= \rho^4\gamma_{v} 
        \\
    {\E}(\Delta^{2}_\rho y_{it} \times (\Delta^{2}_\rho y_{it-2})^2) 
        &= -\rho^2\gamma_{v} 
        \\
    {\E}((\Delta^{2}_\rho y_{it})^4)
        &= (1+\rho^4)\kappa_{\zeta} + (1+\rho^8)\kappa_{v} + 6\rho^2(\sigma_{\zeta}^2)^2 + 6\rho^4(\sigma_{v}^2)^2 + 3(1+\rho^4)^2(\sigma_{u^y}^2)^2 \\
        &+ 6(1+\rho^4)\Big((1+\rho^2)\sigma_{\zeta}^2 \sigma_{v}^2 + (1+\rho^2)\sigma_{\zeta}^2 \sigma_{u^y}^2 + (1+\rho^4)\sigma_{v}^2 \sigma_{u^y}^2\Big) 
        \\
    {\E}((\Delta^{2}_\rho y_{it})^2 \times (\Delta^{2}_\rho y_{it-2})^2)
        &= \rho^4\kappa_{v} + (1+\rho^2)^2(\sigma_{\zeta}^2)^2 + (1+\rho^4+\rho^8)(\sigma_{v}^2)^2 + (1+4\rho^4+\rho^8)(\sigma_{u^y}^2)^2 \\
        &+ 2(1+\rho^2)(1+\rho^4)\sigma_{\zeta}^2 \sigma_{v}^2 + 2(1+\rho^2)(1+\rho^4)\sigma_{\zeta}^2 \sigma_{u^y}^2 + (2(1+\rho^4)^2 + 4\rho^4) \sigma_{v}^2 \sigma_{u^y}^2 
        \\ 
    {\E}((\Delta^{2}_\rho y_{it})^3 \times \Delta^{2}_\rho y_{it-2})
        &= -\rho^6\kappa_{v} -3\rho^2(\sigma_{v}^2)^2 -3\rho^2(1+\rho^4)(\sigma_{u^y}^2)^2 \\
        &-3\rho^2(1+\rho^2)\sigma_{\zeta}^2 \sigma_{v}^2 -3\rho^2(1+\rho^2)\sigma_{\zeta}^2 \sigma_{u^y}^2 -6\rho^2(1+\rho^4)\sigma_{v}^2 \sigma_{u^y}^2 
        \\  
    {\E}(\Delta^{2}_\rho y_{it} \times (\Delta^{2}_\rho y_{it-2})^3)
        &= -\rho^2\kappa_{v} -3\rho^6(\sigma_{v}^2)^2 -3\rho^2(1+\rho^4)(\sigma_{u^y}^2)^2 \\
        &-3\rho^2(1+\rho^2)\sigma_{\zeta}^2 \sigma_{v}^2 -3\rho^2(1+\rho^2)\sigma_{\zeta}^2 \sigma_{u^y}^2 -6\rho^2(1+\rho^4)\sigma_{v}^2 \sigma_{u^y}^2
\end{align*}

\begin{sidewaystable}[] 
\begin{center}
\caption{Targeted joint moments of consumption and income}\label{AppTable::List_of_Moments_C}
\begin{tabular}{l L{3.7cm} L{5.4cm} L{4.8cm} l L{4.8cm}}
\toprule
\emph{Cons. fn.:}       &\mcl{3}{c}{\textbf{Linear}} && \mcl{1}{c}{\textbf{Quadratic}}\\
\cmidrule{2-4}\cmidrule{6-6}
\emph{Moments:}         
    & \mcl{1}{c}{2\textsuperscript{nd}}        &  \mcl{1}{c}{2\textsuperscript{nd}, 3\textsuperscript{rd}} &  \mcl{1}{c}{2\textsuperscript{nd}, 3\textsuperscript{rd}, 4\textsuperscript{th}} && \mcl{1}{c}{2\textsuperscript{nd}, 3\textsuperscript{rd}, 4\textsuperscript{th} (income shocks)} \\
\midrule
1   & ${\E}((\Delta^{2} c_{it})^2)$                                                     & ${\E}((\Delta^{2} c_{it})^2)$                            
    & ${\E}((\Delta^{2} c_{it})^2)$                                                     && ${\E}((\Delta^{2} c_{it})^2)$ \\
2   & ${\E}(\Delta^{2} c_{it} \times \Delta^{2} c_{it-2})$                              & ${\E}(\Delta^{2} c_{it} \times \Delta^{2} c_{it-2})$         
    & ${\E}(\Delta^{2} c_{it} \times \Delta^{2} c_{it-2})$                              && ${\E}(\Delta^{2} c_{it} \times \Delta^{2} c_{it-2})$\\
3   & ${\E}(\Delta^{2}_\rho y_{it} \times \Delta^{2} c_{it})$                    & ${\E}(\Delta^{2}_\rho y_{it} \times \Delta^{2} c_{it})$           
    & ${\E}(\Delta^{2}_\rho y_{it} \times \Delta^{2} c_{it})$                    && ${\E}(\Delta^{2}_\rho y_{it} \times \Delta^{2} c_{it})$\\
4   & ${\E}(\Delta^{2}_\rho y_{it} \times \Delta^{2} c_{it-2})$                  & ${\E}(\Delta^{2}_\rho y_{it} \times \Delta^{2} c_{it-2})$       
    & ${\E}(\Delta^{2}_\rho y_{it} \times \Delta^{2} c_{it-2})$                  && ${\E}(\Delta^{2}_\rho y_{it} \times \Delta^{2} c_{it-2})$ \\
5   &                                                                                   & ${\E}((\Delta^{2} c_{it})^3)$                        
    & ${\E}((\Delta^{2} c_{it})^3)$                                                     && \\
6   &                                                                                   & ${\E}((\Delta^{2} c_{it})^2 \times \Delta^{2} c_{it-2})$   
    & ${\E}((\Delta^{2} c_{it})^2 \times \Delta^{2} c_{it-2})$                          && \\
7   &                                                                                   & ${\E}(\Delta^{2} c_{it} \times (\Delta^{2} c_{it-2})^2)$     
    & ${\E}(\Delta^{2} c_{it} \times (\Delta^{2} c_{it-2})^2)$                          && \\
8   &                                                                                   & ${\E}((\Delta^{2}_\rho y_{it})^2 \times \Delta^{2} c_{it})$       
    & ${\E}((\Delta^{2}_\rho y_{it})^2 \times \Delta^{2} c_{it})$                && ${\E}((\Delta^{2}_\rho y_{it})^2 \times \Delta^{2} c_{it})$ \\
9   &                                                                                   & ${\E}( \Delta^{2}_\rho y_{it} \times (\Delta^{2} c_{it})^2)$       
    & ${\E}( \Delta^{2}_\rho y_{it} \times (\Delta^{2} c_{it})^2)$               && \\
10  &                                                                                   & ${\E}((\Delta^{2}_\rho y_{it})^2 \times \Delta^{2} c_{it-2})$   
    & ${\E}((\Delta^{2}_\rho y_{it})^2 \times \Delta^{2} c_{it-2})$              && ${\E}((\Delta^{2}_\rho y_{it})^2 \times \Delta^{2} c_{it-2})$ \\
11  &                                                                                   & ${\E}(\Delta^{2}_\rho y_{it} \times (\Delta c_{it-2})^2)$       
    & ${\E}(\Delta^{2}_\rho y_{it} \times (\Delta^{2} c_{it-2})^2)$              && \\
12  &                                                                                   & ${\E}((\Delta^{2}_\rho y_{it-2})^2 \times \Delta^{2} c_{it})$     
    & ${\E}((\Delta^{2}_\rho y_{it-2})^2 \times \Delta^{2} c_{it})$              && ${\E}((\Delta^{2}_\rho y_{it-2})^2 \times \Delta^{2} c_{it})$ \\
13  &                   & ${\E}(\Delta^{2}_\rho y_{it} \times \Delta^{2}_\rho y_{it+2} \times \Delta^{2} c_{it})$ 
                        & ${\E}(\Delta^{2}_\rho y_{it} \times \Delta^{2}_\rho y_{it+2} \times \Delta^{2} c_{it})$  
                       && ${\E}(\Delta^{2}_\rho y_{it} \times \Delta^{2}_\rho y_{it+2} \times \Delta^{2} c_{it})$\\
14  &                   & ${\E}(\Delta^{2}_\rho y_{it} \times \Delta^{2}_\rho y_{it+2} \times \Delta^{2} c_{it-2})$ 
                        & ${\E}(\Delta^{2}_\rho y_{it} \times \Delta^{2}_\rho y_{it+2} \times \Delta^{2} c_{it-2})$         
                       && ${\E}(\Delta^{2}_\rho y_{it} \times \Delta^{2}_\rho y_{it+2} \times \Delta^{2} c_{it-2})$\\
15  &                   & ${\E}(\Delta^{2}_\rho y_{it} \times \Delta^{2}_\rho y_{it+2} \times \Delta^{2} c_{it+2})$ 
                        & ${\E}(\Delta^{2}_\rho y_{it} \times \Delta^{2}_\rho y_{it+2} \times \Delta^{2} c_{it+2})$         
                       && ${\E}(\Delta^{2}_\rho y_{it} \times \Delta^{2}_\rho y_{it+2} \times \Delta^{2} c_{it+2})$\\
16  &                                                                   &       
    & ${\E}((\Delta^{2} c_{it})^4)$                                    && \\
17  &                                                                   &       
    & ${\E}((\Delta^{2} c_{it})^2 \times (\Delta^{2} c_{it-2})^2)$     && \\
18  &                                                                   &       
    & ${\E}((\Delta^{2} c_{it})^3 \times \Delta^{2} c_{it-2})$         && \\
19  &                                                                   &       
    & ${\E}(\Delta^{2} c_{it} \times (\Delta^{2} c_{it-2})^3)$         && \\
20  &                                                                   &       
    & ${\E}((\Delta^{2}_\rho y_{it})^2 \times (\Delta^{2} c_{it})^2)$      &&\\
21  &                                                                   &       
    & ${\E}((\Delta^{2}_\rho y_{it})^3 \times \Delta^{2} c_{it})$           &&\\
22  &                                                                   &       
    & ${\E}(\Delta^{2}_\rho y_{it} \times (\Delta^{2} c_{it})^3)$           &&\\
23  &                                                                   &       
    & ${\E}((\Delta^{2}_\rho y_{it})^2 \times (\Delta^{2} c_{it-2})^2)$     &&\\
24  &                                                                   &       
    & ${\E}((\Delta^{2}_\rho y_{it-2})^2 \times (\Delta^{2} c_{it})^2 )$    &&\\
25  &                                                                   &       
    & ${\E}((\Delta^{2}_\rho y_{it})^3 \times \Delta^{2} c_{it-2})$         && \\
26  &                                                                   &      
    & ${\E}(\Delta^{2}_\rho y_{it} \times (\Delta^{2} c_{it-2})^3)$         && \\
\bottomrule
\end{tabular}
\caption*{\fsz\emph{Notes:} The table lists the moments targeted in the estimation of the different specifications of the consumption process. The notation $\Delta^2 x_{t} = x_{t} - x_{t-2}$ indicates a first difference of the generic variable $x$ between periods $t$ and $t-2$; $\Delta^2_\rho x_{t} = x_{t} - \rho^2 x_{t-2}$ indicates the analogous quasi-difference.}
\end{center}
\end{sidewaystable}

\subsubsection{Moments of consumption -- linear function}\label{SubSubAppendix::Identification_Targeted_Moments_LinearFunction}

\begin{align*}
    {\E}((\Delta^2 c_{it})^2) 
        &= 2\sigma_\xi^{2} + (\phi^{(1)})^2(1+\rho^2)\sigma_\zeta^{2} + (\psi^{(1)})^2\sigma_{v}^{2} + 2\sigma_{u^{c}}^{2} 
        \\
    {\E}(\Delta^2 c_{it}\times\Delta^2 c_{it-2}) 
        &= -\sigma_{{u}^{c}}^{2} 
        \\
    {\E}(\Delta^{2}_\rho y_{it}\times\Delta^2 c_{it}) 
        &= \phi^{(1)}(1+\rho^2)\sigma_\zeta^{2} + \psi^{(1)}\sigma_v^{2} 
        \\
    {\E}(\Delta^{2}_\rho y_{it}\times\Delta^2 c_{it-2}) 
        &= -\psi^{(1)}\rho^2\sigma_v^{2} 
        \\
    {\E}((\Delta^2 c_{it})^3)
        &= 2\gamma_{\xi} + (\phi^{(1)})^3 (1+\rho^3) \gamma_{\zeta} + (\psi^{(1)})^3 \gamma_{v} 
        \\
    {\E}((\Delta^2 c_{it})^2 \times \Delta^2 c_{it-2}) 
        &= \gamma_{u^c} 
        \\
    {\E}(\Delta^2 c_{it} \times (\Delta^2 c_{it-2})^2) 
        &= - \gamma_{u^c} 
        \\
    {\E}((\Delta^{2}_\rho y_{it})^2\times\Delta^2 c_{it}) 
        &= \phi^{(1)} (1+\rho^3) \gamma_{\zeta} + \psi^{(1)}\gamma_{v} 
        \\
    {\E}(\Delta^{2}_\rho y_{it} \times (\Delta^2 c_{it})^2) 
        &= (\phi^{(1)})^2 (1+\rho^3) \gamma_{\zeta} + (\psi^{(1)})^2\gamma_{v} 
        \\
    {\E}((\Delta^{2}_\rho y_{it})^2\times \Delta^2 c_{it-2}) 
        &= \psi^{(1)}\rho^4\gamma_{v} 
        \\
    {\E}(\Delta^{2}_\rho y_{it} \times (\Delta^2 c_{it-2})^2) 
        &= -(\psi^{(1)})^2\rho^2\gamma_{v} 
        \\
    {\E}((\Delta^{2}_\rho y_{it-2})^2\times \Delta^2 c_{it}) 
        &= 0 
        \\
    {\E}(\Delta^{2}_\rho y_{it} \times \Delta^{2}_\rho y_{it+2} \times \Delta^2 c_{it}) 
        &= -\psi^{(1)} \rho^2\gamma_{v} 
        \\
    {\E}(\Delta^{2}_\rho y_{it} \times \Delta^{2}_\rho y_{it+2} \times \Delta^2 c_{it-2}) 
        &=0 
        \\ 
    {\E}(\Delta^{2}_\rho y_{it} \times \Delta^{2}_\rho y_{it+2} \times \Delta^2 c_{it+2}) 
        &=0 
        \\
    {\E}((\Delta^2 c_{it})^4)
        &= 2 \kappa_{\xi} + (\phi^{(1)})^4 (1+\rho^4) \kappa_{\zeta} + (\psi^{(1)})^4 \kappa_{v} + 2 \kappa_{u^c} + 6 (\sigma_{\xi}^2)^2 \\
        &+ 6 (\phi^{(1)})^4 \rho^2 (\sigma_{\zeta}^2)^2 + 6 (\sigma_{u^c}^2)^2 + 12 (\phi^{(1)})^2 (1+\rho^2) \sigma_{\xi}^2 \sigma_{\zeta}^2 \\
        &+ 12 (\psi^{(1)})^2 \sigma_{\xi}^2 \sigma_{v}^2 + 24 \sigma_{\xi}^2 \sigma_{u^c}^2 + 6 (\phi^{(1)})^2 (\psi^{(1)})^2 (1+\rho^2)\sigma_{\zeta}^2 \sigma_{v}^2 \\
        &+ 12 (\phi^{(1)})^2 (1+\rho^2) \sigma_{\zeta}^2 \sigma_{u^c}^2 + 12 (\psi^{(1)})^2 \sigma_{v}^2 \sigma_{u^c}^2
        \\
    {\E}((\Delta^2 c_{it})^2 \times (\Delta^2 c_{it-2})^2)
        &= 4 (\sigma_{\xi}^2)^2 + 4 (\phi^{(1)})^2 (1+\rho^2) \sigma_{\xi}^2 \sigma_{\zeta}^2 + 4 (\psi^{(1)})^2 \sigma_{\xi}^2 \sigma_{v}^2 + 8 \sigma_{\xi}^2 \sigma_{u^c}^2 \\
        &+ (\phi^{(1)})^4 (1+\rho^2)^2 (\sigma_{\zeta}^2)^2 + 3 (\sigma_{u^c}^2)^2 + 2(\phi^{(1)})^2 (\psi^{(1)})^2 (1+\rho^2) \sigma_{\zeta}^2 \sigma_{v}^2 \\
        &+ 4 (\phi^{(1)})^2 (1+\rho^2) \sigma_{\zeta}^2 \sigma_{u^c}^2 + (\psi^{(1)})^4 (\sigma_{v}^2)^2 + 4 (\psi^{(1)})^2 \sigma_{v}^2 \sigma_{u^c}^2 + \kappa_{u^c} 
            \\
    {\E}((\Delta^2 c_{it})^3 \times \Delta^2 c_{it-2})
        &= - \kappa_{u^c} -3 (\sigma_{u^c}^2)^2  -6 \sigma_{\xi}^2 \sigma_{u^c}^2 -3 (\phi^{(1)})^2 (1+\rho^2)\sigma_{\zeta}^2 \sigma_{u^c}^2 -3 (\psi^{(1)})^2 \sigma_{v}^2 \sigma_{u^c}^2 
        \\
    {\E}(\Delta^2 c_{it} \times (\Delta^2 c_{it-2})^3)
        &= - \kappa_{u^c} -3 (\sigma_{u^c}^2)^2  -6 \sigma_{\xi}^2 \sigma_{u^c}^2 -3 (\phi^{(1)})^2 (1+\rho^2)\sigma_{\zeta}^2 \sigma_{u^c}^2 -3 (\psi^{(1)})^2 \sigma_{v}^2 \sigma_{u^c}^2 
        \\
    {\E}((\Delta^{2}_\rho y_{it})^2 \times (\Delta^2 c_{it})^2)
        &= 2 (1+\rho^2) \sigma_{\zeta}^2 \sigma_{\xi}^2 + (\phi^{(1)})^2 (1+\rho^4) \kappa_{\zeta} + 6 (\phi^{(1)})^2 \rho^2 (\sigma_{\zeta}^2)^2 \\
        &+ (\psi^{(1)})^2 (1+\rho^2) \sigma_{\zeta}^2 \sigma_{v}^2 + 2 (1+\rho^2) \sigma_{\zeta}^2 \sigma_{u^c}^2 + 2 (1+\rho^4) \sigma_{v}^2 \sigma_{\xi}^2 \\
        &+ (\phi^{(1)})^2 (1+\rho^2)(1+\rho^4)\sigma_{v}^2 \sigma_{\zeta}^2 + (\psi^{(1)})^2 \kappa_{v} + (\psi^{(1)})^2 \rho^4 (\sigma_{v}^2)^2 \\
        &+ 2 (1+\rho^4) \sigma_{v}^2 \sigma_{u^c}^2 + 2 (1+\rho^4) \sigma_{u^y}^2 \sigma_{\xi}^2 + (\phi^{(1)})^2 (1+\rho^2)(1+\rho^4)\sigma_{u^y}^2 \sigma_{\zeta}^2 \\
        &+ (\psi^{(1)})^2 (1+\rho^4)\sigma_{u^y}^2 \sigma_{v}^2 + 2 (1+\rho^4)\sigma_{u^y}^2 \sigma_{u^c}^2 + 4\phi^{(1)} \psi^{(1)} (1+\rho^2)\sigma_{\zeta}^2 \sigma_{v}^2 
        \\
    {\E}((\Delta^{2}_\rho y_{it})^3 \times \Delta^2 c_{it})
        &= \phi^{(1)} (1+\rho^4)\kappa_{\zeta} + 6 \phi^{(1)} \rho^2 (\sigma_{\zeta}^2)^2 + \psi^{(1)} \kappa_{v} + 3 \psi^{(1)} \rho^4(\sigma_{v}^2)^2\\
        &+ 3\psi^{(1)} (1+\rho^2)\sigma_{\zeta}^2 \sigma_{v}^2 + 3 \phi^{(1)} (1+\rho^2)(1+\rho^4)\sigma_{\zeta}^2 \sigma_{v}^2 \\
        &+ 3 \phi^{(1)} (1+\rho^2)(1+\rho^4)\sigma_{\zeta}^2 \sigma_{u^y}^2 + 3 \psi^{(1)} (1+\rho^4)\sigma_{v}^2 \sigma_{u^y}^2 
        \\
    {\E}(\Delta^{2}_\rho y_{it} \times (\Delta^2 c_{it})^3)
        &= (\phi^{(1)})^3 (1+\rho^4)\kappa_{\zeta} + 6 (\phi^{(1)})^3 \rho^2(\sigma_{\zeta}^2)^2 + (\psi^{(1)})^3 \kappa_{v} \\
        &+ 6\phi^{(1)} (1+\rho^2)\sigma_{\xi}^2 \sigma_{\zeta}^2 + 6\psi^{(1)} \sigma_{\xi}^2 \sigma_{v}^2 + 3\phi^{(1)} (\psi^{(1)})^2 (1+\rho^2) \sigma_{\zeta}^2 \sigma_{v}^2 \\
        &+ 3 (\phi^{(1)})^2\psi^{(1)} (1+\rho^2)\sigma_{\zeta}^2 \sigma_{v}^2 + 6 \phi^{(1)} (1+\rho^2)\sigma_{\zeta}^2 \sigma_{u^c}^2 + 6 \psi^{(1)} \sigma_{v}^2 \sigma_{u^c}^2 
        \\
    {\E}((\Delta^{2}_\rho y_{it})^2 \times (\Delta^2 c_{it-2})^2)
        &= 2 (1+\rho^2) \sigma_{\zeta}^2 \sigma_{\xi}^2 + (\phi^{(1)})^2 (1+\rho^2)^2 (\sigma_{\zeta}^2)^2 + (\psi^{(1)})^2 (1+\rho^2) \sigma_{\zeta}^2 \sigma_{v}^2 \\
        &+ 2 (1+\rho^2) \sigma_{\zeta}^2 \sigma_{u^c}^2 + 2 (1+\rho^4) \sigma_{v}^2 \sigma_{\xi}^2 + (\phi^{(1)})^2 (1+\rho^2)(1+\rho^4)\sigma_{v}^2 \sigma_{\zeta}^2 \\
        &+ (\psi^{(1)})^2 \rho^4\kappa_{v} + (\psi^{(1)})^2 (\sigma_{v}^2)^2 + 2 (1+\rho^4) \sigma_{v}^2 \sigma_{u^c}^2 + 2 (1+\rho^4) \sigma_{u^y}^2 \sigma_{\xi}^2 \\
        &+ (\phi^{(1)})^2 (1+\rho^2)(1+\rho^4)\sigma_{u^y}^2 \sigma_{\zeta}^2 + (\psi^{(1)})^2 (1+\rho^4)\sigma_{u^y}^2 \sigma_{v}^2 + 2 (1+\rho^4) \sigma_{u^y}^2 \sigma_{u^c}^2
        \\
    {\E}((\Delta^{2}_\rho y_{it-2})^2 \times (\Delta^2 c_{it})^2)
        &= 2 (1+\rho^2) \sigma_{\zeta}^2 \sigma_{\xi}^2 + (\phi^{(1)})^2 (1+\rho^2)^2 (\sigma_{\zeta}^2)^2 + (\psi^{(1)})^2 (1+\rho^2) \sigma_{\zeta}^2 \sigma_{v}^2 \\
        &+ 2 (1+\rho^2) \sigma_{\zeta}^2 \sigma_{u^c}^2 + 2 (1+\rho^4) \sigma_{v}^2 \sigma_{\xi}^2 + (\phi^{(1)})^2 (1+\rho^2)(1+\rho^4)\sigma_{v}^2 \sigma_{\zeta}^2 \\
        &+ (\psi^{(1)})^2 (1+\rho^4)(\sigma_{v}^2)^2 + 2 (1+\rho^4) \sigma_{v}^2 \sigma_{u^c}^2 + 2 (1+\rho^4) \sigma_{u^y}^2 \sigma_{\xi}^2 \\
        &+ (\phi^{(1)})^2 (1+\rho^2)(1+\rho^4)\sigma_{u^y}^2 \sigma_{\zeta}^2 + (\psi^{(1)})^2 (1+\rho^4)\sigma_{u^y}^2 \sigma_{v}^2 + 2 (1+\rho^4) \sigma_{u^y}^2 \sigma_{u^c}^2
        \\
    {\E}((\Delta^{2}_\rho y_{it})^3 \times \Delta^2 c_{it-2})
        &= -\psi^{(1)} \rho^6 \kappa_{v} -3 \psi^{(1)} \rho^2 (\sigma_{v}^2)^2 - 3\psi^{(1)} (1+\rho^2)\rho^2\sigma_{\zeta}^2 \sigma_{v}^2 - 3\psi^{(1)} (1+\rho^4)\rho^2\sigma_{v}^2 \sigma_{u^y}^2 
        \\
    {\E}(\Delta^{2}_\rho y_{it} \times (\Delta^2 c_{it-2})^3)
        &= -(\psi^{(1)})^3 \rho^2 \kappa_{v} -6 \psi^{(1)} \rho^2 \sigma_{v}^2 \sigma_{\xi}^2 -3 (\phi^{(1)})^2 \psi^{(1)} (1+\rho^2)\rho^2\sigma_{v}^2 \sigma_{\zeta}^2 -6 \psi^{(1)} \rho^2 \sigma_{v}^2 \sigma_{u^c}^2 
\end{align*}

\subsubsection{Moments of consumption -- quadratic function}\label{SubSubAppendix::Identification_Targeted_Moments_QuadrFunction}

\begin{align*}
    {\E}((\Delta^2 c_{it})^2) 
        &= 2\sigma_{\xi}^{2} + (\phi^{(1)})^2(1+\rho^2)\sigma_{\zeta}^{2} + (\psi^{(1)})^2\sigma_{v}^{2} + (\phi^{(2)})^2\Big((1+\rho^4)\kappa_{\zeta} + 6\rho^2(\sigma_{\zeta}^{2})^2\Big) \\
        &+ (\psi^{(2)})^2\kappa_{v} + (\omega^{(22)})^2(1+\rho^2)\sigma_{\zeta}^{2}\sigma_{v}^{2} + 2\sigma_{u^{c}}^{2} \\
        &+ 2\phi^{(1)}\phi^{(2)}(1+\rho^3)\gamma_{\zeta} + 2\psi^{(1)}\psi^{(2)}\gamma_{v} + 2\phi^{(2)}\psi^{(2)}(1+\rho^2)\sigma_{\zeta}^{2} \sigma_{v}^{2} \\
    {\E}(\Delta^2 c_{it}\times\Delta^2 c_{it-2}) 
        &= (\phi^{(2)})^2(1+\rho^2)^2(\sigma_{\zeta}^{2})^2 + 2\phi^{(2)}\psi^{(2)}(1+\rho^2)\sigma_{\zeta}^{2} \sigma^2_{v} + (\psi^{(2)})^2 (\sigma^2_{v})^2 - \sigma_{u^{c}}^{2} \\
    {\E}(\Delta^{2}_\rho y_{it}\times\Delta^2 c_{it}) 
        &= \phi^{(1)}(1+\rho^2)\sigma_{\zeta}^{2} + \psi^{(1)}\sigma_{v}^{2} + \phi^{(2)}(1+\rho^3)\gamma_{\zeta} + \psi^{(2)}\gamma_{v} \\
    {\E}(\Delta^{2}_\rho y_{it}\times\Delta^2 c_{it-2}) 
        &= - \psi^{(1)}\rho^2\sigma_{v}^{2} - \psi^{(2)}\rho^2\gamma_{v} \\
    {\E}((\Delta^{2}_\rho y_{it})^2\times\Delta^2 c_{it}) 
        &= \phi^{(1)}(1+\rho^3)\gamma_{\zeta} + \psi^{(1)}\gamma_{v} \\
        &+ \phi^{(2)} \Big((1+\rho^4)\kappa_{\zeta} + 6\rho^2(\sigma_{\zeta}^2)^2 + (1+\rho^2)(1+\rho^4)\sigma_{\zeta}^2(\sigma_{v}^2 + \sigma_{u^y}^2)\Big)\\ 
        &+ \psi^{(2)} \Big(\kappa_{v} + \rho^4(\sigma_{v}^2)^2 + (1+\rho^2)\sigma_{\zeta}^2 \sigma_{v}^2 + (1+\rho^4)\sigma_{v}^2\sigma_{u^y}^2\Big) + 2\omega^{(22)}(1+\rho^2)\sigma_{\zeta}^2 \sigma_{v}^2 \\
    {\E}((\Delta^{2}_\rho y_{it})^2\times \Delta^2 c_{it-2}) 
        &= \psi^{(1)}\rho^4\gamma_{v} + \phi^{(2)}(1+\rho^2)\sigma_{\zeta}^2 \Big((1+\rho^2)\sigma_{\zeta}^2 + (1+\rho^4)(\sigma_{v}^2 +\sigma_{u^y}^2)\Big) \\
        &+ \psi^{(2)} \Big(\rho^4\kappa_{v} + (\sigma_{v}^2)^2 + (1+\rho^2)\sigma_{\zeta}^2\sigma_{v}^2 + (1+\rho^4)\sigma_{v}^2\sigma_{u^y}^2\Big) \\
    {\E}((\Delta^{2}_\rho y_{it-2})^2\times \Delta^2 c_{it}) 
        &= \phi^{(2)}(1+\rho^2)\sigma_{\zeta}^2 \Big((1+\rho^2)\sigma_{\zeta}^2 + (1+\rho^4)(\sigma_{v}^2+\sigma_{u^y}^2)\Big) \\
        &+ \psi^{(2)} \sigma_{v}^2 \Big((1+\rho^2)\sigma_{\zeta}^2 + (1+\rho^4)(\sigma_{v}^2+\sigma_{u^y}^2)\Big) \\
    {\E}(\Delta^{2}_\rho y_{it} \times \Delta^{2}_\rho y_{it+2} \times \Delta^2 c_{it}) 
        &= -\psi^{(1)}\rho^2\gamma_{v} - \phi^{(2)}\rho^2(1+\rho^2)\sigma_{\zeta}^2 (\sigma_{v}^2 + \sigma_{u^y}^2) \\
        &- \psi^{(2)}\rho^2(\kappa_{v} + \sigma_{v}^2\sigma_{u^y}^2) - \omega^{(22)}\rho^2(1+\rho^2)\sigma_{\zeta}^2 \sigma_{v}^2 \\
    {\E}(\Delta^{2}_\rho y_{it} \times \Delta^{2}_\rho y_{it+2} \times \Delta^2 c_{it-2}) 
        &= - \phi^{(2)}\rho^2(1+\rho^2)\sigma_{\zeta}^2(\sigma_{v}^2 + \sigma_{u^y}^2) - \psi^{(2)}\rho^2\sigma_{v}^2(\sigma_{v}^2+\sigma_{u^y}^2) \\ 
    {\E}(\Delta^{2}_\rho y_{it} \times \Delta^{2}_\rho y_{it+2} \times \Delta^2 c_{it+2}) 
        &= - \phi^{(2)}\rho^2(1+\rho^2)\sigma_{\zeta}^2(\sigma_{v}^2 + \sigma_{u^y}^2) - \psi^{(2)}\rho^2\sigma_{v}^2(\sigma_{v}^2+\sigma_{u^y}^2)        
\end{align*}

\FloatBarrier

\section{Empirical details and additional results}\label{Appendix::Empirics}

\subsection{Summary statistics in the PSID sample}\label{SubAppendix::Summary_Statistics}

Table \ref{AppTable::Descriptives} presents summary statistics in the PSID sample over the period 1999–2019 (columns 1--2), as well as in subsamples sorted by wealth and education (columns 3--6).

\begin{table}[t!]  
\begin{center}
\caption{Descriptive statistics, PSID}\label{AppTable::Descriptives}
\begin{tabular}{L{4.3cm} C{1.5cm} C{1.5cm} C{1.6cm} C{1.6cm} C{1.6cm} C{1.7cm}}
\toprule
                                    &           &                   & \mcl{2}{c}{Low wealth}                        & \mcl{2}{c}{High wealth}                       \\
                                    & \mcl{2}{c}{Baseline sample}   & No college                & Some college      & No college                & Some college      \\
\cmidrule{2-7}
                                    & mean      & median    & mean      & mean      & mean      & mean              \\
                                    & (1)       & (2)       & (3)       & (4)       & (5)       & (6)               \\
\midrule
\emph{Consumption}                  &  &  &  &  &  &  \\
Nondurable+services                 &    26,766 &    22,895 &    21,280 &    24,566 &    23,675 &    31,479 \\
~~Food at home                      &     8,256 &     7,609 &     7,438 &     8,210 &     7,505 &     8,870 \\
~~Utilities                         &     3,515 &     3,238 &     3,351 &     3,205 &     3,471 &     3,787 \\
~~Food out                          &     3,318 &     2,511 &     2,387 &     2,928 &     3,109 &     4,037 \\
~~Health                            &     1,614 &       863 &     1,321 &     1,445 &     1,623 &     1,850 \\
~~Transportation                    &     5,866 &     4,900 &     5,283 &     5,489 &     5,711 &     6,406 \\
~~Education                         &     3,202 &         0 &       957 &     1,986 &     1,810 &     5,352 \\
~~Childcare                         &       977 &         0 &       546 &     1,259 &       537 &     1,136 \\
Rent (or rent eq.)                  &    18,105 &    13,821 &     8,910 &    12,631 &    16,449 &    26,020 \\
~~Home insurance                    &       894 &       747 &       585 &       712 &       908 &     1,142 \\
\noalign{\smallskip}\noalign{\smallskip}
\emph{Income}                       &  &  &  &  &  &  \\
Disposable income                   &   142,910 &   113,581 &    84,414 &   110,176 &   120,931 &   195,158 \\
Taxable+transfer inc.               &   137,277 &   109,003 &    77,707 &   106,276 &   113,789 &   189,446 \\
Earnings                            &   118,423 &    97,154 &    71,504 &    99,798 &    90,950 &   158,616 \\
~~Male                              &    81,857 &    62,185 &    46,025 &    65,725 &    57,519 &   114,545 \\
~~Female                            &    36,567 &    29,183 &    25,479 &    34,074 &    33,430 &    44,070 \\
\noalign{\smallskip}\noalign{\smallskip}
\emph{Wealth}                       &   547,761 &   191,193 &    68,561 &    70,665 &   670,948 & 1,024,577 \\
\noalign{\smallskip}\noalign{\smallskip}
\emph{Demographics}                 &  &  &  &  &  &  \\
\% working (male)                   &      0.89 &      1.00 &      0.89 &      0.94 &      0.82 &      0.89 \\
\% working (female)                 &      0.79 &      1.00 &      0.79 &      0.82 &      0.78 &      0.78 \\
Age (male)                          &        46 &        45 &        44 &        42 &        50 &        48 \\
\% some college (male)              &      0.68 &      1.00 &      0.00 &      1.00 &      0.00 &      1.00 \\
\% some college (female)            &      0.71 &      1.00 &      0.42 &      0.79 &      0.48 &      0.87 \\
\noalign{\smallskip}\noalign{\smallskip}
Periods per household               &      5.63 &      5.00 &      4.90 &      4.88 &      5.86 &      6.62 \\
Observations                        & \mcl{2}{c}{20,866}    &     4,258 &     5,159 &     2,428 &     9,021 \\
\bottomrule
\end{tabular}
\caption*{\fsz\emph{Notes:} The table reports summary statistics for consumption, income, wealth, and demographics in our baseline sample in the PSID and in subsamples sorted by wealth and education, over $1999$--$2019$. Monetary amounts are expressed in \$2018. Rent for owners is calculated at 6\% of the self-reported value of the house.}
\end{center}
\end{table}

\subsection{Comparison with administrative moments}\label{SubAppendix::Empirics_Comparison_Guvenen}

Table \ref{AppTable::Empirical_Moments_Earnings} reports the empirical second, third, and fourth moments of the cross-sectional distribution of unexplained growth in male earnings (panel A) and household earnings (panel B), as opposed to the moments of household disposable income in table \ref{Table::Empirical_Moments_Income_Consumption}. Column 1 presents moments in the PSID sample, while column 2 presents the corresponding statistics, wherever available, from the U.S. Social Security Administration data of \citet{Guvenen_Karahan_Ozkan_Song2021}.

\begin{table}[t]  
\begin{center}
\caption{Empirical moments, male earnings and household earnings}\label{AppTable::Empirical_Moments_Earnings}
\begin{tabular}{L{5.4cm} rl c C{4.4cm}}
\toprule
													& \mcl{2}{c}{Baseline sample} 			&& \citet{Guvenen_Karahan_Ozkan_Song2021} \\
\cmidrule{2-3}\cmidrule{5-5}									
\emph{Income data:}									& \mcl{2}{c}{PSID} 				 		&& Social Security Admin. 	\\
													& \mcl{2}{c}{(1)} 						&& (2) 					\\
\midrule
\mcl{5}{c}{\textbf{Panel A. Moments of male earnings growth}} 				\\
${\Var}(\Delta y_{it})$ 							&0.271 		&(0.010) 					&&0.323 				\\
${\Cov}(\Delta y_{it},\Delta y_{it+1})$				&-0.065 	&(0.004) 					&&--					\\
${\Skew}(\Delta y_{it})$							&-0.569 	&(0.128) 					&&-1.039				\\
${\Cov}((\Delta y_{it})^2,\Delta y_{it+1})$			&0.087 		&(0.010) 					&&--					\\
${\Kurt}(\Delta y_{it})$							&13.711 	&(0.551) 					&&13.494				\\
${\Cov}((\Delta y_{it})^2,(\Delta y_{it+1})^2)$		&0.253 		&(0.022) 					&&--					\\
\noalign{\smallskip}
\mcl{5}{c}{\textbf{Panel B. Moments of household earnings growth}} 				\\
${\Var}(\Delta y_{it})$								&0.233		&(0.009)					&& --	\\
${\Cov}(\Delta y_{it},\Delta y_{it+1})$				&-0.056		&(0.004)					&& -- 	\\
${\Skew}(\Delta y_{it})$							&-0.728		&(0.142)					&& --	\\
${\Cov}((\Delta y_{it})^2,\Delta y_{it+1})$			&0.045		&(0.008)					&& --	\\
${\Kurt}(\Delta y_{it})$							&14.728		&(0.822)					&& --	\\
${\Cov}((\Delta y_{it})^2,(\Delta y_{it+1})^2)$		&0.176		&(0.017)					&& --	\\
\bottomrule
\end{tabular}
\caption*{\fsz\emph{Notes:} The table presents the second, third, and fourth moments of earnings growth. Skewness and kurtosis correspond to the third and fourth standardized moments respectively. Column 1 reports moments of biennial earnings growth in the PSID (we maintain the notation $\Delta x_{t}$ for the first difference of variable $x$ over time, noting that, given the biennial nature of the PSID, this corresponds to a difference over two calendar years), while column 2 reports moments of male earnings growth from the U.S. Social Security Administration data. The latter concern the one-year male earnings growth, unconditional on past earnings, and correspond to averages over ages 30-55 of, respectively, dispersion squared, skewness, and kurtosis, as reported in appendix tables C23-C25 in \cite{Guvenen_Karahan_Ozkan_Song2021}. Block bootstrap standard errors are in parentheses.}
\end{center}
\end{table}

\subsection{Sources of tail income shocks}\label{SubAppendix::Sources_Income_Shocks}

Table \ref{AppTable::Sources_Income_Shocks} presents evidence on the underlying sources of large income reductions that drive the left skewness and excess kurtosis of income growth. Columns 1 \& 2 report average unemployment, disability, bad health, etc, in the bottom 10\% and top 90\% of the distribution of residual income changes $\Delta y_{it}$, respectively. Column 3 reports the $p$-value of the difference between the two; in most cases, the bottom 10\% (effectively households facing large income reductions) experience unemployment, worsening health, large drops in hours, and precarious housing at much higher rates than the rest.

\begin{table}[t]  
\begin{center}
\caption{Sources of tail income shocks}\label{AppTable::Sources_Income_Shocks}
\begin{tabular}{L{6.5cm} C{2.3cm} C{2.3cm} C{3.0cm}}
\toprule
                                                    & \mcl{2}{c}{Disposable income growth}      &          \\                                    
                                                    & Bottom 10\%       & Top 90\%              &               \\
                                                    \cmidrule{2-3}
                                                    &mean               & mean                  & $p$-value $\Delta$mean    \\
                                                    & (1)               & (2)                   & (3)    \\
\midrule
\% unemployed $ t$ (male)                           &     0.082 &     0.034 &     0.000 \\
\% unemployed $ t-1$ (male)                         &     0.084 &     0.036 &     0.000 \\
\% unemployed $ t$ (female)                         &     0.081 &     0.037 &     0.000 \\
\% unemployed $ t-1$ (female)                       &     0.081 &     0.041 &     0.000 \\
\% main job laid off (male)                         &     0.054 &     0.039 &     0.003 \\
\% second job laid off (male)                       &     0.153 &     0.117 &     0.000 \\
\% main job laid off (female)                       &     0.073 &     0.059 &     0.023 \\
\% second job laid off (female)                     &     0.096 &     0.095 &     0.921 \\
\% physical disability (male)                       &     0.130 &     0.103 &     0.001 \\
\% physical disability (female)                     &     0.121 &     0.120 &     0.840 \\
\% physical disability (both)                       &     0.032 &     0.025 &     0.077 \\
\% overall health got worse (male)                  &     0.129 &     0.110 &     0.036 \\
\% overall health got worse (female)                &     0.121 &     0.104 &     0.052 \\
\% eviction from home                               &     0.049 &     0.031 &     0.000 \\
Hours growth (male)                                 &    -0.309 &     0.031 &     0.000 \\
Hours growth (female)                               &    -0.228 &     0.023 &     0.000 \\
\bottomrule
\end{tabular}
\caption*{\fsz\emph{Notes:} The table reports the sample mean (mostly empirical probabilities) of the various events listed in the rows of the table in the bottom 10\% and top 90\% of the distribution of residual income growth $\Delta y_{it}$. Given the biennial nature of the PSID, $\Delta y_{it}$ corresponds to a difference over two calendar years.}
\end{center}
\end{table}

\subsection{CEX data and imputation in the PSID}\label{SubAppendix::Empirics_CEX}

The Consumer Expenditure Survey (CEX) accounts for about $95\%$ of all household expenditures from a highly disaggregated list of consumption goods and services \citep[see][for a recent overview]{MeyerSullivan2023onsumptionIncomeInequality}. The CEX is run by the Census Bureau and the Bureau of Labor Statistics and forms the basis for the calculation of the consumer's price index.

\

\noindent \textbf{Sample selection and variables.} To parallel the design of the PSID sample, we use CEX interview data between 1999 and 2019. We select a sample of households that mimics closely our baseline selection in the PSID.\footnote{To improve the quality of the food demand estimation subsequently, we keep \emph{all} calendar years between 1999 and 2019, i.e., not just the odd ones as in the PSID.} Specifically, we select continuously married couples with the male spouse aged 30 to 65. We require non-missing data on expenditure and basic demographics, and we drop those with zero food expenditure given the nature of the imputation subsequently. As in the PSID, we define consumption as the sum of real expenditure on nondurable goods and services, namely food (at home and outside), utilities, out-of-pocket health expenses, public transport, vehicle expenses, education, and daycare. As the CEX collects quarterly data over four rolling quarters, we generate annual consumption as the sum of consumption over four quarters. We assign the final observation to the calendar year to which the underlying quarterly data mostly correspond to. Our final sample consists of 31,751 observations.

Table \ref{AppTable::PSID_CEX_sample} presents a comparison of means between the PSID and CEX samples for a host of household characteristics over the period 1999--2019. The two samples are very close with respect to age, family size, number of children, race, education, region of residence, and labor market participation. Household earnings are higher in the PSID than in the CEX; BPP noted this also for the earlier period (1980--1992) and argued that it is due to the more comprehensive definition of income in the PSID. Food expenditure in the PSID is higher, on average, by about 60\% than food expenditure in the CEX. This is in contrast to BPP who find that food expenditure is similar across the two surveys over 1980--1992. This results in consumption being about 25\% higher in the PSID than in the CEX.

\begin{table}[t]  
\begin{center}
\caption{Comparison of sample means, PSID and CEX}\label{AppTable::PSID_CEX_sample}
\begin{tabular}{l cc c cc c cc}
\toprule
										& \mcl{2}{c}{1999} 				&& \mcl{2}{c}{2009} 				&& \mcl{2}{c}{2019}				\\
\cmidrule{2-3}\cmidrule{5-6}\cmidrule{8-9}									
										& PSID 			& CEX 			&& PSID 			& CEX 			&& PSID 			& CEX		\\
\midrule
Earnings 								& 72,829 		& 59,011 		&& 109,047 			& 84,383 		&& 130,488 			& 106,316 	\\
Consumption 							& 16,115 		& 10,631 		&& 22,317 			& 15,454 		&& 27,315 			& 18,592 	\\
Food 									& 7,434 		& 4,461 		&& 9,306 			& 6,090 		&& 12,575 			& 7,636 	\\
Age 									& 44.70 		& 45.76 		&& 46.40 			& 47.59 		&& 46.69 			& 48.40 	\\
Family size 							& 3.39 			& 3.54 			&& 3.24 			& 3.48 			&& 3.42 			& 3.44 		\\
\# children 							& 1.13 			& 1.21 			&& 1.01 			& 1.11 			&& 1.19 			& 1.07 		\\
\% white 								& 0.93 			& 0.88 			&& 0.91 			& 0.86 			&& 0.91 			& 0.84 		\\
\% high school dropout 					& 0.09 			& 0.12 			&& 0.06 			& 0.12 			&& 0.07 			& 0.10 		\\
\% high school graduate 				& 0.28 			& 0.27 			&& 0.25 			& 0.25 			&& 0.22 			& 0.21 		\\
\% college dropout 						& 0.62 			& 0.62 			&& 0.69 			& 0.63 			&& 0.71 			& 0.69 		\\
\% Northeast 							& 0.17 			& 0.18 			&& 0.18 			& 0.19 			&& 0.16 			& 0.17 		\\
\% Midwest 								& 0.32 			& 0.24 			&& 0.30 			& 0.25 			&& 0.31 			& 0.21 		\\
\% South 								& 0.33 			& 0.32 			&& 0.32 			& 0.34 			&& 0.33 			& 0.33 		\\
\% West 								& 0.18 			& 0.25 			&& 0.20 			& 0.22 			&& 0.20 			& 0.28 		\\
\% working (male) 						& 0.89 			& 0.88 			&& 0.91 			& 0.93 			&& 0.90 			& 0.92 		\\
\% working (female) 					& 0.81 			& 0.77 			&& 0.79 			& 0.81 			&& 0.79 			& 0.79 		\\
\bottomrule
\end{tabular}
\caption*{\fsz\emph{Notes:} The table presents a comparison of means in the PSID and CEX samples in 1999, 2009, and 2019.}
\end{center}
\end{table}

\begin{figure}[]
    \centering
    \caption{Moments of log consumption in the PSID and CEX}\label{AppFigure::Comparison_PSID_CEX}
    \begin{subfigure}[t]{0.57\textwidth}
        \centering
        \includegraphics[width=\textwidth]{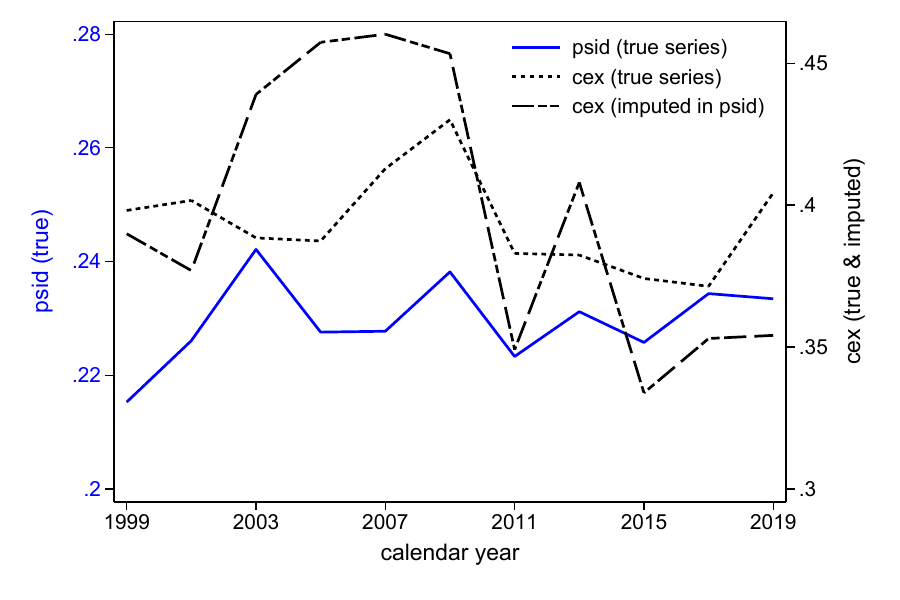}
        \caption{Variance of log consumption}\label{AppFigure::Comparison_PSID_CEX_a}
    \end{subfigure}\\[4pt]
    \begin{subfigure}[t]{0.57\textwidth}
        \centering
        \includegraphics[width=\textwidth]{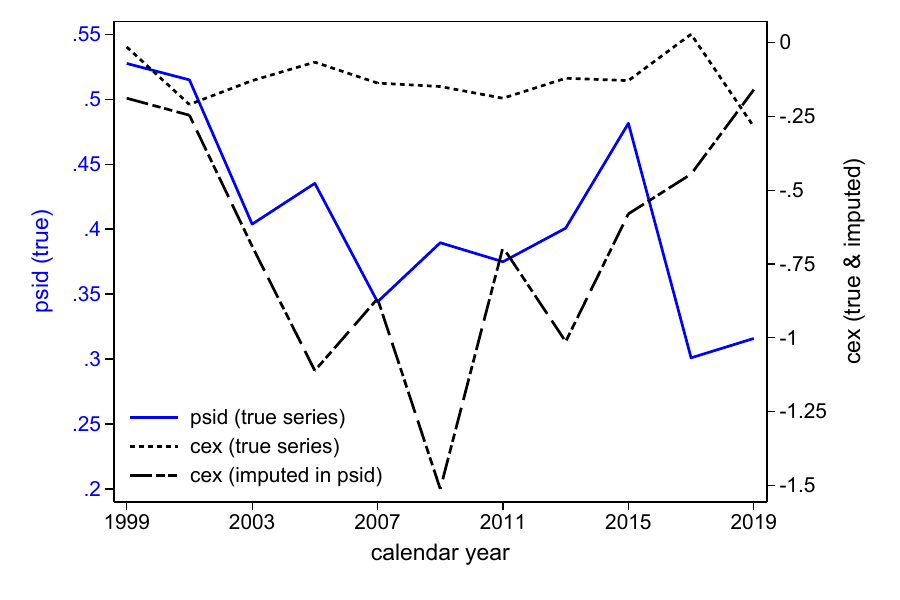}
        \caption{Skewness of log consumption}\label{AppFigure::Comparison_PSID_CEX_b}
    \end{subfigure}\\[4pt]
    \begin{subfigure}[t]{0.57\textwidth}
        \centering
        \includegraphics[width=\textwidth]{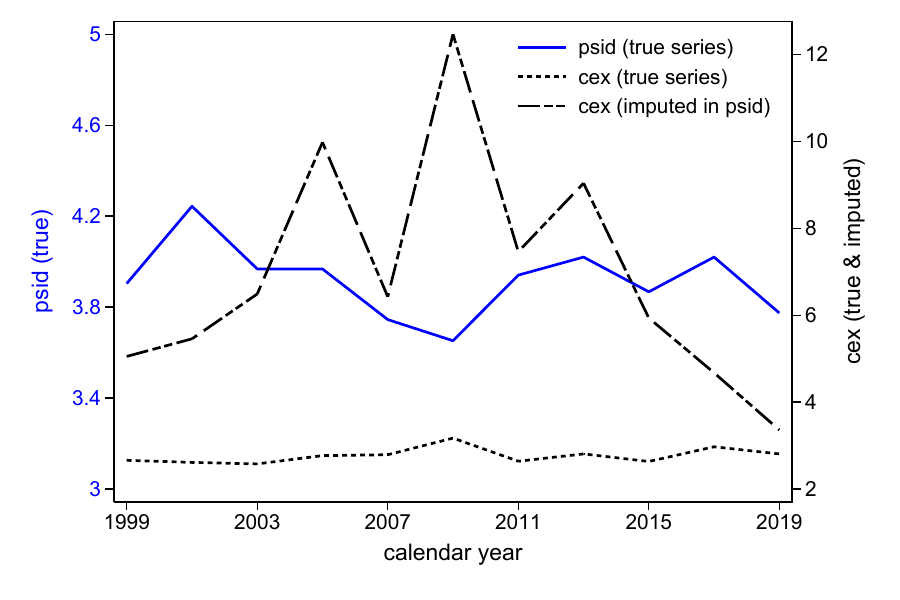}
        \caption{Kurtosis of log consumption}\label{AppFigure::Comparison_PSID_CEX_c}
    \end{subfigure}%
    \caption*{\fsz\emph{Notes:} The figure plots the moments of log consumption in the PSID (blue solid line; left), the CEX (black dotted line; right), and imputed from the CEX into the PSID (black dashed-dotted line; right axis).}
\end{figure}

Figure \ref{AppFigure::Comparison_PSID_CEX} plots the variance, skewness, and kurtosis of log consumption in the PSID (blue solid line) and CEX samples (black dotted line). The patterns for the variance and kurtosis are quite similar in the two surveys over time, although consumption in the PSID is less volatile but slightly more leptokurtic. For skewness, there is disagreement both in the trends over time and the sign. The PSID features positive while the CEX weakly negative skewness. Overall, these statistics (in particular skewness) suggest that consumption across otherwise similarly selected samples in the PSID and CEX exhibits some notable differences.

\

\noindent \textbf{Consumption imputation from CEX into PSID.} The previous comparison is not directly informative about the moments of consumption \emph{growth}, which is the relevant variable in our model. The CEX lacks the panel dimension needed for our analysis, so it cannot be used to obtain measures of annual (or biennial) consumption growth.\footnote{The CEX consists of repeated cross sections, in which households are interviewed over 4 quarters. There is also an initial training quarter, which is rarely used in empirical analyses.} We thus follow BPP and impute nondurable consumption from the CEX into the PSID. 

The imputation involves two distinct steps. In the first step, we estimate food demand in the CEX. Specifically, we let real food expenditure be a log-linear function of the log of nondurable expenditure $c_{it}$, relative prices $\mathbf{p}_t$, and demographics $\mathbf{W}_{it}$, namely
\begin{equation}\label{AppEq::Food_Demand}
    f_{it} = \beta (D_{it}) c_{it} + \mathbf{p}_{t}^{\prime}\boldsymbol{\gamma} + \mathbf{W}_{it}^{\prime}\boldsymbol{\mu} + e_{it},
\end{equation}
where $f_{it}$ is the log of real annual food expenditure for household $i$ in the CEX in year $t$, $\beta$ is the budget elasticity of food demand, which we allow to shift with time and household characteristics $D_{it}$, and $e_{it}$ captures unobserved heterogeneity in food demand.\footnote{Prices $\mathbf{p}_t$ include the relative prices of food, public transportation, utilities, medical services, and childcare. Demographics $\mathbf{W}_{it}$ include a quadratic polynomial in age and dummy variables for the number of children, family size, education, region of residence, race, and cohort. $D_{it}$ includes year dummies and dummies for education and the number of children.} To address possible endogeneity of nondurable expenditure, we follow an instrumental variables approach using as instruments the mean and standard deviation of household disposable income and the standard deviation of spousal wages by cohort, education, year, the former interacted with time, education, and number of children dummies.\footnote{These instruments are different from those used in BPP, namely the cohort-education-year specific averages of male and female hourly wages, interacted with time, education, and kids dummies. The use of their instruments over our sample period often resulted in convergence problems.}

In the second step, we invert \eqref{AppEq::Food_Demand} under the assumption that food demand is monotonic in nondurable expenditure. We can then use the inverted equation to predict $c_{it}$ for a given level of food expenditure, given prices and demographics. Since the CEX and PSID samples are representative of the same underlying population, and since food expenditure, prices, and demographics are available in both surveys, we can use the inverted \eqref{AppEq::Food_Demand} to impute consumption into the PSID. This allows us to obtain an alternative, external measure of consumption in the PSID, and thus of consumption growth.

Figure \ref{AppFigure::Comparison_PSID_CEX} plots the variance, skewness, and kurtosis of log consumption imputed from the CEX (black dashed-dotted line). The magnitudes of the various moments are quite different between the true PSID and the imputed series, as are also the trends. Imputed consumption is more volatile than the true PSID, skewness is negative (as opposed to positive in the PSID) and peaks down in 2007--2011, kurtosis is often more than twice as high as kurtosis in the PSID and peaks upwards in 2007--2011. 

Table \ref{Table::Empirical_Moments_Income_Consumption} in the text reports the second and higher-moments of consumption growth in the imputed series, residualized through similar first-stage regressions as in the true data. The main qualitative features of the distributions remain similar across the true PSID data and the imputed series, although the moments of imputed consumption are several-fold accentuated relative to the true data. The imputation likely imparts substantial measurement/imputation error to consumption, which casts doubt on the reliability of the partial insurance estimates derived from imputed consumption data from the CEX. 

\

\noindent \textbf{An explanation for why the imputation inflates $\boldsymbol{\phi_t^{(1)}}$.} Our empirical findings suggest that estimation of the linear specification using imputed consumption data inflates $\phi_t^{(1)}$ relative to estimation using the true PSID series. One reason behind this is neglected correlation between measurement error in income and expenditure.

Suppose that consumption and income measurement errors are positively correlated, i.e., ${\Cov}(u_{it}^c, u_{it}^y)=\sigma_{u_t}>0$. The positive correlation may be the result of, e.g., recall bias that affects similarly all monetary amounts. This is something we have not allowed for in the paper, and neither have BPP. While $\phi_t^{(1)}$ is identified via ${\E}(\Delta c_{it}(\Delta y_{it-1}+\Delta y_{it}+\Delta y_{it+1}))/\sigma_{\zeta_t}^2$ in the baseline permanent-transitory income process, in practice BPP and virtually all follow-up papers estimate it through $({\E}(\Delta c_{it}\Delta y_{it}) + {\E}(\Delta c_{it}\Delta y_{it+1}))/\sigma_{\zeta_t}^2$, assuming that $\sigma_{u_t}=0$ (uncorrelated errors). This is what we also do in our baseline estimation of $\phi_t^{(1)}$ in the linear specification. However, if the errors \emph{are} correlated, then this target moment in fact identifies $\phi_t^{(1)\text{psid*}} = \phi_t^{(1)} + \sigma_{u_t}/\sigma_{\zeta_t}^2$. Clearly, correlation between consumption and income errors causes positive bias as $\phi_t^{(1)\text{psid*}}>\phi_t^{(1)}$. This is true in the baseline, i.e., without the use of imputed consumption, but the bias becomes multiplicatively larger with imputed consumption, as we show below.

In the simplest form without covariates, imputed log consumption in the PSID is given by $c_{it}^\text{imp} = \kappa_t (f_{it} - \pi_t)$, where $f_{it}$ is log food expenditure in the PSID, $\pi_t=\mathbf{p}_{t}^{\prime}\boldsymbol{\gamma}$ is an intercept that depends on prices, and $\kappa_t = \beta (D_{it})^{-1}$ is the inverse budget elasticity of food consumption. Suppose that $\pi_t=0$ for simplicity, so $c_{it}^\text{imp} = \kappa_t f_{it}$. As $\beta (D_{it})$ is typically less than 1 \citep[e.g.,][BPP]{BlundellPistaferriPreston_2004_Imputation}, it follows that $\kappa_t>1$. \citet{BlundellPistaferriPreston_2004_Imputation} discuss extensively the demand estimation in the CEX, and provide an in-depth characterization of the bias induced by endogenous expenditure or non-classical measurement error in the CEX. However, they do not discuss measurement error in food expenditure \emph{in the PSID}, and in particular the case when such error may correlate with error in income.

Suppose that food expenditure in the PSID is measured with error, so imputed consumption is $c_{it}^\text{imp*} = \kappa_t (f_{it} + u_{it}^f)$. Suppose that food consumption and income error are positively correlated, i.e., ${\Cov}(u_{it}^f, u_{it}^y)=\sigma_{u_t}>0$. Because $\kappa_t>1$, the consumption imputation magnifies the role of the consumption measurement error. One can show that, with imputed consumption, the moment used to identify $\phi_t^{(1)}$ in fact identifies $\phi_t^{(1)\text{imp*}} = \phi_t^{(1)} + \kappa_t\sigma_{u_t}/\sigma_{\zeta_t}^2$. As a result, $\phi_t^{(1)\text{imp*}} > \phi_t^{(1)\text{psid*}}$ and the imputation inflates the transmission parameter of permanent shocks vis-\`a-vis the true consumption series.

\FloatBarrier
\subsection{Estimation using the original BPP data}\label{SubAppendix::Empirics_ComparisonBPP}

Table \ref{AppTable::Replication_BPP_19801992} uses the original BPP data over 1980--1992 \citep[taken from the replication package of][available online]{Blundell_Pistaferri_Preston2008} and does three things. 

First, we replicate BPP over the original data/period of time, using our code but BPP's \emph{annual} growth rates of income and consumption. This is column 1 in table \ref{AppTable::Replication_BPP_19801992}. We obtain numerically similar results to BPP, namely $\phi^{(1)}=0.64$ while $\psi^{(1)}$ is statistically zero. 

Second, we calculate \emph{biennial} growth rates of income and consumption and re-estimate the linear model. This is column 2 in table \ref{AppTable::Replication_BPP_19801992}; this estimation is analogous to the linear function estimation in column 1 of table \ref{Table::Consumption_function}, albeit using the original BPP sample. Note that despite the use of biennial data, $\phi^{(1)}$ and $\psi^{(1)}$ still express the pass-through at annual rates (see appendix \ref{Appendix::Identification}). $\phi^{(1)}$ drops by 20\% to $0.518$, while $\psi^{(1)}$ remains small but turns statistically significant. With biennial data, we only identify the pass-though of an `aggregate' permanent shock over two years. Some higher frequency shocks are muted at the observed lower frequency, given that we can only observe an average of two one-year permanent shocks over the two-year period. In other words, the biennial data allow us to measure only an `aggregate' or average pass-through within a given period, and some higher-frequency shocks are inevitably smoothed out at this observed lower frequency.

Third, we re-estimate the linear model assuming that the target second-order income and consumption moments are time-invariant. This differs from the original BPP exercise where the moments vary with time, but it is otherwise similar to our main exercise in the paper given that we cannot precisely estimate higher-order moments on a year-by-year basis. The results are in columns 3-4 in table \ref{AppTable::Replication_BPP_19801992}. Time-invariance further reduces the transmission parameter of permanent shocks ($\phi^{(1)}$ drops to $0.404$), bringing the pass-through closer to the value we estimate in the modern data. Taking averages of moments over time attenuates the largest income-consumption co-movements; consequently, the model can fit the data using, ceteris paribus, smaller transmission parameters. This is similar to \citet{ChatterjeeMorleySingh2021_BPP} who also find smaller pass-through rates (higher degree of partial insurance) when they impose time-invariance in BPP's target moments.

There are some caveats comparing the exercise in table \ref{AppTable::Replication_BPP_19801992} and our results from the recent PSID in table \ref{Table::Consumption_function}. First, our main exercise uses consumption data internally available in the PSID, while table \ref{AppTable::Replication_BPP_19801992} uses consumption imputed from the CEX. We showed earlier that the imputation imparts substantial imputation error. Second, while the estimation in table \ref{AppTable::Replication_BPP_19801992} does not correct the income moments for measurement error (in line with BPP), our main exercise in table \ref{Table::Consumption_function} does so \citep[in line with][]{Blundell_Pistaferri_Saporta-Eksten2016}. Finally, the use of biennial growth rates does not allow the identification of $\theta$, the moving average parameter of the transitory shock, so the transitory shock is $MA(0)$ in our baseline ($\theta=0$), while it is $MA(1)$ in the original BPP exercise. 

\begin{table}[]  
\begin{center}
\caption{Replication of BPP, original BPP data 1980-1992}\label{AppTable::Replication_BPP_19801992}
\begin{tabular}{l C{2.3cm} C{2.3cm} c C{2.3cm} C{2.4cm}}
\toprule
\emph{Consumption fn.:}                 & \mcl{5}{c}{\textbf{Linear}}  	\\
\cmidrule{2-6}
\emph{Income moments:}                  & 2\textsuperscript{nd}         & 2\textsuperscript{nd}      
                                        												&& 2\textsuperscript{nd} 		& 2\textsuperscript{nd}      	\\                                        
\emph{Consumption data:}                & CEX           & CEX           				&& CEX           & CEX          				\\[5pt]
\emph{Time variability:}              	& \mcl{2}{c}{\textbf{Time-varying moments}}    		&& \mcl{2}{c}{\textbf{Time-invariant moments}}     			\\
\cmidrule{2-3}\cmidrule{5-6}
\emph{Growth rates:}                	& annual        & biennial      				&& annual       & biennial     					\\
										& \ssz (BPP$^*$) \nsz & 					&&				& 								\\
                                        & (1)           & (2)           				&& (3)          & (4)          					\\
\midrule                                        
$\phi^{(1)}$							&0.644			&0.518							&&0.504			&0.404							\\
										&(0.079)		&(0.068)						&&(0.076)		&(0.057)						\\
$\psi^{(1)}$							&0.024			&0.117							&&0.043			&0.158							\\
										&(0.044)		&(0.050)						&&(0.044)		&(0.051)						\\               
\midrule     
$\theta^\#$								&0.112			&0								&&0.115			&0								\\
										&(0.025)		&\ssz (not identified) \nsz		&&(0.026)		&\ssz (not identified) \nsz		\\
                   
\bottomrule
\end{tabular}
\caption*{\fsz\emph{Notes:} The table presents the estimates of the parameters of the linear consumption function, using the original BPP data over 1980--1992 (PSID income data, consumption data imputed from the CEX). The data are available through \citet{Blundell_Pistaferri_Preston2008}'s online replication package. Columns 1-2 allow the target second-order income and consumption moments to vary with calendar time, while columns 3-4 target said moments imposing time-invariance. Columns 1 and 3 estimate the model using annual growth rates of income and consumption, while columns 2 and 4 use biennial rates. In line with BPP, estimation is done via diagonally weighted GMM; GMM standard errors are in parentheses.\\
$^*$The small unimportant numerical differences between column 1 herein and the original BPP estimates are likely due to different software and optimization algorithms (our code uses Matlab and \texttt{fmincon} while BPP use Gauss). \\
$^\#$$\theta$ is the moving average parameter of the transitory shock. BPP assume that the transitory component of income is given by $v_{it}=\epsilon_{it}+\theta\epsilon_{it-1}$, where $\epsilon_{it}$ is the transitory shock. Given their use of annual growth rates, $\theta$ is identified by the first-order income autocovariance. Our use of biennial data in this paper rules out identification of $\theta$, which we thus set to 0.}
\end{center}
\end{table}

\FloatBarrier
\subsection{Additional results}\label{SubAppendix::Empirics_Additonal_Results}

Tables \ref{AppTable::Income_process_OptimalMD} and \ref{AppTable::Consumption_function_OptimalMD} present the estimates of the parameters of the income process and the consumption function using the empirical covariance matrix of the target moments as weighting matrix (optimal GMM). As such, these tables accompany tables \ref{Table::Income_process} and \ref{Table::Consumption_function} in the paper, which report results from equally weighted GMM. 

Table \ref{AppTable::Consumption_function_CEX} presents the estimates of the parameters of the consumption function using imputed consumption from the CEX. As such, this table accompanies table \ref{Table::Consumption_function} in the paper, which reports results using consumption internally available in the PSID. In the linear specification, $\phi^{(1)}$ is consistently higher than that obtained from the consumption data internally available in the PSID, regardless of whether we target only second- or also higher-order moments. This echoes our earlier discussion about the imputation inflating the pass-through of shocks. We also find that $\psi^{(1)}$ is negative in all cases. Moreover, when up to fourth-order moments are targeted (column 3), the point estimate is particularly large in magnitude ($-0.418$). This is counterintuitive, raising further concerns about the reliability of the imputation procedure, particularly when higher-order moments are involved. We draw similar conclusions about $\phi^{(1)}$ and $\psi^{(1)}$ in the context of the quadratic function. $\phi^{(2)}<0$ in that case, with a magnitude similar to our baseline results from the PSID ($-0.036$ versus $-0.04$; albeit not significant). Once again, this is indicative of the asymmetric pass-through of negative versus positive permanent shocks. Similar to our baseline, we find $\psi^{(2)}>0$.

Table \ref{AppTable::Income_process_age_lagY} presents the estimates of the income process by age and the household's position in the income distribution. Skewness of all shocks is more negative at older ages. Skewness of permanent shocks is weakly positive in the bottom income tertile but turns very negative in the top tertile. The opposite is true for transitory shocks; skewness is negative in the bottom income tertile but strongly positive in the top.

Tables \ref{AppTable::Linear_Consumption_function_age} and \ref{AppTable::Linear_Consumption_function_lagY} present results from the linear consumption function by age and the household's position in the income distribution, respectively. $\phi^{(1)}$ exhibits a non-monotonic pattern over the three age brackets -- decreasing from $0.154$ over ages 30--40 to $0.126$ over 41--50 before increasing to $0.180$ over ages 51--65. A similar non-monotonic pattern appears in column 2, where third moments are also targeted, and is further accentuated in column 3 where all moments up to fourth-order are targeted. This is indicative of a misspecification in the linear function, in the sense that its theoretical restrictions (higher wealth insures income fluctuations better) are not supported by the empirical results.\footnote{This conclusion is consistent with BPP, who find a decreasing pattern with age only when they impose a linear age trend on the transmission parameters. By contrast, when they generalize this to a quadratic trend, they report a worsening of their results without including further details.} By contrast, $\psi^{(1)}$ is statistically indistinguishable from zero in all three age brackets. $\phi^{(1)}$ is uniformly lower in the top income tertile vis-\`a-vis the bottom tertile, suggesting that the income-rich, who are also wealthier, have more insurance against permanent income shocks than the poor.

\begin{table}[t!]  
\begin{center}
\caption{Estimates of the income process, optimal weighting}\label{AppTable::Income_process_OptimalMD}
\begin{tabular}{ll L{2.5cm} R{1.3cm} L{1.3cm} R{1.3cm} L{1.3cm} R{1.5cm} L{1.5cm}}
\toprule
\mcl{3}{l}{\emph{Income moments:}}                      & \mcl{2}{c}{2\textsuperscript{nd}} & \mcl{2}{c}{2\textsuperscript{nd}, 3\textsuperscript{rd}}  & \mcl{2}{c}{2\textsuperscript{nd}, 3\textsuperscript{rd}, 4\textsuperscript{th}}   \\
\mcl{3}{l}{\emph{Income data:}}                         & \mcl{2}{c}{PSID}                  & \mcl{2}{c}{PSID}                                          & \mcl{2}{c}{PSID} \\
&&                                                      & \mcl{2}{c}{(1)}                   & \mcl{2}{c}{(2)}                                           & \mcl{2}{c}{(3)}  \\
\midrule
\mcl{9}{c}{\textbf{Panel A. Permanent shocks}}\\
Variance    & $\sigma_{\zeta}^2$        &               &0.027          &(0.002)            &0.027          &(0.002)                                    &0.025      &(0.002)    \\
Skewness    & $\gamma_{\zeta}$          & central       &               &                   &-0.004         &(0.003)                                    &-0.004     &(0.003)    \\
            &                           & standardized  &               &                   &-0.895         &(0.633)                                    &-0.997     &(0.760)    \\
Kurtosis    & $\kappa_{\zeta}$          & central       &               &                   &               &                                           &0.037      &(0.007)    \\
            &                           & standardized  &               &                   &               &                                           &61.636     &(14.018)   \\
\noalign{\smallskip}
\mcl{9}{c}{\textbf{Panel B. Transitory shocks}}\\
Variance    & $\sigma_{v}^2$            &               &0.033          &(0.002)            &0.033          &(0.002)                                    &0.034      &(0.002)    \\
Skewness    & $\gamma_{v}$              & central       &               &                   &-0.008         &(0.003)                                    &-0.008     &(0.003)    \\
            &                           & standardized  &               &                   &-1.351         &(0.517)                                    &-1.324     &(0.485)    \\
Kurtosis    & $\kappa_{v}$              & central       &               &                   &               &                                           &0.049      &(0.005)    \\
            &                           & standardized  &               &                   &               &                                           &41.579     &(7.052)    \\
\bottomrule
\end{tabular}
\caption*{\fsz\emph{Notes:} The table presents the estimates of the parameters of the income process, imposing stationarity over time. Column 1 targets only the second moments of income growth; column 2 targets also third moments; column 3 targets all moments up to fourth-order. Estimation is via GMM using the optimal weighting matrix; block bootstrap standard errors are in parentheses. This table accompanies table \ref{Table::Income_process} in the paper, which shows the baseline results from equally weighted GMM.}
\end{center}
\end{table}

\begin{table}[t!]  
\begin{center}
\caption{Estimates of the consumption function, optimal weighting}\label{AppTable::Consumption_function_OptimalMD}
\begin{tabular}{l C{1.8cm} C{1.8cm} C{1.9cm} C{2.1cm} c C{2.4cm}}
\toprule
\emph{Consumption fn.:}             & \mcl{4}{c}{\textbf{Linear}}                           && \textbf{Quadratic} \\
\cmidrule{2-5}\cmidrule{7-7}
\emph{Income moments:}                  & 2\textsuperscript{nd}     & 2\textsuperscript{nd}     & 2\textsuperscript{nd}, 3\textsuperscript{rd}      
                                        & 2\textsuperscript{nd}, 3\textsuperscript{rd}, 4\textsuperscript{th} 
                                        && 2\textsuperscript{nd}, 3\textsuperscript{rd}, 4\textsuperscript{th}      \\                                        
\emph{Consumption data:}                & PSID          & CEX           & PSID          & PSID          && PSID     \\
                                        & (1)           & (2)           & (3)           & (4)           && (5)      \\
\midrule                                        
$\phi^{(1)}$                            &0.152          &0.288          &0.149          &0.146          &&0.169     \\
                                        &(0.028)        &(0.044)        &(0.025)        &(0.029)        &&(0.030)   \\
$\psi^{(1)}$                            &-0.006         &-0.109         &-0.015         &-0.004         &&-0.018    \\
                                        &(0.045)        &(0.078)        &(0.037)        &(0.041)        &&(0.047)   \\
$\phi^{(2)}$                            &               &               &               &               &&-0.037    \\
                                        &               &               &               &               &&(0.022)   \\
$\psi^{(2)}$                            &               &               &               &               &&-0.007    \\
                                        &               &               &               &               &&(0.030)   \\
$\omega^{(22)}$                         &               &               &               &               &&0.740     \\
                                        &               &               &               &               &&(0.731)   \\            
\midrule
$\sigma^2_{\xi}$                        &0.019          &0.019          &0.019          &0.018          &&0.018     \\
                                        &(0.001)        &(0.004)        &(0.001)        &(0.001)        &&(0.001)   \\
$\gamma_{\xi}$                          &               &               &-0.001         &-0.001         &&\\
                                        &               &               &(0.001)        &(0.001)        &&\\
$\kappa_{\xi}$                          &               &               &               &0.005          &&\\
                                        &               &               &               &(0.001)        &&\\
\midrule
$\sigma^2_{u_c}$                        &0.044          &0.140          &0.043          &0.043          &&0.044     \\
                                        &(0.002)        &(0.009)        &(0.001)        &(0.001)        &&(0.001)   \\
$\gamma_{u_c}$                          &               &               &0.002          &0.002          &&\\
                                        &               &               &(0.001)        &(0.001)        &&\\
$\kappa_{u_c}$                          &               &               &               &0.012          &&\\
                                        &               &               &               &(0.001)        &&\\                    
\bottomrule
\end{tabular}
\caption*{\fsz\emph{Notes:} The table presents the estimates of the parameters of the consumption function, imposing homogeneous transmission parameters over the lifecycle/in the cross-section and stationarity of taste heterogeneity and consumption measurement error. Columns 1-4 present parameter estimates in the linear function, while column 5 presents estimates in the quadratic case; the order of moments targeted in each case is shown at the top of the table. Except for column 2 where we use imputed data from the CEX, all other columns use consumption data internally available in the PSID. Estimation is via GMM using the optimal weighting matrix; block bootstrap standard errors are in parentheses. This table accompanies table \ref{Table::Consumption_function} in the paper, which shows the baseline results from equally weighted GMM.}
\end{center}
\end{table}

\begin{table}[t!]  
\begin{center}
\caption{Estimates of the consumption function, imputed consumption}\label{AppTable::Consumption_function_CEX}
\begin{tabular}{l C{2.3cm} C{2.3cm} C{2.3cm} c C{2.4cm}}
\toprule
\emph{Consumption fn.:}                 & \mcl{3}{c}{\textbf{Linear}}                   && \textbf{Quadratic}       \\
\cmidrule{2-4}\cmidrule{6-6}
\emph{Income moments:}                  & 2\textsuperscript{nd}         & 2\textsuperscript{nd}, 3\textsuperscript{rd}      
                                        & 2\textsuperscript{nd}, 3\textsuperscript{rd}, 4\textsuperscript{th} 
                                        && 2\textsuperscript{nd}, 3\textsuperscript{rd}, 4\textsuperscript{th}      \\                                        
\emph{Consumption data:}                & CEX           & CEX           & CEX           && CEX                      \\
                                        & (1)           & (2)           & (3)           && (4)                      \\
\midrule                                        
$\phi^{(1)}$                            &0.288          &0.339          &0.404          &&0.260\\
                                        &(0.044)        &(0.074)        &(0.118)        &&(0.043)\\
$\psi^{(1)}$                            &-0.109         &-0.155         &-0.418         &&-0.108\\
                                        &(0.078)        &(0.098)        &(0.247)        &&(0.086)\\
$\phi^{(2)}$                            &               &               &               &&-0.036\\
                                        &               &               &               &&(0.038)\\
$\psi^{(2)}$                            &               &               &               &&0.025\\
                                        &               &               &               &&(0.045)\\
$\omega^{(22)}$                         &               &               &               &&-0.157\\
                                        &               &               &               &&(1.195)\\                 
\midrule        
$\sigma^2_{\xi}$                        &0.019          &0.018          &0.000          &&0.019\\
                                        &(0.004)        &(0.004)        &(0.004)        &&(0.004)\\
$\gamma_{\xi}$                          &               &-0.016         &-0.016         &&  \\
                                        &               &(0.007)        &(0.007)        &&  \\
$\kappa_{\xi}$                          &               &               &0.000          &&  \\
                                        &               &               &(0.000)        &&  \\
\midrule        
$\sigma^2_{u_c}$                        &0.140          &0.140          &0.152          &&0.140\\
                                        &(0.009)        &(0.009)        &(0.012)        &&(0.009)\\
$\gamma_{u_c}$                          &               &-0.185         &-0.185         &&\\
                                        &               &(0.037)        &(0.037)        &&\\
$\kappa_{u_c}$                          &               &               &0.816          &&\\
                                        &               &               &(0.183)        &&\\                        
\bottomrule
\end{tabular}
\caption*{\fsz\emph{Notes:} The table presents the estimates of the parameters of the consumption function, imposing homogeneous transmission parameters over the lifecycle/in the cross-section and stationarity of taste heterogeneity and consumption measurement error. Columns 1-3 present parameter estimates in the linear function, while column 4 presents estimates in the quadratic case; the order of moments targeted in each case is shown at the top of the table. All columns use consumption imputed from the CEX (see appendix \ref{SubAppendix::Empirics_CEX}). Estimation is via equally weighted GMM; block bootstrap standard errors are in parentheses. This table accompanies table \ref{Table::Consumption_function} in the paper, which shows the baseline results using consumption internally available in the PSID (column 2 in table \ref{Table::Consumption_function} is the same as column 1 in the present table).}
\end{center}
\end{table}

Table \ref{AppTable::Linear_Consumption_function_subsamples} presents estimation results for the linear consumption function in the subsamples formed by wealth and education. The odd-numbered columns present parameter estimates when only second-order moments are targeted. $\phi^{(1)}$ becomes monotonically smaller as we move from group 1 (low wealth, no college) to group 4 (high wealth, some college), reflecting the insurance role of wealth, which increases monotonically across groups. This is similar to what we find in the quadratic specification. However, as soon as we target up to fourth-order moments in the even-numbered columns, this monotonic pattern is lost. Transitory shocks transmit strongly in group 1, and with a negative coefficient.

Table \ref{AppTable::Income_process_rho} presents the estimates of the income process, when permanent income follows an AR(1) process, i.e., it features linear persistence. Table \ref{AppTable::Consumption_function_rho} presents the associated estimates of the parameters of the linear and quadratic consumption functions.

\begin{sidewaystable}[t!]    
\begin{center}
\caption{Estimates of the income process, by age and position in income distribution}\label{AppTable::Income_process_age_lagY}
\begin{tabular}{ll L{2.5cm} C{2.1cm} C{2.1cm} C{2.1cm} c C{3.0cm} C{3.0cm}}
\toprule
            &                           &               & \mcl{3}{c}{Age}                                   && \mcl{2}{c}{Position in income distribution $y_{it-1}$}  \\
            &                           &               & 30--40        & 41--50            & 51--65        && Bottom tertile   & Top tertile   \\
\cmidrule{4-6}\cmidrule{8-9}
\mcl{3}{l}{\emph{Income moments:}}                      & 2\textsuperscript{nd}, 3\textsuperscript{rd}, 4\textsuperscript{th}
                                                        & 2\textsuperscript{nd}, 3\textsuperscript{rd}, 4\textsuperscript{th}
                                                        & 2\textsuperscript{nd}, 3\textsuperscript{rd}, 4\textsuperscript{th} &
                                                        & 2\textsuperscript{nd}, 3\textsuperscript{rd}, 4\textsuperscript{th}
                                                        & 2\textsuperscript{nd}, 3\textsuperscript{rd}, 4\textsuperscript{th} \\
\mcl{3}{l}{\emph{Income data:}}                         & PSID          & PSID              & PSID          && PSID             & PSID          \\
&&                                                      & (1)           & (2)               & (3)           && (4)              & (5)           \\
\midrule
\mcl{9}{c}{\textbf{Panel A. Permanent shocks}}              \\
Variance    & $\sigma_{\zeta}^2$        &               &0.028          &0.026              &0.034          &&0.037             &0.032          \\
            &                           &               &(0.004)        &(0.004)            &(0.005)        &&(0.004)           &(0.003)        \\
Skewness    & $\gamma_{\zeta}$          & central       &-0.003         &0.004              &-0.015         &&0.010             &-0.028         \\
            &                           &               &(0.006)        &(0.005)            &(0.006)        &&(0.005)           &(0.004)        \\
            &                           & standardized  &-0.653         &0.926              &-2.386         &&1.344             &-4.851         \\
            &                           &               &(1.291)        &(1.114)            &(1.167)        &&(0.783)           &(0.987)        \\
Kurtosis    & $\kappa_{\zeta}$          & central       &0.043          &0.030              &0.058          &&0.052             &0.039          \\
            &                           &               &(0.014)        &(0.013)            &(0.016)        &&(0.013)           &(0.008)        \\
            &                           & standardized  &56.275         &45.537             &50.945         &&37.627            &36.974         \\
            &                           &               &(23.787)       &(22.709)           &(21.605)       &&(13.058)          &(10.586)       \\
\noalign{\smallskip}
\mcl{9}{c}{\textbf{Panel B. Transitory shocks}}                 \\
Variance    & $\sigma_{v}^2$            &               &0.033          &0.027              &0.036          &&0.020             &0.020          \\
            &                           &               &(0.005)        &(0.004)            &(0.005)        &&(0.004)           &(0.003)        \\
Skewness    & $\gamma_{v}$              & central       &-0.008         &-0.005             &-0.010         &&-0.024            &0.021          \\
            &                           &               &(0.006)        &(0.005)            &(0.005)        &&(0.005)           &(0.004)        \\
            &                           & standardized  &-1.427         &-1.178             &-1.453         &&-8.246            &7.333          \\
            &                           &               &(1.057)        &(1.088)            &(0.782)        &&(3.074)           &(2.235)        \\
Kurtosis    & $\kappa_{v}$              & central       &0.056          &0.040              &0.050          &&0.014             &0.011          \\
            &                           &               &(0.013)        &(0.007)            &(0.008)        &&(0.008)           &(0.006)        \\
            &                           & standardized  &52.863         &53.730             &38.832         &&33.974            &27.286         \\
            &                           &               &(19.670)       &(16.619)           &(12.566)       &&(24.531)          &(16.935)       \\
\bottomrule
\end{tabular}
\caption*{\fsz\emph{Notes:} The table presents the estimates of the parameters of the income process, allowing them to vary over ages 30--40, 41--50, and 51--65 of the male spouse, and, separately, by the household's position in the income distribution (bottom and top tertiles of the distribution of lag income $y_{it-1}$). We target all income moments up to fourth-order; results when targeting only lower order moments are available upon request. Estimation is via equally weighted GMM; GMM standard errors clustered at the household level are in parentheses. This table accompanies tables \ref{Table::Consumption_function_age} and \ref{Table::Consumption_function_lagY} in the paper, as well as the following tables \ref{AppTable::Linear_Consumption_function_age} and \ref{AppTable::Linear_Consumption_function_lagY}.}
\end{center}
\end{sidewaystable}

\begin{table}[t!]  
\begin{center}
\caption{Estimates of the linear consumption function, by age}\label{AppTable::Linear_Consumption_function_age}
\begin{tabular}{l c R{1.4cm} L{1.4cm} R{1.4cm} L{1.4cm} R{1.4cm} L{1.4cm}}
\toprule
\mcl{2}{l}{\emph{Consumption fn.:}}     & \mcl{6}{c}{\textbf{Linear}}                   \\
\cmidrule{3-8}
\mcl{2}{l}{\emph{Income moments:}}      & \mcl{2}{c}{2\textsuperscript{nd}}             & \mcl{2}{c}{2\textsuperscript{nd}, 3\textsuperscript{rd}}     
                                        & \mcl{2}{c}{2\textsuperscript{nd}, 3\textsuperscript{rd}, 4\textsuperscript{th}}   \\                                        
\mcl{2}{l}{\emph{Consumption data:}}    & \mcl{2}{c}{PSID}          & \mcl{2}{c}{PSID}          & \mcl{2}{c}{PSID}          \\
&                                       & \mcl{2}{c}{(1)}           & \mcl{2}{c}{(2)}           & \mcl{2}{c}{(3)}           \\
\midrule                                        
$\phi^{(1)}$        & 30-40             &0.154&(0.045)              &0.164&(0.051)              &0.089&(0.058)              \\
                    & 41-50             &0.126&(0.051)              &0.146&(0.055)              &0.079&(0.089)              \\
                    & 51-65             &0.180&(0.049)              &0.175&(0.051)              &0.136&(0.061)              \\
$\psi^{(1)}$        & 30-40             &0.094&(0.079)              &0.083&(0.073)              &0.095&(0.107)              \\
                    & 41-50             &0.010&(0.079)              &-0.016&(0.072)             &-0.075&(0.061)             \\
                    & 51-65             &-0.046&(0.070)             &-0.044&(0.066)             &0.037&(0.101)              \\
\midrule                        
$\sigma^2_{\xi}$    & 30-40             &0.014&(0.002)              &0.014&(0.002)              &0.016&(0.002)              \\
                    & 41-50             &0.022&(0.002)              &0.022&(0.002)              &0.022&(0.002)              \\
                    & 51-65             &0.021&(0.002)              &0.021&(0.002)              &0.023&(0.002)              \\
$\gamma_{\xi}$      & 30-40             &&                          &-0.001&(0.001)             &-0.001&(0.001)             \\
                    & 41-50             &&                          &0.006&(0.001)              &0.006&(0.002)              \\
                    & 51-65             &&                          &-0.006&(0.002)             &-0.006&(0.002)             \\
$\kappa_{\xi}$      & 30-40             &&                          &&                          &0.004&(0.002)              \\
                    & 41-50             &&                          &&                          &0.007&(0.002)              \\
                    & 51-65             &&                          &&                          &0.006&(0.002)              \\
\midrule                        
$\sigma^2_{u_c}$    & 30-40             &0.046&(0.003)              &0.046&(0.003)              &0.046&(0.003)              \\
                    & 41-50             &0.040&(0.002)              &0.040&(0.002)              &0.040&(0.002)              \\
                    & 51-65             &0.043&(0.002)              &0.043&(0.002)              &0.043&(0.002)              \\
$\gamma_{u_c}$      & 30-40             &&                          &0.000&(0.002)              &0.000&(0.002)              \\
                    & 41-50             &&                          &0.002&(0.002)              &0.002&(0.002)              \\
                    & 51-65             &&                          &0.004&(0.002)              &0.004&(0.002)              \\
$\kappa_{u_c}$      & 30-40             &&                          &&                          &0.014&(0.003)              \\
                    & 41-50             &&                          &&                          &0.012&(0.002)              \\
                    & 51-65             &&                          &&                          &0.012&(0.002)              \\
\bottomrule
\end{tabular}
\caption*{\fsz\emph{Notes:} The table presents the estimates of the parameters of the linear consumption function, allowing them to vary over ages 30--40, 41--50, and 51--65 of the male spouse. The underlying income moments also vary over these age brackets (table \ref{AppTable::Income_process_age_lagY}). All columns use consumption data internally available in the PSID. Estimation is via equally weighted GMM; block bootstrap standard errors are in parentheses. This table accompanies table \ref{Table::Consumption_function_age} in the paper, which shows results from the quadratic specification.}
\end{center}
\end{table}

\begin{table}[t!]  
\begin{center}
\caption{Estimates of the linear consumption function, over position in income distribution}\label{AppTable::Linear_Consumption_function_lagY}
\begin{tabular}{l C{1.3cm} C{1.5cm} C{2.1cm} c C{1.3cm} C{1.5cm} C{2.1cm}}
\toprule
\emph{Consumption fn.:}                 & \mcl{7}{c}{\textbf{Linear}} \\
\cmidrule{2-8}
\emph{Lag income $y_{it-1}$:}           & \mcl{3}{c}{Bottom tertile}                    && \mcl{3}{c}{Top tertile} \\
\cmidrule{2-4}\cmidrule{6-8}
\emph{Income moments:}                  & 2\textsuperscript{nd} & 2\textsuperscript{nd}, 3\textsuperscript{rd}  & 2\textsuperscript{nd}, 3\textsuperscript{rd}, 4\textsuperscript{th} &
                                        & 2\textsuperscript{nd} & 2\textsuperscript{nd}, 3\textsuperscript{rd}  & 2\textsuperscript{nd}, 3\textsuperscript{rd}, 4\textsuperscript{th}\\
\emph{Consumption data:}                & PSID          & PSID          & PSID          && PSID         & PSID          & PSID          \\
                                        & (1)           & (2)           & (3)           && (4)          & (5)           & (6)           \\
\midrule                                        
$\phi^{(1)}$                            &0.103          &0.083          &0.063          &&0.102         &0.060          &0.036          \\
                                        &(0.031)        &(0.032)        &(0.035)        &&(0.036)       &(0.034)        &(0.041)        \\
$\psi^{(1)}$                            &-0.045         &0.013          &0.025          &&-0.004        &0.101          &0.071          \\
                                        &(0.086)        &(0.038)        &(0.054)        &&(0.093)       &(0.048)        &(0.073)        \\
\midrule
$\sigma^2_{\xi}$                        &0.014          &0.014          &0.014          &&0.025         &0.025          &0.026          \\
                                        &(0.002)        &(0.002)        &(0.002)        &&(0.002)       &(0.002)        &(0.002)        \\
$\gamma_{\xi}$                          &               &-0.001         &-0.001         &&              &0.002          &0.002          \\
                                        &               &(0.001)        &(0.001)        &&              &(0.001)        &(0.001)        \\
$\kappa_{\xi}$                          &               &               &0.003          &&              &               &0.009          \\
                                        &               &               &(0.002)        &&              &               &(0.002)        \\
\midrule
$\sigma^2_{u_c}$                        &0.044          &0.044          &0.044          &&0.041         &0.041          &0.041          \\
                                        &(0.002)        &(0.002)        &(0.002)        &&(0.002)       &(0.002)        &(0.002)        \\
$\gamma_{u_c}$                          &               &0.003          &0.003          &&              &0.003          &0.003          \\
                                        &               &(0.002)        &(0.002)        &&              &(0.002)        &(0.002)        \\
$\kappa_{u_c}$                          &               &               &0.011          &&              &               &0.010          \\
                                        &               &               &(0.002)        &&              &               &(0.002)        \\
\bottomrule
\end{tabular}
\caption*{\fsz\emph{Notes:} The table presents the estimates of the parameters of the linear consumption function in the bottom and top tertiles of the distribution of lag income $y_{it-1}$. The underlying income moments also vary over the income distribution (table \ref{AppTable::Income_process_age_lagY}). All columns use consumption data internally available in the PSID. Estimation is via equally weighted GMM; GMM standard errors clustered at the household level are in parentheses. This table accompanies table \ref{Table::Consumption_function_lagY} in the paper, which shows results from the quadratic specification.}
\end{center}
\end{table}

\begin{sidewaystable}[t!]  
\begin{center}
\caption{Estimates of the linear consumption function, by wealth and education}\label{AppTable::Linear_Consumption_function_subsamples}
\begin{tabular}{l C{1.4cm} C{2.1cm} C{1.4cm} C{2.1cm} c C{1.4cm} C{2.1cm} C{1.4cm} C{2.1cm}}
\toprule
\emph{Consumption fn.:}         & \mcl{8}{c}{\textbf{Linear}}                                   \\
\cmidrule{2-10}
                                & \mcl{4}{c}{Low wealth}        && \mcl{4}{c}{High wealth}      \\
\emph{Subsample:}               & \mcl{2}{c}{No college}        & \mcl{2}{c}{Some college}      && \mcl{2}{c}{No college}   & \mcl{2}{c}{Some college}  \\
\cmidrule{2-5}\cmidrule{7-10}
\emph{Income moments:}          &  2\textsuperscript{nd}        &  2\textsuperscript{nd}, 3\textsuperscript{rd}, 4\textsuperscript{th}
                                &  2\textsuperscript{nd}        &  2\textsuperscript{nd}, 3\textsuperscript{rd}, 4\textsuperscript{th}     
                                && 2\textsuperscript{nd}        &  2\textsuperscript{nd}, 3\textsuperscript{rd}, 4\textsuperscript{th} 
                                &  2\textsuperscript{nd}        &  2\textsuperscript{nd}, 3\textsuperscript{rd}, 4\textsuperscript{th} \\                                        
\emph{Consumption data:}        & PSID          & PSID          & PSID          & PSID          && PSID         & PSID          & PSID          & PSID          \\
                                & (1)           & (2)           & (3)           & (4)           && (5)          & (6)           & (7)           & (8)           \\
\midrule                                        
$\phi^{(1)}$                    &0.242          &0.143          &0.198          &0.153          &&0.171         &0.028          &0.099          &0.106          \\
                                &(0.053)        &(0.042)        &(0.053)        &(0.056)        &&(0.071)       &(0.089)        &(0.041)        &(0.055)        \\
$\psi^{(1)}$                    &-0.178         &-0.164         &0.015          &0.073          &&0.056         &0.142          &0.003          &0.004          \\
                                &(0.107)        &(0.077)        &(0.089)        &(0.091)        &&(0.089)       &(0.131)        &(0.059)        &(0.071)        \\          
\midrule                                
$\sigma^2_{\xi}$                &0.014          &0.016          &0.013          &0.014          &&0.014         &0.015          &0.025          &0.026          \\
                                &(0.003)        &(0.003)        &(0.002)        &(0.002)        &&(0.002)       &(0.002)        &(0.002)        &(0.002)        \\
$\gamma_{\xi}$                  &               &-0.003         &               &-0.001         &&              &-0.003         &               &0.000          \\
                                &               &(0.002)        &               &(0.001)        &&              &(0.002)        &               &(0.001)        \\
$\kappa_{\xi}$                  &               &0.005          &               &0.000          &&              &0.000          &               &0.009          \\
                                &               &(0.002)        &               &(0.002)        &&              &(0.002)        &               &(0.002)        \\
\midrule                                
$\sigma^2_{u_c}$                &0.045          &0.045          &0.045          &0.045          &&0.038         &0.038          &0.043          &0.043          \\
                                &(0.003)        &(0.003)        &(0.003)        &(0.003)        &&(0.004)       &(0.004)        &(0.002)        &(0.002)        \\
$\gamma_{u_c}$                  &               &-0.001         &               &0.004          &&              &0.010          &               &0.002          \\
                                &               &(0.003)        &               &(0.002)        &&              &(0.005)        &               &(0.001)        \\
$\kappa_{u_c}$                  &               &0.013          &               &0.013          &&              &0.018          &               &0.011          \\
                                &               &(0.003)        &               &(0.002)        &&              &(0.006)        &               &(0.002)        \\
\bottomrule
\end{tabular}
\caption*{\fsz\emph{Notes:} The table presents the estimates of the parameters of the linear consumption function, allowing them to vary by wealth and education. The underlying income moments also vary by wealth and education (results not shown for brevity). We do not report results when up to third-order moments of income and consumption are targeted, but these are available upon request. Low (high) wealth is defined on the basis of household wealth being less (more) than median real wealth in the sample over $1999$--$2019$. No college versus some college is defined on the basis of the highest level of education attained by the male spouse (cutoff is 12 years of schooling). All columns use consumption data internally available in the PSID. Estimation is via equally weighted GMM; GMM standard errors clustered at the household level are in parentheses. This table accompanies the quadratic specification estimates in table \ref{Table::Quadratic_Consumption_function_subsamples} in the paper.}
\end{center}
\end{sidewaystable}

\begin{table}[t!]  
\begin{center}
\caption{Estimates of the income process, income persistence}\label{AppTable::Income_process_rho}
\begin{tabular}{ll L{2.5cm} R{1.3cm} L{1.3cm} R{1.3cm} L{1.3cm} R{1.5cm} L{1.5cm}}
\toprule
\mcl{3}{l}{\emph{Income moments:}}                      & \mcl{2}{c}{2\textsuperscript{nd}} & \mcl{2}{c}{2\textsuperscript{nd}, 3\textsuperscript{rd}}  & \mcl{2}{c}{2\textsuperscript{nd}, 3\textsuperscript{rd}, 4\textsuperscript{th}}   \\
\mcl{3}{l}{\emph{Income data:}}                         & \mcl{2}{c}{PSID}                  & \mcl{2}{c}{PSID}                                          & \mcl{2}{c}{PSID} \\
&&                                                      & \mcl{2}{c}{(1)}                   & \mcl{2}{c}{(2)}                                           & \mcl{2}{c}{(3)}  \\
\midrule
\mcl{9}{c}{\textbf{Panel A. Permanent shocks}}\\
Variance    & $\sigma_{\zeta}^2$        &               &0.058      &(0.016)                &0.056                  &(0.016)                            &0.036      &(0.006)    \\
Skewness    & $\gamma_{\zeta}$          & central       &           &                       &-0.007                 &(0.003)                            &-0.006     &(0.003)    \\
            &                           & standardized  &           &                       &-0.503                 &(0.337)                            &-0.910     &(0.512)    \\
Kurtosis    & $\kappa_{\zeta}$          & central       &           &                       &                       &                                   &0.039      &(0.007)    \\
            &                           & standardized  &           &                       &                       &                                   &29.891     &(11.459)   \\
\noalign{\smallskip}
\mcl{9}{c}{\textbf{Panel B. Transitory shocks}}\\
Variance    & $\sigma_{v}^2$            &               &0.011      &(0.015)                &0.014                  &(0.014)                            &0.030      &(0.005)    \\
Skewness    & $\gamma_{v}$              & central       &           &                       &-0.005                 &(0.003)                            &-0.006     &(0.003)    \\
            &                           & standardized  &           &                       &-3.272                 &(5.415)                            &-1.178     &(0.694)    \\
Kurtosis    & $\kappa_{v}$              & central       &           &                       &                       &                                   &0.047      &(0.006)    \\
            &                           & standardized  &           &                       &                       &                                   &52.091     &(19.426)   \\
\noalign{\smallskip}
\mcl{9}{c}{\textbf{Panel C. Linear persistence}}\\
AR(1)       & $\rho$                    &               &0.846      &(0.038)                &0.854                  &(0.038)                            &0.931      &(0.028)    \\
\bottomrule
\end{tabular}
\caption*{\fsz\emph{Notes:} The table presents the estimates of the parameters of the AR(1) income process. Column 1 targets only the second moments of income growth; column 2 targets also third moments; column 3 targets all moments up to fourth-order. Estimation is via equally weighted GMM; block bootstrap standard errors are in parentheses. This table accompanies table \ref{Table::Consumption_function_rho} in the paper, as well as table \ref{AppTable::Consumption_function_rho} that follows.}
\end{center}
\end{table}

\begin{table}[t!]  
\begin{center}
\caption{Estimates of the consumption function, income persistence}\label{AppTable::Consumption_function_rho}
\begin{tabular}{l C{1.8cm} C{1.9cm} C{2.1cm} c C{2.4cm}}
\toprule
\emph{Consumption fn.:}             & \mcl{3}{c}{\textbf{Linear}}                           && \textbf{Quadratic} \\
\cmidrule{2-4}\cmidrule{6-6}
\emph{Income moments:}                  & 2\textsuperscript{nd}     & 2\textsuperscript{nd}, 3\textsuperscript{rd}      
                                        & 2\textsuperscript{nd}, 3\textsuperscript{rd}, 4\textsuperscript{th} 
                                        && 2\textsuperscript{nd}, 3\textsuperscript{rd}, 4\textsuperscript{th}      \\                                        
\emph{Consumption data:}                & PSID          & PSID          & PSID          && PSID     \\
                                        & (1)           & (2)           & (3)           && (4)      \\
\midrule                                        
$\phi^{(1)}$                            &0.089          &0.090          &0.110          &&0.112     \\
                                        &(0.029)        &(0.029)        &(0.041)        &&(0.024)   \\
$\psi^{(1)}$                            &-0.139         &-0.077         &-0.013         &&-0.010    \\
                                        &(0.265)        &(0.083)        &(0.077)        &&(0.066)   \\
$\phi^{(2)}$                            &               &               &               &&-0.049    \\
                                        &               &               &               &&(0.026)   \\
$\psi^{(2)}$                            &               &               &               &&0.019     \\
                                        &               &               &               &&(0.037)   \\
$\omega^{(22)}$                         &               &               &               &&0.430     \\
                                        &               &               &               &&(0.847)   \\            
\midrule
$\sigma^2_{\xi}$                        &0.019          &0.019          &0.020          &&0.019     \\
                                        &(0.001)        &(0.001)        &(0.001)        &&(0.001)   \\
$\gamma_{\xi}$                          &               &-0.001         &-0.001         &&\\
                                        &               &(0.001)        &(0.001)        &&\\
$\kappa_{\xi}$                          &               &               &0.031          &&\\
                                        &               &               &(0.002)        &&\\
\midrule
$\sigma^2_{u_c}$                        &0.044          &0.044          &0.044          &&0.044     \\
                                        &(0.002)        &(0.002)        &(0.001)        &&(0.002)   \\
$\gamma_{u_c}$                          &               &0.002          &0.002          &&\\
                                        &               &(0.001)        &(0.001)        &&\\
$\kappa_{u_c}$                          &               &               &0.013          &&\\
                                        &               &               &(0.001)        &&\\                   
\bottomrule
\end{tabular}
\caption*{\fsz\emph{Notes:} The table presents the estimates of the parameters of the consumption function, when income features linear persistence. Columns 1-3 present parameter estimates in the linear function, while column 4 presents estimates in the quadratic case; the order of moments targeted in each case is shown at the top of the table. Estimation is via equally weighted GMM; block bootstrap standard errors are in parentheses. This table accompanies table \ref{Table::Consumption_function_rho} in the paper.}
\end{center}
\end{table}

\FloatBarrier
\section{Some welfare costs of tail income risk}\label{Appendix::WelfareCosts}

Given our empirical, quasi reduced-form approach to measuring partial insurance, we calculate the welfare costs of income risk outside of a fully specified structural model. Specifically, we draw a sequence of income shocks over the lifetime, we use the different consumption specifications to simulate responses to those shocks, and we use parametric utility to calculate welfare based on those consumption responses. Our focus is on the costs of non-Gaussian relative to (counterfactual) Gaussian income risk, and we are particularly interested in contrasting the implications of linear versus nonlinear transmission.

Normal mixtures offer a flexible way of modeling non-Gaussian distributions and have by now become a standard choice in quantitative income dynamics \citep[][]{Guvenen_Karahan_Ozkan_Song2021,GuvenenOzkanMadera2024NonGaussianRisk}. We draw permanent and transitory income shocks for 50,000 households from a mixture of two normal distributions, namely
\begin{equation*}
    \zeta_{it} \sim
        \begin{cases}
            {\cal N}(\mu_{\zeta,1},\sigma_{\zeta,1}) \text{ with prob. } p_{\zeta} \\
            {\cal N}(\mu_{\zeta,2},\sigma_{\zeta,2}) \text{ with prob. } 1-p_{\zeta} \\
        \end{cases}
    \text{ and }~
    v_{it} \sim
        \begin{cases}
            {\cal N}(\mu_{v,1},\sigma_{v,1})  \text{ with prob. } p_{v} \\
            {\cal N}(\mu_{v,2},\sigma_{v,2})  \text{ with prob. } 1-p_{v}. \\
        \end{cases}
\end{equation*}
 There are five parameters per shock $(p, \mu_{1}, \sigma_{1}, \mu_{2}, \sigma_{2})$, which are estimated using the method of simulated moments to match the first four moments of shocks in the PSID (column 3, table \ref{Table::Income_process}). We assume $\mu_{1}<0$ for identification. Table \ref{AppTable::NormalMixturesModel} reports the parameter estimates and the (excellent) model fit. In the counterfactual scenario with Gaussian risk, we fix $p=0$, $\mu_{2}=0$, and set $\sigma_{2}^2$ equal to the empirical variance of each shock.

We simulate alternative consumption profiles in response to the income shocks using the linear and nonlinear specifications
\begin{align*}
    \Delta \ln C_{it} &= \mu^{c}_t + \phi_{it}^{(1)}\zeta_{it} + \psi_{it}^{(1)}v_{it},\\
    \Delta \ln C_{it} &= \mu^{c}_t + \phi_{it}^{(1)}\zeta_{it} + \psi_{it}^{(1)}v_{it} + \phi_{it}^{(2)}\zeta_{it}^2 + \psi_{it}^{(2)}v_{it}^2 + \omega_{it}^{(22)}\zeta_{it}v_{it}.
\end{align*}
These reflect the linear and quadratic consumption functions \eqref{Eq::Consumption_Function_Linear} and \eqref{Eq::Consumption_Function_Quadratic} in the paper, with $\mu^{c}_t = f(\Delta \mathbf{Z}_{it})$ (deterministic profile) and $\xi_{it} = 0$ (mean taste shock). In practice, we let $\mu^{c}_t$ equal the average (non-residualized) consumption growth by year, which we estimate in the data. We simulate consumption over a working life of 35 years, starting at age 30.\footnote{We also simulate a retirement period of 20 years, starting at age 65, in which absence of income shocks implies $\Delta \ln C_{it} = \mu^{c}_t$. We fix the initial condition at age 30 ($\ln C_{it=30}$) to its mean value in the sample. In all the simulations, we restrict $\omega_{it}^{(22)}=0$ as this is never precisely estimated in any cut of the sample.} In the baseline, we use the time-invariant homogeneous partial insurance parameters from table \ref{Table::Consumption_function}, but we subsequently relax time-invariance and homogeneity across households.

\begin{table}[t!]  
\begin{center}
\caption{Normal mixtures, parameter estimates and model fit}\label{AppTable::NormalMixturesModel}
\begin{tabular}{L{2.5cm} L{2cm} L{2.5cm} C{2cm} C{2.5cm} C{2.5cm}}
\toprule
\mcl{6}{c}{\textbf{Panel A. Parameter estimates}}               \\
Permanent   & $p_{\zeta}$               & $\mu_{\zeta,1}$   & $\sigma_{\zeta,1}$    & $\mu_{\zeta,2}$   & $\sigma_{\zeta,2}$    \\
            & 0.058                     & -0.049            & 0.683                 & 0.003             & 0.053                 \\
Transitory  & $p_{v}$                   & $\mu_{v,1}$       & $\sigma_{v,1}$        & $\mu_{v,2}$       & $\sigma_{v,2}$        \\
            & 0.059                     & -0.093            & 0.713                 & 0.006             & 0.020                 \\
\midrule
\mcl{6}{c}{\textbf{Panel B. Model fit permanent shocks}}  \\
            &                           &               &                           & \textbf{\emph{Model}}     & \textbf{\emph{Data}}  \\
Variance    & $\sigma_{\zeta}^2$        &               &                           &0.030              & 0.030     \\
            &                           &               &                           &                   & (0.002)   \\
Skewness    & $\gamma_{\zeta}$          & central       &                           &-0.004             & -0.004    \\
            &                           &               &                           &                   & (0.003)   \\
            &                           & standardized  &                           &-0.757             & -0.831    \\
            &                           &               &                           &                   & (0.571)   \\
Kurtosis    & $\kappa_{\zeta}$          & central       &                           & 0.038             & 0.039     \\
            &                           &               &                           &                   & (0.007)   \\
            &                           & standardized  &                           & 43.146            & 44.393    \\
            &                           &               &                           &                   & (11.080)  \\
\noalign{\smallskip}
\mcl{6}{c}{\textbf{Panel C. Model fit transitory shocks}} \\
            &                           &               &                           & \textbf{\emph{Model}}     & \textbf{\emph{Data}} \\
Variance    & $\sigma_{v}^2$            &               &                           &0.031              & 0.031     \\
            &                           &               &                           &                   &(0.003)\\
Skewness    & $\gamma_{v}$              & central       &                           &-0.008             &-0.008\\
            &                           &               &                           &                   &(0.003)\\
            &                           & standardized  &                           &-1.520             &-1.431\\
            &                           &               &                           &                   &(0.566)\\
Kurtosis    & $\kappa_{v}$              & central       &                           & 0.047             &0.048\\
            &                           &               &                           &                   &(0.005)\\
            &                           & standardized  &                           & 49.630            &49.226\\
            &                           &               &                           &                   &(10.119)\\
\bottomrule
\end{tabular}
\caption*{\fsz\emph{Notes:} The table reports the parameter estimates (panel A) and the model fit (panels B \& C) of the quantitative model for income shocks (mixture of normal distributions). Estimation is done via the method of simulated moments, based on 50,000 random draws for each type of income shock.}
\end{center}
\end{table}

We calculate the (ex-ante) welfare implied by a given consumption profile ${\cal C}_{i} = \{C_{it}\}_{t=0}^{T}$ using constant coefficient of relative risk aversion utility ${\cal U}({\cal C}_{i}) = \sum_{t=0}^{T}\beta^{t}\frac{C_{it}^{1-\gamma}-1}{1-\gamma}$ if $\gamma\neq1$, and logarithmic utility ${\cal U}({\cal C}_{i}) = \sum_{t=0}^{T}\beta^{t}\log (C_{it})$ if $\gamma=1$. We then calculate the welfare cost of tail risk as the percentage reduction in lifetime consumption that a household who faces Gaussian income risk is willing to accept in order to avoid tail (namely, non-Gaussian) risk. This is given by $\alpha_{i}$ for a given household $i$, which, similar to the welfare calculations in \citet{WuKrueger2021_ConsumptionInsurance}, satisfies the indifference equation ${\cal U}\big((1-\alpha_{i}){\cal C}_{i}^{*}\big) = {\cal U}\big({\cal C}_{i}\big)$. Here, ${\cal C}_{i}^{*}$ and ${\cal C}_{i}$ are alternative consumption profiles that reflect \emph{less} (Gaussian) and \emph{more} (non-Gaussian) income risk, respectively. We consider three alternative scenarios over which we calculate the welfare costs of non-Gaussian income risk. 

In a first scenario, both ${\cal C}_{i}^{*}$ and ${\cal C}_{i}$ are simulated from the \emph{linear} specification in response to Gaussian and non-Gaussian income shocks, respectively. The estimate of $\alpha_{i}$ in this case reflects the welfare cost of tail risk in the context of linear transmission. Tail risk is still costly in this context because the probability of very negative shocks (reflecting the long left fat tail) is higher relative to a Gaussian density.

In a second scenario, both ${\cal C}_{i}^{*}$ and ${\cal C}_{i}$ are simulated from the \emph{quadratic} specification in response to Gaussian and non-Gaussian income shocks, respectively. The estimate of $\alpha_{i}$ here reflects the welfare cost of tail risk in the presence of nonlinear transmission. Given the evidence that larger negative shocks have a stronger impact on consumption when nonlinear transmission is allowed for, we expect larger welfare costs of tail risk in this scenario.

In a third scenario, the less risky profile ${\cal C}_{i}^{*}$ reflects the \emph{linear} specification in response to Gaussian risk, while the riskier profile ${\cal C}_{i}$ is simulated from the \emph{quadratic} specification in response to non-Gaussian shocks. Here $\alpha_{i}$ reflects both the welfare cost of tail risk \emph{and} the bias induced by neglecting the possibility of nonlinear transmission of such risk.

Table \ref{AppTable::WelfareCosts} averages $\alpha_{i}$ across households, for $\gamma=2$. The first row presents the unconditional average cost across all households. The following rows present the average costs over subsets of households defined by the magnitude of the worst permanent shock they experience. We also count the fraction of households who face a shock at least as bad, drawn from the normal mixture, \emph{at least once} in their lifetime. Two main points stand out.

\begin{table}[t!]  
\begin{center}
\caption{Welfare costs of non-Gaussian relative to Gaussian income risk}\label{AppTable::WelfareCosts}
\begin{tabular}{L{4cm} c C{2.4cm} C{2.4cm} C{2.4cm}}
\toprule
                                        &                           & \mcl{3}{c}{Average welfare costs ${\E}(\alpha_i)$, $\gamma=2$}             \\
\cmidrule{3-5}
                                        &                           & ${\cal C}_{i}^{*}$ linear     & ${\cal C}_{i}^{*}$ quadratic          & ${\cal C}_{i}^{*}$ linear\\
\emph{Specification:}                   &                           & ${\cal C}_{i}$ linear         & ${\cal C}_{i}$ quadratic              & ${\cal C}_{i}$ quadratic\\
                                        & \emph{\% households}$^\#$ & (1)                           & (2)                                   & (3) \\
\midrule
All households                          & 100\%                     & -0.81                         & -0.88                                 & 0.66                    \\
$\min\{\zeta_{it}\}<-0.50$              & 40.31\%                   & 4.80                          & 6.39                                  & 7.82                    \\
$\min\{\zeta_{it}\}<-0.70$              & 29.11\%                   & 6.36                          & 8.65                                  & 10.04                   \\
$\min\{\zeta_{it}\}<-0.95$              & 17.07\%                   & 8.54                          & 11.89                                 & 13.23                   \\   
\bottomrule
\end{tabular}
\caption*{\fsz\emph{Notes:} The table reports the welfare costs of non-Gaussian relative to Gaussian income risk, for $\gamma=2$. We simulate consumption for 50,000 households using the time-invariant transmission parameters of table \ref{Table::Consumption_function}. The first row presents the average costs across all households. The next rows present the average costs over subsets of households defined by the magnitude of the worst permanent shock they experience. $^\#$\% of households who face a shock at least as bad, drawn from the normal mixture, at least once in their lifetime.}
\end{center}
\end{table}

First, the unconditional cost is small but negative in both the first (linear) and second (nonlinear) scenarios, suggesting households may in fact prefer tail \emph{over} Gaussian risk. While this may seem counterintuitive, table \ref{AppTable::NormalMixturesModel} shows that tail risk entails a small probability (about 5\%) of shocks from a distribution with negative mean and large variance, and a high probability of shocks from one with a small positive mean and very low variance. Most households never actually experience shocks from the former; instead, they face the latter, whose variance is much smaller than that of Gaussian risk. Counterfactually, Gaussian risk implies larger variability for many, so households are \emph{on average} willing to pay to move to the non-Gaussian setting. However, the average cost turns positive in scenario 3, which evaluates the consumption loss of tail risk and the bias from a misspecified consumption function. That is, neglecting non-Gaussian features and nonlinear transmission leads to an overestimation of consumer welfare by 0.66\% of lifetime consumption. 

Second, the welfare costs become substantial for households who actually experience large negative shocks at least once in life -- those drawing from the first distribution in the mixture. Households facing a permanent shock at least $3\sigma$ below the mean ($\min\{\zeta_{it}\}<-0.50$; 40.31\% of households experience this at least once) give up 4.8\% of lifetime consumption to avoid tail risk when risk is transmitted linearly (column 1, second row). This rises to 6.39\% in the presence of nonlinear transmission (column 2, second row), as bad shocks transmit more strongly in this case. For worse events, costs increase further to almost 1/8 or 11.89\% of lifetime consumption among households with at least one $5\sigma$ shock below the mean (column 2, fourth row; $\min\{\zeta_{it}\}<-0.95$; 17.07\% of households experience this at least once), and are higher still in scenario 3 where we evaluate the welfare cost of tail risk and the bias from a misspecified consumption function. They are also (much) higher under a counterfactual \emph{zero risk} scenario, like that explored in \citet{GuvenenOzkanMadera2024NonGaussianRisk}.

These findings recur when we account for heterogeneity in income risk, consumption transmission, or wealth. Table \ref{AppTable::WelfareCostTimeVaryingPhiPsi} calculates the welfare costs allowing for age-specific risk and consumption transmission; the costs are in the same ballpark as the baseline described above. Table \ref{AppTable::WelfareCostWealth} further calculates the welfare costs by wealth. The least wealthy have costs almost twice as large as the baseline, while the wealthier have smaller costs. This gradient is explained by the different degrees of consumption insurance between the two groups. As the likelihood that a borrowing constraint binds depends on wealth and age, the wealth- and age-specific results are indirectly reflective also of such constraints. Without a structural model, we cannot allow households to switch wealth groups endogenously (e.g., in response to an income shock), yet the baseline costs (table \ref{AppTable::WelfareCosts}) lie in the middle of the wealth-specific costs (table \ref{AppTable::WelfareCostWealth}), so exogenous wealth is unlikely to have caused a serious bias in our baseline calculations. The welfare estimates are also robust to small variations in $\gamma$ (table \ref{AppTable::WelfareCostGamma}).

\begin{table}[t!]  
\begin{center}
\caption{Welfare costs of non-Gaussian relative to Gaussian income risk, age-specific income risk and transmission parameters}\label{AppTable::WelfareCostTimeVaryingPhiPsi}
\begin{tabular}{L{4cm} c C{2.4cm} C{2.4cm} C{2.4cm}}
\toprule
                                        &                           & \mcl{3}{c}{Average welfare costs ${\E}(\alpha_i)$, $\gamma=2$}             \\
\cmidrule{3-5}
                                        &                           & ${\cal C}_{i}^{*}$ linear     & ${\cal C}_{i}^{*}$ quadratic          & ${\cal C}_{i}^{*}$ linear\\
\emph{Specification:}                   &                           & ${\cal C}_{i}$ linear         & ${\cal C}_{i}$ quadratic              & ${\cal C}_{i}$ quadratic\\
                                        & \emph{\% households}$^\#$ & (1)                           & (2)                                   & (3) \\
\midrule
All households                          & 100\%                     & -1.02                         & -1.06                                 & 0.31     \\
$\min\{\zeta_{it}\}<-0.50$              & 39.18\%                   & 3.87                          &  6.07                                 & 7.35     \\
$\min\{\zeta_{it}\}<-0.70$              & 29.23\%                   & 5.14                          & 8.29                                  & 9.54     \\
$\min\{\zeta_{it}\}<-0.95$              & 18.28\%                   & 6.79                          & 11.55                                 & 12.75    \\
\bottomrule
\end{tabular}
\caption*{\fsz\emph{Notes:} The table reports the welfare costs of non-Gaussian relative to Gaussian income risk, for $\gamma=2$. We simulate consumption for 50,000 households using age-specific income parameters (table \ref{AppTable::Income_process_age_lagY}) and consumption transmission parameters (table \ref{Table::Consumption_function_age}). The first row presents the average costs across all households. The next rows present the average costs over subsets of households defined by the magnitude of the worst permanent shock they experience. $^\#$\% of households who face a shock at least as bad, drawn from the normal mixture, at least once in their lifetime.}
\end{center}
\end{table}

\begin{table}[b!]  
\begin{center}
\caption{Welfare costs of non-Gaussian relative to Gaussian income risk, by wealth}\label{AppTable::WelfareCostWealth}
\begin{tabular}{L{4cm} c C{2.4cm} C{2.4cm} C{2.4cm}}
\toprule
                                        &                           & \mcl{3}{c}{Average welfare costs ${\E}(\alpha_i)$, $\gamma=2$}             \\
\cmidrule{3-5}
                                        &                           & ${\cal C}_{i}^{*}$ linear     & ${\cal C}_{i}^{*}$ quadratic          & ${\cal C}_{i}^{*}$ linear\\
\emph{Specification:}                   &                           & ${\cal C}_{i}$ linear         & ${\cal C}_{i}$ quadratic              & ${\cal C}_{i}$ quadratic\\
                                        & \emph{\% households}$^\#$ & (1)                           & (2)                                   & (3) \\
\midrule
\mcl{5}{c}{\textbf{Panel A. Low wealth}}\\
All households                          & 100\%                     & -1.58                         & -1.80                                 & 6.14    \\
$\min\{\zeta_{it}\}<-0.50$              & 40.96\%                   & 6.78                          & 7.43                                  & 14.64   \\
$\min\{\zeta_{it}\}<-0.70$              & 26.87\%                   & 9.61                          & 10.63                                 & 17.60   \\
$\min\{\zeta_{it}\}<-0.95$              & 12.99\%                   & 13.12                         & 14.72                                 & 21.36   \\
\noalign{\smallskip}
\mcl{5}{c}{\textbf{Panel B. High wealth} }\\
All households                          & 100\%                     & -0.62                         & -0.69                                 & -0.16   \\
$\min\{\zeta_{it}\}<-0.50$              & 47.18\%                   & 3.06                          & 4.59                                  & 5.09    \\
$\min\{\zeta_{it}\}<-0.70$              & 35.12\%                   & 4.18                          & 6.42                                  & 6.91    \\
$\min\{\zeta_{it}\}<-0.95$              & 21.91\%                   & 5.73                          & 9.06                                  & 9.54    \\
\bottomrule
\end{tabular}
\caption*{\fsz\emph{Notes:} The table reports the welfare costs of non-Gaussian relative to Gaussian income risk, for $\gamma=2$ and low (panel A) versus high wealth (panel B). Low (high) wealth is defined on the basis of household wealth being less (more) than median real wealth over 1999–2019. We simulate consumption for 50,000 households using wealth-specific income and consumption transmission parameters, which, for reasons of brevity, are not reported in the main text. The consumption parameters can be proxied by table \ref{Table::Quadratic_Consumption_function_subsamples}, which reports wealth-and-education-specific results, i.e., an even finer cut of the sample. The first row presents the average costs across all households in the wealth group. The next rows present the average costs over subsets of households defined by the magnitude of the worst permanent shock they experience. $^\#$\% of households who face a shock at least as bad, drawn from the normal mixture, at least once in their lifetime.}
\end{center}
\end{table}

\begin{table}[t!]  
\begin{center}
\caption{Welfare costs of non-Gaussian relative to Gaussian income risk, varying $\gamma$}\label{AppTable::WelfareCostGamma}
\begin{tabular}{L{4cm} c C{2.4cm} C{2.4cm} C{2.4cm}}
\toprule
                                        &                           & \mcl{3}{c}{Average welfare costs ${\E}(\alpha_i)$}             \\
\cmidrule{3-5}
                                        &                           & ${\cal C}_{i}^{*}$ linear     & ${\cal C}_{i}^{*}$ quadratic          & ${\cal C}_{i}^{*}$ linear\\
\emph{Specification:}                   &                           & ${\cal C}_{i}$ linear         & ${\cal C}_{i}$ quadratic              & ${\cal C}_{i}$ quadratic\\
                                        & \emph{\% households}$^\#$ & (1)                           & (2)                                   & (3) \\
\midrule
\mcl{5}{c}{\textbf{Panel A.} $\boldsymbol{\gamma=1}$}\\
All households                          & 100\%                     & -0.86                         & -0.95                                 & 0.64  \\
$\min\{\zeta_{it}\}<-0.50$              & 40.31\%                   & 4.93                          & 6.46                                  & 7.94  \\
$\min\{\zeta_{it}\}<-0.70$              & 29.11\%                   & 6.50                          & 8.71                                  & 10.15 \\
$\min\{\zeta_{it}\}<-0.95$              & 17.07\%                   & 8.69                          & 11.90                                 & 13.29 \\
\noalign{\smallskip}
\mcl{5}{c}{\textbf{Panel B.} $\boldsymbol{\gamma=3}$}\\
All households                          & 100\%                     & -0.76                         & -0.81                                 & 0.66  \\
$\min\{\zeta_{it}\}<-0.50$              & 40.31\%                   & 4.68                          & 6.32                                  & 7.69  \\
$\min\{\zeta_{it}\}<-0.70$              & 29.11\%                   & 6.22                          & 8.59                                  & 9.93 \\
$\min\{\zeta_{it}\}<-0.95$              & 17.07\%                   & 8.40                          & 11.89                                 & 13.17 \\
\bottomrule
\end{tabular}
\caption*{\fsz\emph{Notes:} The table reports the welfare costs of non-Gaussian relative to Gaussian income risk, for $\gamma=1$ (panel A) and $\gamma=3$ (panel B). We simulate consumption for 50,000 households using the time-invariant consumption transmission parameters of table \ref{Table::Consumption_function}. The first row presents the average costs across all households. The next rows present the average costs over subsets of households defined by the magnitude of the worst permanent shock they experience. $^\#$\% of households who face a shock at least as bad, drawn from the normal mixture, at least once in their lifetime.}
\end{center}
\end{table}

\end{document}